\documentclass[prd,preprint,eqsecnum,nofootinbib,amsmath,amssymb,
               tightenlines,superscriptaddress,floatfix]{revtex4}
\usepackage{hyperref}
\hypersetup{
    colorlinks=true,
    linkcolor={red!50!black},
    citecolor={blue!70!black},
    urlcolor={blue!80!black}
}

\usepackage{graphicx}
\usepackage{bm}
\usepackage{amsfonts}
\usepackage{booktabs}

\usepackage{xcolor}

\usepackage{dsfont}
\def\openone{\mathds{1}}

\def\p{{\bm p}}
\def\q{{\bm q}}

\def\P{{\bm P}}

\def\bcalP{{\bm{\mathcal P}}}
\def\bcalT{{\bm{\mathcal T}}}
\def\uOmega{\underline{\Omega}}
\def\Ybar{\overline{Y}}
\def\eps{\epsilon}

\def\Nc{N_{\rm c}}
\def\Nf{N_{\rm f}}
\def\CA{C_{\rm A}}
\def\dA{d_{\rm A}}

\def\alphas{\alpha_{\rm s}}
\def\alphaqed{\alpha_{\scriptscriptstyle\rm EM}}
\def\Re{\operatorname{Re}}

\def\tr{\operatorname{tr}}
\def\sgn{\operatorname{sgn}}
\def\sh{\operatorname{sh}}

\def\cth{\operatorname{cth}}

\def\ix{{\rm i}}
\def\fx{{\rm f}}
\def\xx{{\rm x}}
\def\xbx{{\bar{\rm x}}}
\def\yx{{\rm y}}
\def\ybx{{\bar{\rm y}}}
\def\zx{{\rm z}}

\def\Ax{\ybx}
\def\bx{\yx}

\def\yfrak{{\mathfrak y}}

\def\seq{{\rm seq}}

\def\new{{\rm new}}

\def\gammaE{\gamma_{\rm\scriptscriptstyle E}}

\def\Beta{\operatorname{B}}

\def\xe{x_e}
\def\ye{y_e}

\def\yfrake{{\mathfrak y}_e}
\def\MSbar{\overline{\mbox{MS}}}

\def\bbI{\mathbb{I}}

\def\MSbar{\overline{\mbox{MS}}}
\def\barOmega{\bar\Omega}
\def\qhatA{\hat q_{\rm A}}

\def\virtI{{\rm virt\,I}}
\def\virtIc{{\rm virt\,Ic}}
\def\virtIs{{\rm virt\,Is}}
\def\virtII{{\rm virt\,II}}
\def\virtIIs{{\rm virt\,IIs}}

\def\V{v}
\def\R{r}
\def\sigren{\sigma_{\rm ren}}
\def\sigbare{\sigma_{\rm bare}}
\def\renlog{{\rm ren\,log}}
\def\Abar{{\bar A}}
\def\altT{\bar\tau}
\def\altxi{\bar\xi}
\def\LObar{\overline{\rm LO}}
\def\NLObar{\overline{\rm NLO}}

\newcounter{savefootnote}
\newcommand{\astfootnote}[1]{%
  \setcounter{savefootnote}{\value{footnote}}%
  \setcounter{footnote}{0}%
  \let\oldthefootnote=\thefootnote
  \renewcommand{\thefootnote}{\fnsymbol{footnote}}%
  \footnote{#1}%
  \let\thefootnote=\oldthefootnote%
  \setcounter{footnote}{\value{savefootnote}}%
}

\begin {document}



\title
    {
      The LPM effect in sequential bremsstrahlung: nearly complete results
      for QCD
    }

\author{Peter Arnold}
\affiliation
    {%
    Department of Physics,
    University of Virginia,
    Charlottesville, Virginia 22904-4714, USA
    \medskip
    }%
\author{Tyler Gorda%
\footnote{
  Current address:
  Technische Universit{\"a}t Darmstadt, Department of
  Physics, 64289 Darmstadt, Germany
}}
\affiliation
    {%
    Department of Physics,
    University of Virginia,
    Charlottesville, Virginia 22904-4714, USA
    \medskip
    }%
\author{Shahin Iqbal%
\footnote{
  This work was completed while Shahin Iqbal was on leave from
  the National Centre for Physics, Quaid-i-Azam University Campus,
  Islamabad, Pakistan.
}}
\affiliation
    {%
    Institute of Particle Physics, \\
    Central China Normal University, Wuhan, 430079, China
    \medskip
    }%

\date {\today}

\begin {abstract}%
{%
  The splitting processes of bremsstrahlung and pair production in a medium
  are coherent over large distances in the very high energy limit,
  which leads to a suppression known as the Landau-Pomeranchuk-Migdal
  (LPM) effect.  We continue study of the case when the coherence
  lengths of two consecutive splitting processes overlap (which is
  important for understanding corrections to standard treatments
  of the LPM effect in QCD), avoiding soft-emission approximations.
  Previous work has computed overlap effects for double splitting
  $g \to gg \to ggg$.  To make use of those results, one also needs
  calculations of related virtual loop corrections to
  single splitting $g \to gg$ in order to cancel severe (power-law)
  infrared (IR) divergences.  This paper provides calculations
  of nearly all such processes involving gluons and discusses
  how to organize the results to demonstrate the cancellation.
  In the soft emission
  limit, our results reproduce the known double-log behavior of earlier
  authors who worked in leading-log approximation.
  We also present
  a first (albeit numerical and not yet analytic) investigation of
  sub-leading, single IR logarithms.
  Ultraviolet divergences appearing in our calculations correctly renormalize
  the coupling $\alphas$ in
  the usual LPM result for leading-order $g \to gg$.
}%
\end {abstract}

\maketitle
\thispagestyle {empty}

{\def\boldmath{}\tableofcontents}
\newpage


\section{Introduction}
\label{sec:intro}

When passing through matter, high energy particles lose energy by
showering, via the splitting processes of hard bremsstrahlung and pair
production.  At very high energy, the quantum mechanical duration of
each splitting process, known as the formation time, exceeds the mean
free time for collisions with the medium, leading to a significant
reduction in the splitting rate known as the Landau-Pomeranchuk-Migdal
(LPM) effect \cite{LP,Migdal}.%
\footnote{
  The papers of Landau and Pomeranchuk \cite{LP} are also available in
  English translation \cite{LPenglish}.
  The generalization to QCD was originally carried out by
  Baier, Dokshitzer, Mueller, Peigne, and Schiff \cite{BDMPS12,BDMPS3}
  and by Zakharov \cite{Zakharov}
  (BDMPS-Z).
}
A long-standing problem in field theory has
been to understand how to implement this effect in cases where
the formation times of two consecutive splittings overlap.
The goal of this paper is to (i) present nearly complete results for
the case of two overlapping gluon splittings (e.g.\ $g \to gg \to ggg$)
and (ii)
confirm that earlier leading-log results for these effects
\cite{Blaizot,Iancu,Wu} are reproduced
by our more-complete results in the appropriate soft limit.
As a necessary step, we discuss how to combine the effects of
overlapping real double splitting ($g \to gg \to ggg$) with corresponding
virtual corrections to single splitting (e.g.\ $g \to gg^* \to ggg^* \to gg$)
to cancel spurious infrared (IR) divergences.
In our analysis of virtual corrections, we will also verify that
we reproduce the correct ultraviolet (UV)
renormalization and running of the QCD
coupling $\alphas$ associated with the high-energy
vertex for single splitting.

In this paper, we will present the formulas for the building blocks
just discussed, but we leave application of those formulas to later
work.  In particular, one of the ultimate motivations \cite{qedNfstop}
of our study is to eventually investigate whether the size of overlap
effects is small enough to justify a picture of parton showers, inside
a quark-gluon plasma, as composed of individual high-energy partons; or
whether the splitting of high-energy partons is so strongly-coupled
that high-energy partons lose their individual identity, similar to
gauge-gravity duality studies
\cite{GubserGluon,HIM,CheslerQuark,adsjet12}
of energy loss.  But, as will be discussed
in our conclusion, further work will be needed to answer that question.

As a technical matter, our calculations are organized \cite{QEDnf} using
Light-Cone Perturbation Theory (LCPT) \cite{LB,BL,BPP}.%
\footnote{
  For readers not familiar with time-ordered
  LCPT who would like
  the simplest possible example of how it reassuringly
  reproduces the results of
  ordinary Feynman diagram calculations,
  we recommend section
  1.4.1 of Kovchegov and Levin's monograph \cite{KL}.
}
As we will explain below, the ``nearly'' in our claim of ``nearly complete
results'' refers to the fact that we have not yet calculated,
for QCD, contributions from diagrams that involve ``instantaneous''
interactions in Light-Cone Perturbation Theory.
The effects of such diagrams have been numerically small in
earlier studies of overlap effects in QED \cite{QEDnf}, and they
do not contribute to our check that our results agree with earlier
leading-log calculations.  For these reasons, and because analysis of
the non-instantaneous diagrams is already complicated, we leave the
calculation of instantaneous diagrams for QCD to later work.
For similar reasons, we also leave to later work the effect of
diagrams involving 4-gluon vertices, like those computed for
real double gluon splitting in ref.\ \cite{4point}.

We make a number of simplifying assumptions also
made in the sequence of earlier papers \cite{2brem,seq,dimreg,QEDnf}
leading up to this work:
We take
the large-$\Nc$ limit, assume that the medium is thick compared to
formation lengths, and use the multiple-scattering ($\hat q$)
approximation appropriate to elastic scattering of high-energy partons
from the (thick) medium.  All of these simplifications could be relaxed in
the context of the underlying formalism used for calculations,%
\footnote{
  In particular, for a discussion of how one could in principle
  eliminate the large-$\Nc$ approximation, see refs.\ \cite{color,Vqhat}.
}
but
practical calculations would then be quite considerably harder; so we
focus on the simplest situation here.


\subsection {The diagrams we compute}

\begin {figure}[t]
\begin {center}
  \includegraphics[scale=0.55]{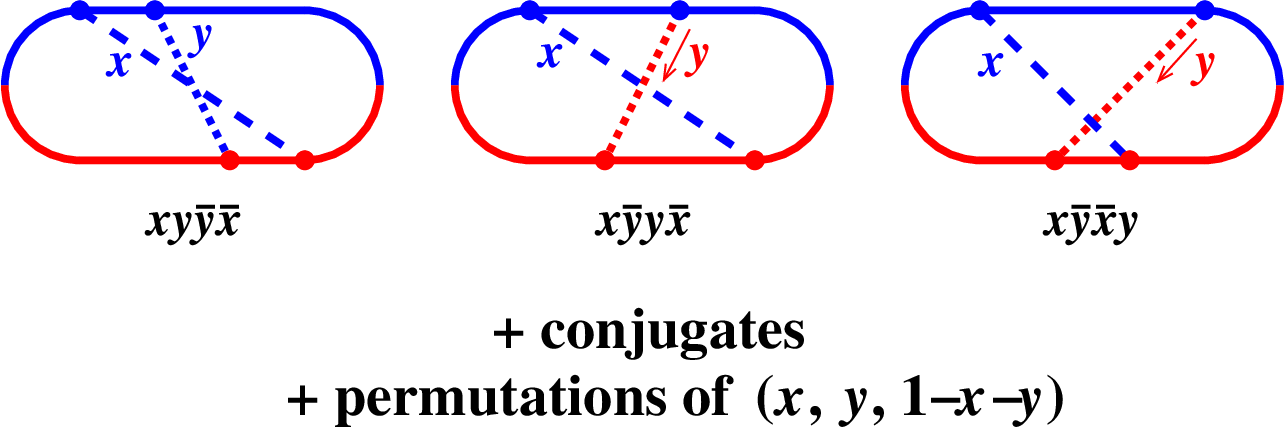}
  \caption{
     \label{fig:crossed}
     ``Crossed'' time-ordered diagrams for the real double gluon splitting rate.
     Labeling of diagrams ($x y \bar y \bar x$, etc.)\ is as in
     ref.\ \cite{2brem}.  All lines in this and other figures represent
     high-energy gluons.
  }
\end {center}
\end {figure}

\begin {figure}[t]
\begin {center}
  \includegraphics[scale=0.55]{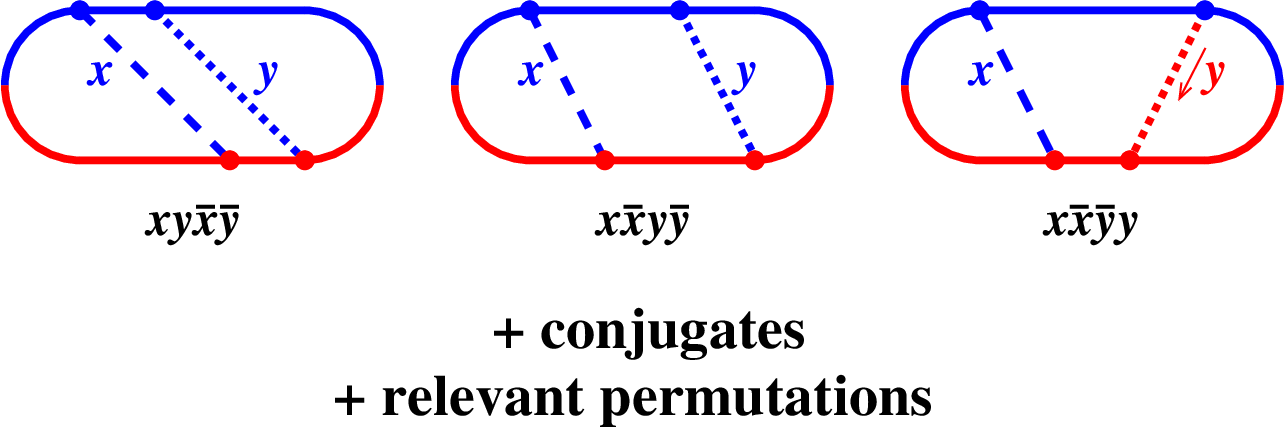}
  \caption{
     \label{fig:seq}
     ``Sequential'' time-ordered diagrams for the real double gluon
     splitting rate \cite{seq}.
  }
\end {center}
\end {figure}

Previous work \cite{2brem,seq,dimreg} has computed overlap effects
for real double gluon splitting ($g \to gg \to ggg$) depicted by the
interference diagrams of figs.\ \ref{fig:crossed} and \ref{fig:seq}.
Each diagram is time-ordered from left to right and
has the following interpretation: The blue (upper)
part of the diagram represents a contribution to the amplitude for
$g \to ggg$, the red (lower) part represents a contribution to the
conjugate amplitude, and the two together represent a particular
contribution to the {\it rate}.  Only high-energy particle lines are
shown explicitly, but each such line is implicitly
summed over an arbitrary number
of interactions with the medium, and the diagram is averaged over
the statistical fluctuations of the medium.
See ref.\ \cite{2brem} for details.
For real double gluon splitting, we will refer to the longitudinal
momentum fractions of the three final-state gluons as $x$, $y$,
and
\begin {equation}
   z \equiv 1{-}x{-}y
\end {equation}
relative to the initial gluon.
Also, our nomenclature is that figs.\ \ref{fig:crossed} and \ref{fig:seq}
are respectively called ``crossed'' and ``sequential'' diagrams
because of the way they are drawn.

For the case of sequential diagrams (fig.\ \ref{fig:seq}),
it is possible for the two consecutive splittings to be arbitrarily
far separated in time, in which case their formation times do not
overlap.  The effect of overlapping formation times in this case
is then determined by subtracting from the sequential diagrams the
corresponding results one would have gotten by treating the two
splittings as independent splittings.  Details are given in
ref.\ \cite{seq}, along with discussion of physical interpretation
and application.%
\footnote{
  See in particular the discussion of section 1.1 of ref.\ \cite{seq}.
}
Whenever such a subtraction needs to be made
on a double-splitting differential rate $d\Gamma$, we will use the
symbol $\Delta\,d\Gamma$ to refer to the subtracted version that
isolates the effect of overlapping formation times.

\begin {figure}[t]
\begin {center}
  \includegraphics[scale=0.55]{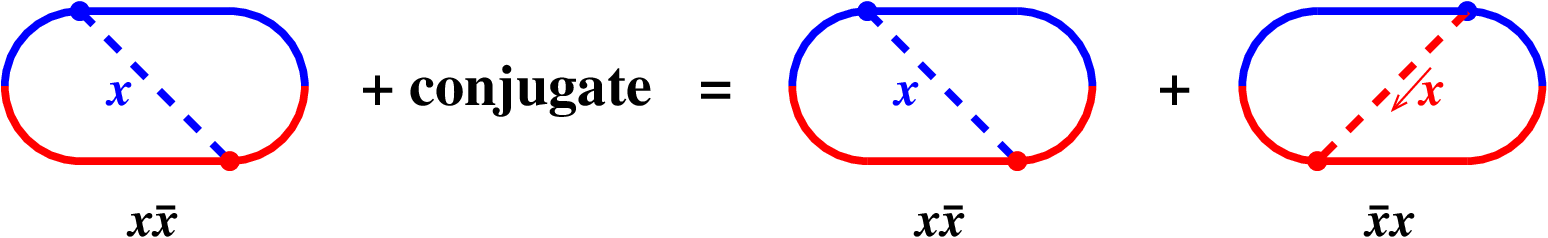}
  \caption{
     \label{fig:LO}
     Time-ordered diagrams for the leading-order rate for
     single gluon splitting.
     \cite{seq}.
  }
\end {center}
\end {figure}

\begin {figure}[t]
\begin {center}
  \includegraphics[scale=0.55]{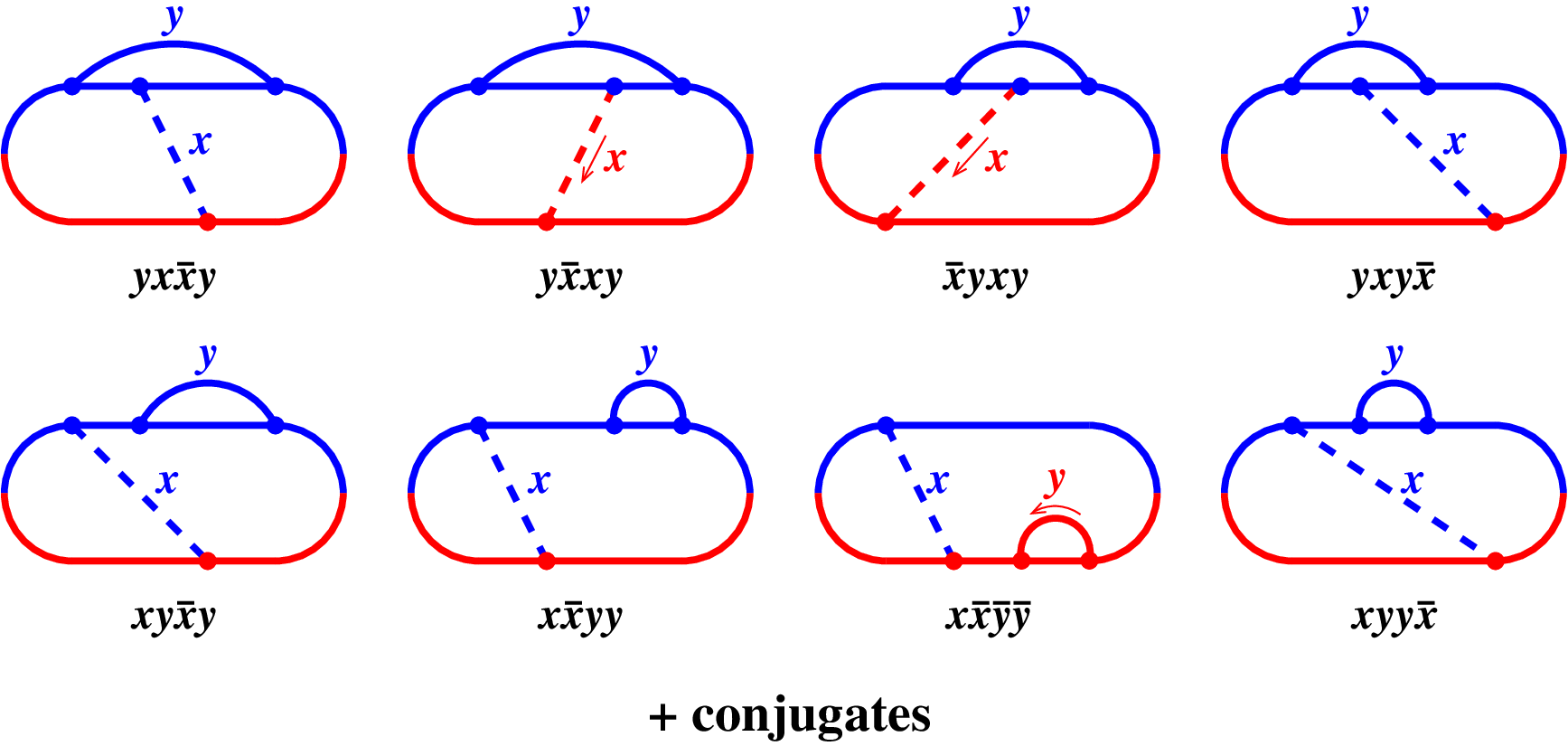}
  \caption{
     \label{fig:virtI}
     Class I one-loop virtual corrections to fig.\ \ref{fig:LO}.
     As with previous figures, not all possible time orderings
     of the diagrams have been shown explicitly but
     the missing orderings are all
     included when one adds in the complex conjugates (``+ conjugates'')
     of the diagrams explicitly shown above.
     Graphically, taking the conjugate flips a diagram about its
     horizontal axis while swapping the colors red and blue.
  }
\end {center}
\end {figure}

\begin {figure}[t]
\begin {center}
  \includegraphics[scale=0.55]{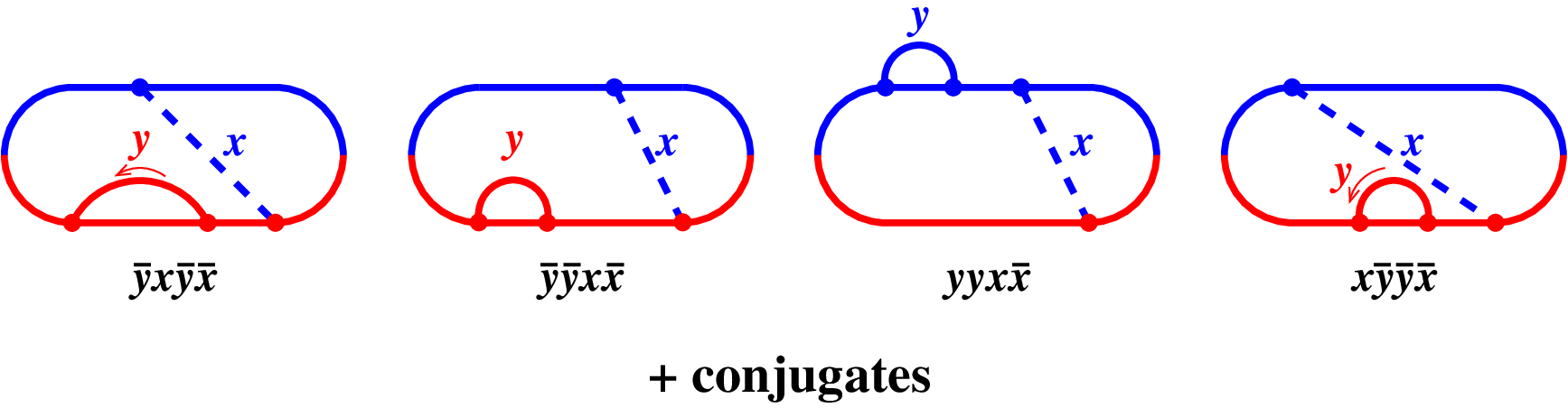}
  \caption{
     \label{fig:virtII}
     Class II one-loop virtual corrections to fig.\ \ref{fig:LO}.
  }
\end {center}
\end {figure}

\begin {figure}[t]
\begin {center}
  \includegraphics[scale=0.55]{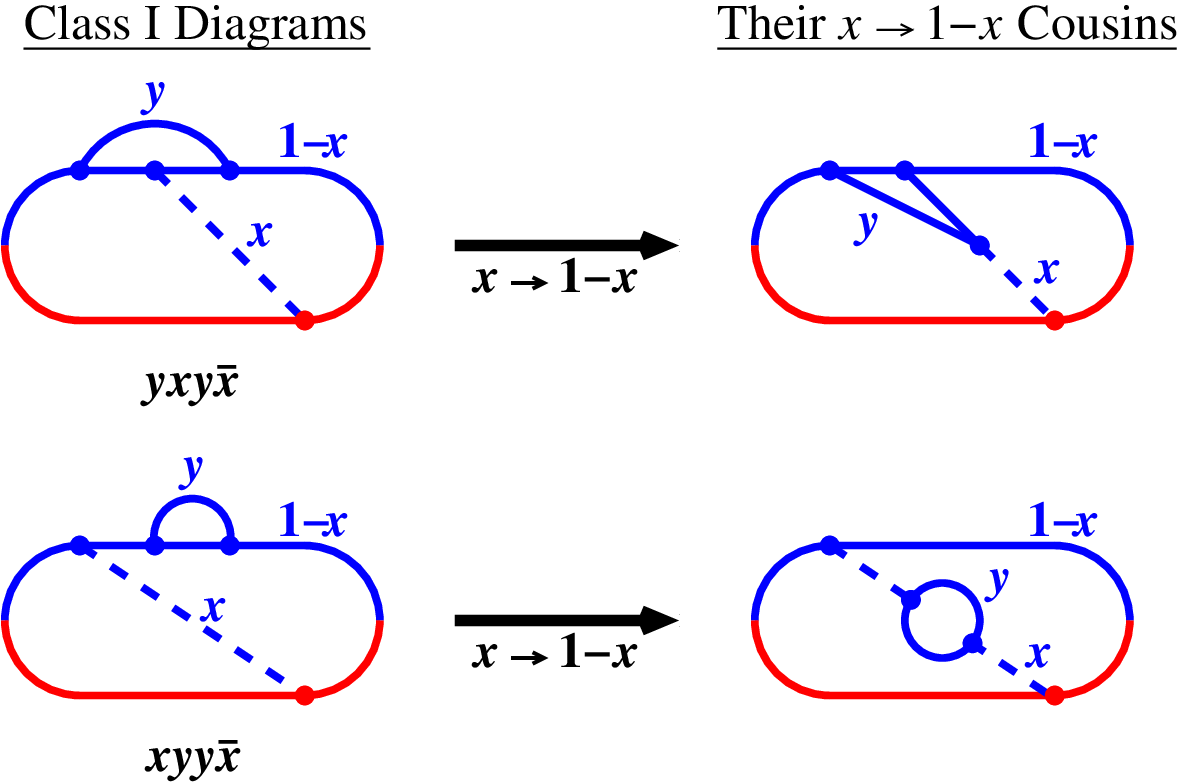}
  \caption{
     \label{fig:virtIb}
     Two examples of diagrams (right) generated by swapping the
     two final-state gluons in Class I diagrams (left) from
     fig.\ \ref{fig:virtI}.  The swap is equivalent to
     replacing $x \to 1{-}x$ in the results for Class I diagrams.
  }
\end {center}
\end {figure}

In the limit that one of the three final-state gluons---say $y$---is soft,
it was found \cite{seq} that the overlap effect on real double splitting
behaves parametrically as%
\footnote{
   See section 1.4 of ref.\ \cite{seq} for a back-of-the-envelope explanation
   of why (\ref{eq:realscaling}) is to be expected.
}
\begin {equation}
   \left[ \Delta \frac{d\Gamma}{dx\,dy} \right]_{g\to ggg}
   \sim \frac{\CA^2 \alphas^2}{x y^{3/2}}
        \left( \frac{\hat q}{E} \right)^{1/2} 
   \qquad \mbox{for $y \ll x < z$.}
\label {eq:realscaling}
\end {equation}
As we'll review later, the $y^{-3/2}$ behavior would lead to
{\it power-law} infrared divergences in energy loss calculations.
Very crudely analogous to what happens in vacuum bremsstrahlung in
QED, where there are (logarithmic) infrared divergences that cancel
in inclusive calculations between real and virtual emissions, we need
to supplement the real double emission processes ($g \to ggg$)
by a calculation of
corresponding virtual corrections to the single emission process
($g \to gg$) of
fig.\ \ref{fig:LO}.  The virtual processes that we calculate in this
paper are shown in fig.\ \ref{fig:virtI} (which we call Class I)
and fig.\ \ref{fig:virtII} (which we call Class II).
There are also cousins of the Class I diagrams
generated by swapping the two final state gluons ($x \to 1{-}x$),
two examples of which are shown in
fig.\ \ref{fig:virtIb}.  For Class II diagrams, such a swap does
not generate a new diagram.

\begin {figure}[t]
\begin {center}
  \includegraphics[scale=0.55]{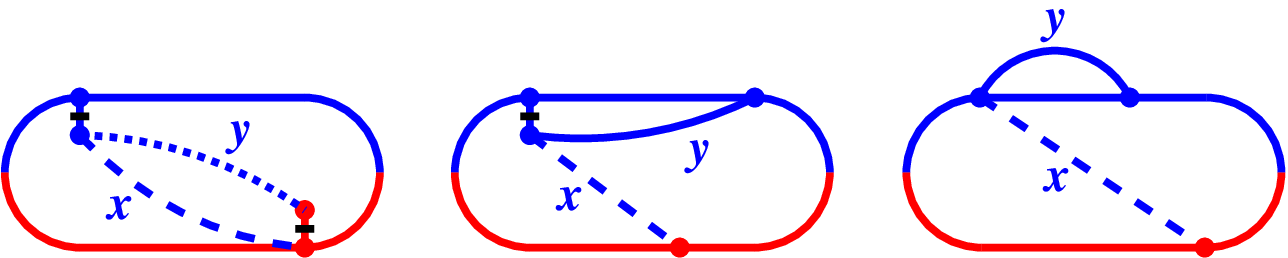}
  \caption{
     \label{fig:later}
     A few examples of diagrams involving either (i) instantaneous interactions
     via longitudinal gluon exchange or (ii) fundamental 4-gluon vertices.
     Longitudinal gluon exchange is represented by a vertical
     (i.e.\ instantaneous) line that is crossed by a black bar, following
     the diagrammatic notation of Light-Cone Perturbation Theory.
  }
\end {center}
\end {figure}

In total, these sets of virtual diagrams include all one-loop
virtual corrections to
single splitting except for processes involving
instantaneous interactions or fundamental 4-gluon vertices.
As mentioned previously, we leave the latter for future work.
A few examples are shown in
fig.\ \ref{fig:later}.  The ``instantaneous'' interactions
(indicated by a propagator crossed by a bar) are instantaneous in
light-cone time and correspond to the exchange of a
longitudinally-polarized gluon in light-cone gauge.
See ref.\ \cite{QEDnf} for examples of such diagrams evaluated in QED.

We should clarify that, physically,
the power-law divergences of (\ref{eq:realscaling})
as $y {\to} 0$ are not actually infinite.  The scaling (\ref{eq:realscaling})
depends on the $\hat q$ approximation, which breaks down when the
soft gluon energy $yE$ becomes as small as the plasma temperature $T$.%
\footnote{
  This will be discussed again later, in section \ref{sec:scales}.
}
In the high-energy limit, however, the cancellation of such
power-law contributions to shower development, even if
only a cancellation of contributions that are
parametrically large in energy rather than truly infinite,
will be critical to extracting the relevant physics that survives after
the cancellation.  In this paper, we will be able to ignore the
far-infrared physics (meaning scale $T \ll E$) that regulates the
power-law divergences and can simply analyze the cancellation
of power-law divergences in the
context of the $\hat q$ approximation appropriate for the
high-energy behavior.


\subsection {Infrared Divergences}
\label {sec:introIR}

We will later discuss the calculation of the differential rates
\begin {equation}
   \left[ \Delta \frac{d\Gamma}{dx\>dy} \right]_{g\to ggg} ,
   \qquad
   \left[ \Delta \frac{d\Gamma}{dx} \right]_{\virtI} ,
   \qquad
   \left[ \Delta \frac{d\Gamma}{dx} \right]_{\virtII} ,
\end {equation}
associated respectively with the real
double emission diagrams
of fig.\ \ref{fig:crossed} plus fig.\ \ref{fig:seq},
the Class I virtual correction diagrams of fig.\ \ref{fig:virtI}, and
the Class II virtual correction diagrams of fig.\ \ref{fig:virtII}.
But here we first preview some results concerning
infrared divergences.

In the virtual diagrams of figs.\ \ref{fig:virtI} and \ref{fig:virtII},
the virtual loop longitudinal momentum fraction $y$
in the amplitude or conjugate
amplitude needs to be integrated over, and it will be convenient to
introduce the notation $[d\Gamma/dx\,dy]_{\virtI}$ and
$[d\Gamma/dx\,dy]_{\virtII}$ for the corresponding integrands of that
$y$ integration.  Our calculations are performed in Light Cone Perturbation
Theory, in which every particle line (virtual as well as real) is
restricted to {\it positive} longitudinal momentum fraction.
The structure of the Class I diagrams of fig.\ \ref{fig:virtI} then
forces $0 < y < 1{-}x$, whereas  the structure of the Class II diagrams
of fig.\ \ref{fig:virtII} forces $0 < y < 1$ instead.
So, in our notation,
\begin {equation}
  \left[ \Delta \frac{d\Gamma}{dx} \right]_{\virtI}
  = \int_0^{1-x} dy \>
      \left[ \Delta \frac{d\Gamma}{dx\,dy} \right]_{\virtI}
  \quad \text{and} \quad
  \left[ \Delta \frac{d\Gamma}{dx} \right]_{\virtII}
  = \int_0^1 dy \>
      \left[ \Delta \frac{d\Gamma}{dx\,dy} \right]_{\virtII} .
\label {eq:virtIntegrands}
\end {equation}

We will later give detailed discussion of how infrared divergences appear
in various calculations associated with shower development,
but a good starting point is to consider the net rate
$[d\Gamma/dx]_{\rm net}$ at which all
of the processes represented by figs.\ \ref{fig:crossed}--\ref{fig:virtIb}
produce one daughter of energy $xE$ (plus any other daughters)
from a particle of energy $E$, for a given $x$.
That's given by
\begin {subequations}
\label {eq:dGnet}
\begin {equation}
   \left[ \frac{d\Gamma}{dx} \right]_{\rm net}
   =
   \left[ \frac{d\Gamma}{dx} \right]^{\LObar}
   +
   \left[ \frac{d\Gamma}{dx} \right]_{\rm net}^{\NLObar}
\end {equation}
where the first term is the rate of the leading-order (LO)
$g\to gg$ process of fig.\ \ref{fig:LO}, and where
the next-to-leading-order (NLO) contribution is%
\footnote{
  Here and throughout, the terms leading-order and next-to-leading-order
  refer to expansion in the $\alphas(Q_\perp)$ associated with each
  splitting vertex for high-energy partons and not to the $\alphas(T)$ that
  controls whether the quark-gluon plasma is strongly or weakly coupled.
}
\begin {align}
   \left[ \frac{d\Gamma}{dx} \right]_{\rm net}^{\NLObar}
   &=
   \left[ \frac{d\Gamma}{dx} \right]_{g\to gg}^{\NLObar}
   +
   \frac12
   \int_0^{1{-}x} dy \>
      \left[ \Delta \frac{d\Gamma}{dx\,dy} \right]_{g\to ggg}
\nonumber\\[5pt]
   &=
     \biggl(
       \int_0^{1-x} dy \, \left[ \Delta \frac{d\Gamma}{dx\,dy} \right]_\virtI
     \biggr)
     + (x \to 1{-}x)
\nonumber\\ & \hspace{10em}
   +
   \int_0^1 dy \, \left[ \Delta \frac{d\Gamma}{dx\,dy} \right]_\virtII
   +
   \frac12
   \int_0^{1{-}x} dy \>
      \left[ \Delta \frac{d\Gamma}{dx\,dy} \right]_{g\to ggg} .
\label {eq:dGnetNLOdef}
\end {align}
\end {subequations}
[See appendix \ref{app:details} for more discussion.]
The bars above $\LObar$ and $\NLObar$ in (\ref{eq:dGnet}) are
a technical distinction that will
be discussed later and can be ignored for now.

In the integrals above, some virtual or final particle has zero
energy at both the lower {\it and} upper limits of the
$y$ integrations, and so both limits are associated with
infrared divergences.  In order to see how divergences behave,
it is convenient to use symmetries and/or change of integration
variables to rewrite the integrals so that the infrared divergences
of $[d\Gamma/dx]_{\rm net}^{\rm NLO}$ are associated {\it only} with
$y \to 0$ (for fixed non-zero $x < 1$).
In particular, (\ref{eq:dGnetNLOdef}) can be rewritten
[see appendix \ref{app:details} for details] as
\begin {align}
   \left[ \frac{d\Gamma}{dx} \right]_{\rm net}^{\NLObar}
   &=
     \biggl(
       \int_0^{1-x} dy \,
       \left\{
         \left[ \Delta \frac{d\Gamma}{dx\,dy} \right]_\virtI
         +
         \left[ \Delta \frac{d\Gamma}{dx\,dy} \right]_\virtII
       \right\}
     \biggr)
     + (x \to 1{-}x)
\nonumber\\ & \hspace{10em}
   +
   \frac12
   \int_0^{1{-}x} dy \>
      \left[ \Delta \frac{d\Gamma}{dx\,dy} \right]_{g\to ggg}
\label {eq:dGnetNLO1}
\end {align}
and thence [appendix \ref{app:details}]
\begin {align}
   \left[ \frac{d\Gamma}{dx} \right]_{\rm net}^{\NLObar}
   &=
   \int_0^{1/2} dy \>
   \Bigl\{
      \bigl[ \V(x,y) \, \theta(y<\tfrac{1-x}{2}) \bigr]
      + [x \to 1-x]
      + \R(x,y) \, \theta(y<\tfrac{1-x}{2})
   \Bigr\}
\nonumber\\
  &=
   \int_0^{1/2} dy \>
   \Bigl\{
      \V(x,y) \, \theta(y<\tfrac{1-x}{2})
      + \V(1{-}x,y) \, \theta(y<\tfrac{x}{2})
      + \R(x,y) \, \theta(y<\tfrac{1-x}{2})
   \Bigr\}
 ,
\label {eq:dGnetNLO}
\end {align}
where contributions from virtual and real double splitting
processes appear in the respective combinations
\begin {subequations}
\label {eq:VRdef}
\begin {align}
  \V(x,y) &\equiv 
  \left(
     \left[ \Delta \frac{d\Gamma}{dx\,dy} \right]_\virtI
     + \left[ \Delta \frac{d\Gamma}{dx\,dy} \right]_\virtII
  \right)
  + ( y \leftrightarrow z ) ,
\label {eq:Vdef}
\\
  \R(x,y) & \equiv
  \left[ \Delta \frac{d\Gamma}{dx\,dy} \right]_{g\to ggg} .
\label {eq:Rdef}
\end {align}
\end {subequations}
The $\theta(\cdots)$ in (\ref{eq:dGnetNLO}) represent unit step
functions [$\theta(\mbox{true})=1$ and $\theta(\mbox{false})=0$],
and they just implement upper limits on the $y$ integration.
The advantage of using the $\theta$ functions is so that we
can combine all the integrals: the
integrals for the separate terms each have power-law
IR divergences, but whether or not those divergences cancel
is now just a question of the $y\to 0$ behavior of the combined
integrand of (\ref{eq:dGnetNLO}).

In the limit $y\to 0$ for fixed $x$, the integrand of
(\ref{eq:dGnetNLO}) approaches
\begin {equation}
   \V(x,y) + \V(1{-}x,y) + \R(x,y) .
\label {eq:VVR}
\end {equation}
Using the symmetry of the $g\to ggg$ rate (\ref{eq:Rdef})
under permutations of $x$, $y$, and $z=1{-}x{-}y$,
we have
$\R(x,y) = \R(1{-}x{-}y,y) \simeq \R(1{-}x,y)$ for small $y$, and
so (\ref{eq:VVR}) approaches
\begin {equation}
   \bigl[ \V(x,y) + \tfrac12 \R(x,y) \bigr] + [x \to 1{-}x] .
\label {eq:VR}
\end {equation}
By (\ref{eq:realscaling}), $\R(x,y) \sim y^{-3/2}$ for small $y$,
and so the integral of $\R(x,y)$ in (\ref{eq:dGnetNLO}) has
a power-law IR divergence proportional to $\int_0 dy/y^{3/2}$.
From the full results for rates that we calculate in this paper, we find
that the $y^{-3/2}$ behavior cancels in the combination
$\V(x,y)+\tfrac12 \R(x,y)$ appearing in (\ref{eq:VR}).
We also find that left behind after this cancellation is,
at leading logarithmic order,
\begin {equation}
   \V(x,y) + \tfrac12 \R(x,y)
   \approx
   -\frac{\CA\alphas}{8\pi}
   \left[ \frac{d\Gamma}{dx} \right]^{\rm LO} \frac{\ln y}{y}
   \,,
\label {eq:VRlimit}
\end {equation}
which generates an IR double log divergence when integrated over
$y$.  As we discuss later, this result, applied to (\ref{eq:dGnetNLO}),
exactly matches leading-log
results derived earlier in the literature \cite{Blaizot,Iancu,Wu}
and so provides a crucial check of our calculations.

Though it should be possible to extract (\ref{eq:VRlimit}) from
our results analytically, so far we have only checked numerically.%
\footnote{
  Analytic extraction of double and single IR
  logs directly from our full rate formulas
  is
  complicated because diagram by diagram the logs are
  subleading to the power-law IR divergences, and the latter
  are already complicated to extract analytically from our
  results.  Interested readers can see a painful
  example in appendix \ref{app:Gxi}.
} 
Fig.\ \ref{fig:dbllogCheck} shows a plot of our full results for
\begin {equation}
  \frac{
     \V(x,y) + \tfrac12 \R(x,y)
     \vphantom{\Big|}
  }{
     \frac{\CA\alphas}{8\pi}
     \left[ \frac{d\Gamma}{dx} \right]^{\rm LO} \frac{1}{y}
     \vphantom{\Big|}
  }
\label {eq:VRratio}
\end {equation}
vs.\ $\ln y$ for a sample value of $x$.
According to (\ref{eq:VRlimit}),
the slope of (\ref{eq:VRratio}) vs.\ $\ln y$ should approach $-1$
as $\ln y \to -\infty$, which we show in
fig.\ \ref{fig:dbllogCheck} by comparison
to the straight line.
We hope in the future to also provide exact analytic results for
single-log divergences that are subleading to the double-log
divergence.  For now we only have numerical results for
those, which we present later with an examination of how
well those numerical results fit an educated guess for their analytic
form.

\begin {figure}[t]
\begin {center}
  \includegraphics[scale=0.55]{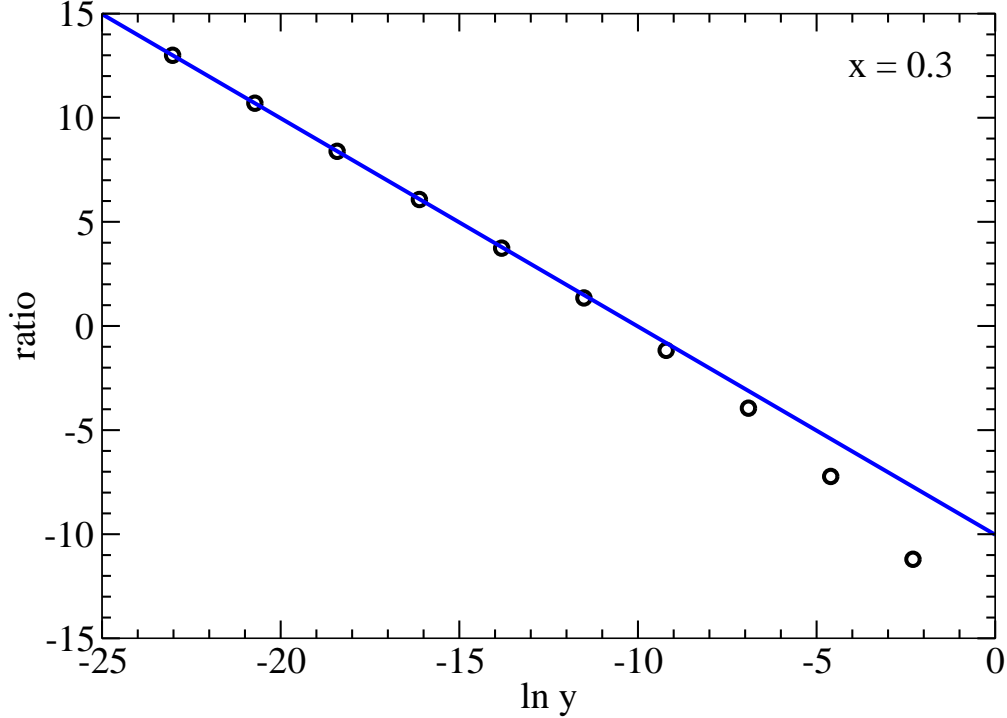}
  \caption{
     \label{fig:dbllogCheck}
     Full numerical results (circular data points)
     for the ratio (\ref{eq:VRratio}) plotted vs.\ $\ln y$
     for the case $x=0.3$.  The blue straight line shows a line of
     slope $-1$ for comparison, showing that our numerical results
     confirm the leading-log behavior (\ref{eq:VRlimit}).
  }
\end {center}
\end {figure}


\subsection {Outline}

The new diagrams needed for this paper are the virtual diagrams of
figs.\ \ref{fig:virtI} and \ref{fig:virtII}.
In the next section, we discuss how we can avoid calculating any of
these diagrams from scratch.  All of the $g \to gg$ QCD virtual diagrams
can be obtained by either (i) transformation from known results for the
$g \to ggg$ QCD diagrams of figs.\ \ref{fig:crossed} and \ref{fig:seq}
or (ii) by adapting the known result for one QED virtual diagram.

In section \ref{sec:IR}, we go into much more detail about how to organize
IR divergences in calculations related to energy loss.
We also show that the double-log behavior (\ref{eq:VRlimit}) is
equivalent to earlier leading-log results.

Section \ref{sec:SingleLog} presents numerical results for sub-leading
single-log divergences and shows that the numerics fit very well, but not
quite perfectly, a form one might guess based on the physics of
double-log divergences.

The formalism and calculations that have led to our results for rates
have spanned many papers, and one can reasonably worry about the
possibility of error somewhere along the way.
Section \ref{sec:error} provides a compendium of several non-trivial
cross-checks of our results.

Section \ref{sec:conclusion} offers our conclusion and our outlook
for what needs to be done in future work.
Appendix \ref{app:summary} contains a complete summary of
all our final formulas for rates.  Many technical
issues, derivations, and side investigations 
are left for the other appendices.


\section{Method for computing diagrams}
\label {sec:diagrams}

\subsection{Symmetry Factor Conventions}
\label{sec:SymFactor}

Before discussing how to find formulas for differential
rates, we should clarify some conventions.
Note each virtual diagram in fig.\ \ref{fig:virtII}, as well as
the second row of
fig.\ \ref{fig:virtI}, has a loop in the amplitude (an all-blue loop)
or conjugate amplitude (an all-red loop) that should be associated
with a diagrammatic loop symmetry factor of $\tfrac12$.
Our convention in this paper
is that any such diagrammatic symmetry factor associated
with an internal loop is already included in the formula for
what we call $\Delta\,d\Gamma/dx\,dy$ in
(\ref{eq:virtIntegrands}).
Note that the loops in the first row of fig.\ \ref{fig:virtI} do
{\it not}\/ have an associated symmetry factor.

In contrast, we do not include any identical-particle final-state
symmetry factors in our formulas for differential rates.
These must be included by hand whenever integrating over the
longitudinal momentum fractions of daughters if the integration
region double-counts final states.  For example, the total rate for
real double-splitting $g\to ggg$ is formally given by
\begin {equation}
  \Delta\Gamma_{g\to ggg} =
  \frac{1}{3!} \int_0^1 dx \int_0^{1{-}x} dy \>
    \left[ \Delta \frac{d\Gamma}{dx\,dy} \right]_{g\to ggg}
\end {equation}
because the integration region used above covers all $3!$ permutations
of possible momentum fractions $x$, $y$, and $z=1{-}x{-}y$ for
the three daughter gluons.
Similarly, for $g \to gg$ processes, formally
\begin {equation}
   \Gamma_{g\to gg}^{\rm LO} =
   \frac{1}{2!} \int_0^1 dx
     \left[ \frac{d\Gamma}{dx} \right]_{g\to gg}^{\rm LO} ,
   \qquad
   \Delta\Gamma_{g\to gg}^{\rm NLO} =
   \frac{1}{2!} \int_0^1 dx
     \left[ \Delta \frac{d\Gamma}{dx} \right]_{g\to gg}^{\rm NLO} .
\end {equation}
We use the caveat ``formally'' because the total splitting rates
$\Gamma$ and $\Delta\Gamma$ above are infrared divergent, but they
provide simple examples for explaining our conventions.


\subsection{Relating virtual diagrams to previous work}

\begin {figure}[t]
\begin {center}
  \includegraphics[scale=0.4]{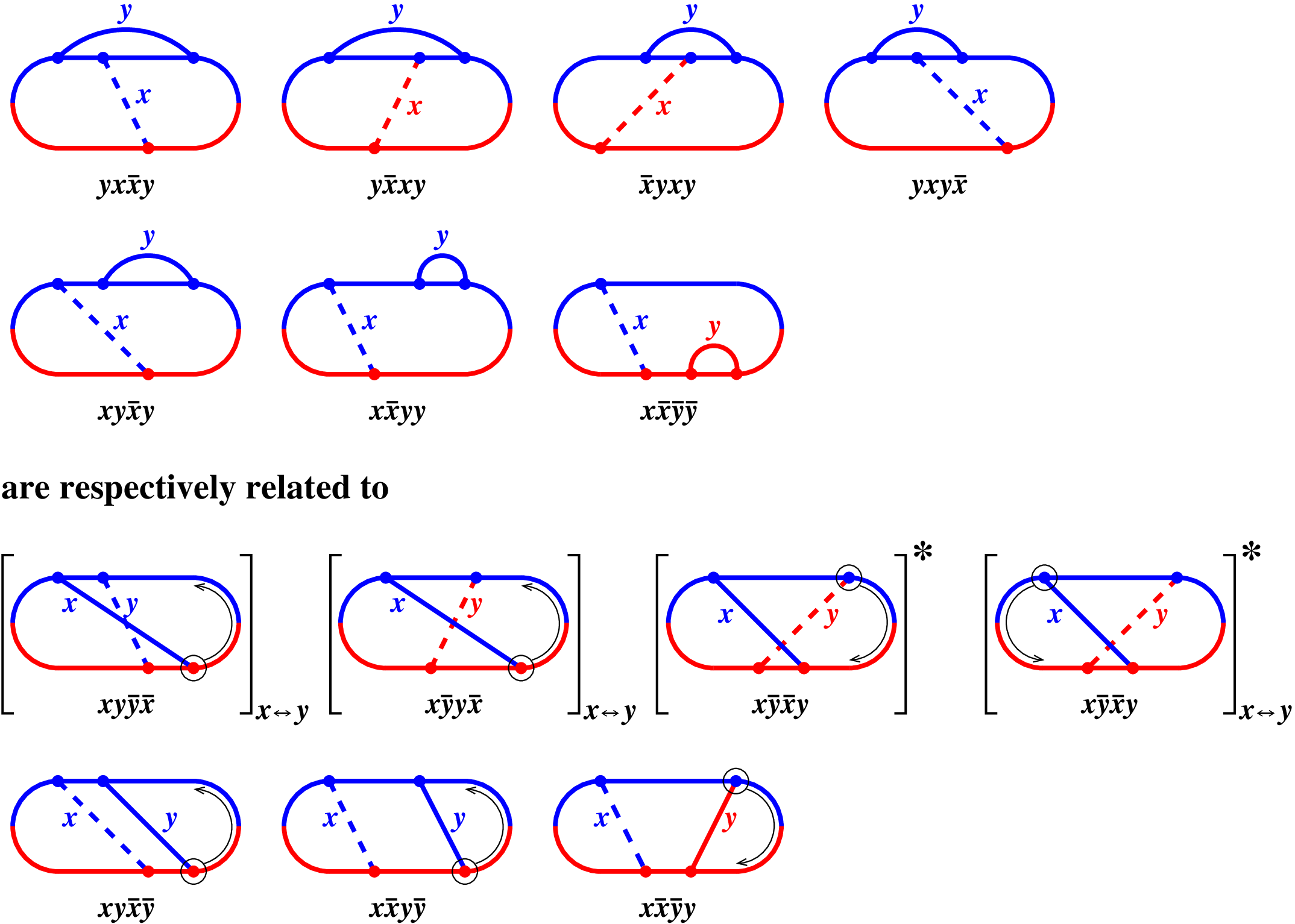}
  \caption{
     \label{fig:relateI}
     Relation of all but one Class I virtual diagram (fig.\ \ref{fig:virtI})
     to real $g \to ggg$ diagrams.
     The black arrows indicate moving the latest-time
     (or earliest-time) vertex using a back-end (or front-end)
     transformation \cite{QEDnf}.
  }
\end {center}
\end {figure}

\begin {figure}[t]
\begin {center}
  \includegraphics[scale=0.4]{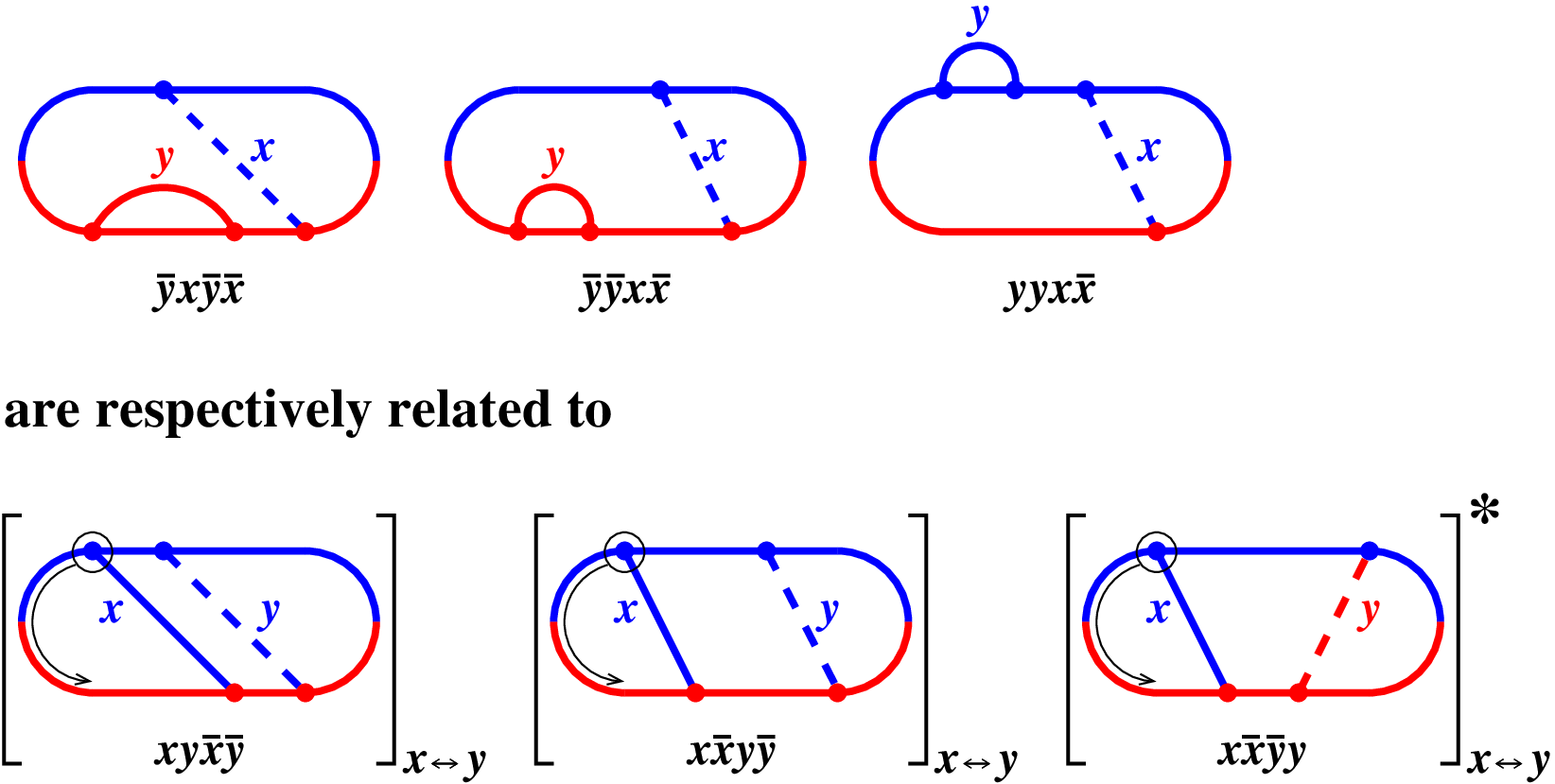}
  \caption{
     \label{fig:relateII}
     Relation of all but one Class II virtual diagram (fig.\ \ref{fig:virtII})
     to real $g \to ggg$ diagrams via front-end transformations.
  }
\end {center}
\end {figure}

In the context of (large-$\Nf$) QED, ref.\ \cite{QEDnf} showed
how many diagrams needed for virtual corrections to single splitting
could be obtained from results for real double splitting via
what were named back-end and front-end transformations.
For the current context of QCD,
figs.\ \ref{fig:relateI} and \ref{fig:relateII}
depict diagrammatically how all
but two of the Class I and II virtual diagrams we need
(figs.\ \ref{fig:virtI} and \ref{fig:virtII}) can be related to known results
for crossed and sequential $g \to ggg$ diagrams
(figs.\ \ref{fig:crossed} and \ref{fig:seq}) using
back-end and front-end transformations, sometimes accompanied by
switching the variable names $x$ and $y$ and/or complex conjugation.
Diagrammatically, a back-end transformation corresponds to taking
the {\it latest}-time splitting vertex in one of our rate diagrams
and sliding it around the back end of the diagram from the amplitude
to the conjugate-amplitude or vice versa.
Diagrammatically, a front-end transformation corresponds to taking
the {\it earliest}-time splitting vertex and sliding it around the
front end of the diagram.

In terms of formulas, the only effect of a back-end transformation
is to introduce an overall minus sign in the corresponding formula
for $d\Gamma/dx\,dy$  \cite{QEDnf}.  So, for example,
fig.\ \ref{fig:relateI} tells us that
\begin {equation}
  \left[ \frac{d\Gamma}{dx\,dy} \right]_{\bar x y x y}
  =
  - \left[ \frac{d\Gamma}{dx\,dy} \right]_{x \bar y \bar x y}^*
\end {equation}
and so
\begin {equation}
  2\Re \left[ \frac{d\Gamma}{dx} \right]_{\bar x y x y}
  =
  - \int_0^{1-x} dy \> 2\Re\left[ \frac{d\Gamma}{dx\,dy} \right]_{x \bar y \bar x y}
  .
\end {equation}
Similarly,
\begin {equation}
  2\Re \left[ \frac{d\Gamma}{dx} \right]_{y x \bar x y}
  =
  - \int_0^{1-x} dy \>
  \left\{
    \mbox{Replace $x\leftrightarrow y$ in formula for}~
    2\Re\left[ \frac{d\Gamma}{dx\,dy} \right]_{x y \bar y \bar x}
  \right\}
  .
\end {equation}
When making a back-end transformation, one may also have to include
a loop symmetry factor if the resulting virtual diagram has one,
which the original $g{\to}ggg$ processes do not.

Front-end transformations are more complicated.
In the cases where it is an $x$ emission at the earliest vertex that is
being moved between the amplitude and conjugate amplitude, requiring
the longitudinal momentum fractions of the lines of the diagrams
to match up requires replacing
\begin {equation}
  (x,y,E) \longrightarrow
  \Bigl( \frac{{-}x}{1{-}x} \,,\, \frac{y}{1{-}x} \,,\, (1{-}x)E \Bigr) ,
\end {equation}
where $E$ is the energy of the initial particle in the real or virtual
double-splitting process.
See section 4.2 of ref.\ \cite{QEDnf} for a more detailed discussion.
There is also an overall normalization factor associated with the
transformation that, for our case here where all the particles are
gluons, amounts to%
\footnote{
  See appendix H of ref.\ \cite{QEDnf}, especially eqs.\ (H.13) and (H.14)
  there.  In (H.13) of ref.\ \cite{QEDnf} there was additionally
  an overall factor of $2\Nf {\cal N}_e/{\cal N}_\gamma$ that arose because
  that front-end transformation related a diagram with an initial electron
  to one with an initial photon, and the $2\Nf {\cal N}_e/{\cal N}_\gamma$
  reflected the different factors associated with averaging over initial
  flavors and helicities.  In our case, the initial particle is always
  a gluon, so no such adjustment is necessary.
  Also, eqs.\ (H.13) and (H.14) of ref.\ \cite{QEDnf} do not have the
  overall minus sign of our (\ref{eq:frontend}) above because they
  included a back-end transformation in addition to the front-end
  transformation.
  Note that those equations have also implemented $x\leftrightarrow y$ in
  addition to the front-end transformation (\ref{eq:frontend}) above.
}
\begin {equation}
   \frac{d\Gamma}{dx\,dy}
   \xrightarrow{\rm front-end}
   - (1{-}x)^{-\eps}
   \left\{
     \frac{d\Gamma}{dx\,dy}
     ~\mbox{with}~
     (x,y,E) \longrightarrow
     \Bigl( \frac{{-}x}{1{-}x} \,,\, \frac{y}{1{-}x} \,,\, (1{-}x)E \Bigr)
   \right\}
\label {eq:frontend}
\end {equation}
in $4{-}\eps$ spacetime dimensions.
The overall factor $(1{-}x)^{-\eps}$ will be relevant because we will use
dimensional regularization to handle and renormalize
UV divergences in our calculation.
We should note that there are a few additional subtleties in
practically implementing front-end transformations, which we leave
to appendix \ref{app:method}.
As an example of (\ref{eq:frontend}),
the relation depicted by the first case of fig.\ \ref{fig:relateII}
gives
\begin {multline}
   2\Re \left[ \frac{d\Gamma}{dx} \right]_{\bar y x \bar y \bar x}
   =
   - \tfrac12 (1{-}x)^{-\eps} \int_0^1 dy 
   \biggl(
    \mbox{Replace $x\leftrightarrow y$ in result of}~
\\
    2\Re
    \biggl\{
      \left[ \frac{d\Gamma}{dx\,dy} \right]_{xy\bar x\bar y}
      ~\mbox{with substitution (\ref{eq:frontend})}
    \biggr\}
  \biggr) .
\end {multline}
The overall factor of $\tfrac12$ is included because of
the loop symmetry factor associated with the (red) loop
in the $\bar y x \bar y \bar x$ virtual diagram.

\begin {figure}[t]
\begin {center}
  \includegraphics[scale=0.4]{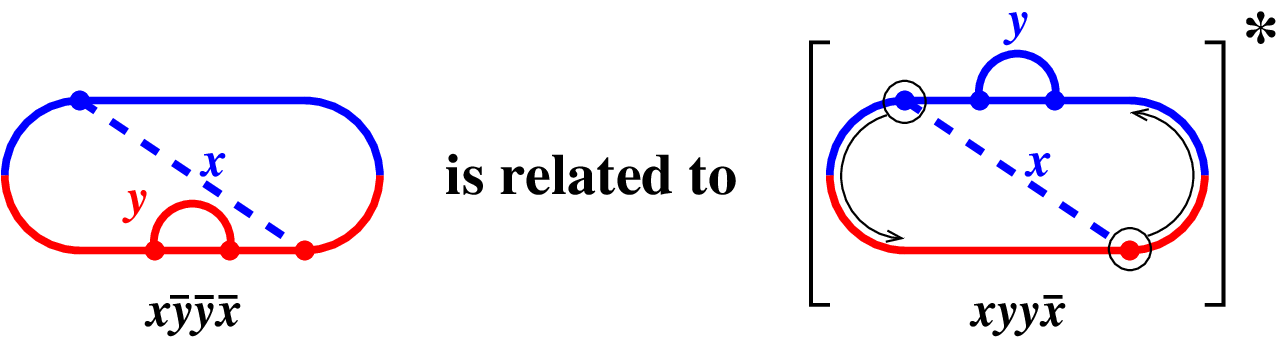}
  \caption{
     \label{fig:relateFund}
     Relation to each other of the two virtual diagrams of
     figs.\ \ref{fig:virtI} and \ref{fig:virtII} that are
     not covered by the relations of
     figs.\ \ref{fig:relateI} and \ref{fig:relateII}.
  }
\end {center}
\end {figure}

\begin {figure}[t]
\begin {center}
  \includegraphics[scale=0.6]{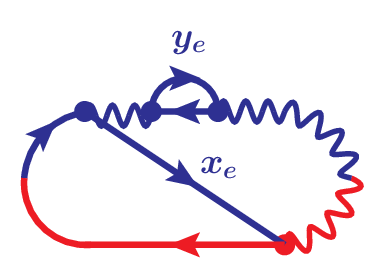}
  \caption{
     \label{fig:QEDfund}
     A QED version \cite{QEDnf} of the $xyy\bar x$ diagram of
     fig.\ \ref{fig:virtI}.
  }
\end {center}
\end {figure}

The only two virtual diagrams not covered by figures
\ref{fig:relateI} and \ref{fig:relateII}
are $xyy\bar x$ and $x\bar y\bar y\bar x$.
But these diagrams are related to
each other by combined front-end and back-end transformations,
as depicted in fig.\ \ref{fig:relateFund}.
That means that transformations have given us a short-cut for determining
all virtual diagrams except for one, which we take to be
$xyy\bar x$.
Fortunately, the $xyy\bar x$ diagram has the same form as
the QED diagram of fig.\ \ref{fig:QEDfund}
previously computed in ref.\ \cite{QEDnf},
and the QED result can be easily adapted to QCD.
One just needs to include QCD group factors associated with
splitting vertices; use QCD instead of QED
Dokshitzer-Gribov-Lipatov-Altarelli-Parisi (DGLAP) splitting functions;
correctly account for identical-particle symmetry factors; and
use QCD rather than QED results for the complex frequencies and
normal modes associated with the $\hat q$ approximation to
the propagation of the high-energy particles through the medium.
Details of the conversion are given in appendix \ref{app:methodFund}.

We give more detail on implementing the above methods in appendix
\ref{app:method}, and final results for unrenormalized diagrams are
given in appendix \ref{app:summary}
[with $\sigren{=}0$ and $\sigbare{=}1$ in section \ref{app:NLOsummary}].


\subsection{UV divergences, renormalization, and running of \boldmath$\alphas$}

The virtual diagrams of figs.\ \ref{fig:virtI} and \ref{fig:virtII} contain
UV-divergent loops in the amplitude or conjugate amplitude.
It may seem surprising that most of them can be related via
figs.\ \ref{fig:relateI} and \ref{fig:relateII} to real double splitting
($g \to ggg$) diagrams that involve only tree-level diagrams in the
amplitude and conjugate amplitude.
This is possible because we are working with {\it time-ordered}\/ diagrams:
individual time-ordered interferences of
tree-level diagrams are UV-divergent even
though the sum of all the different time-orderings is not.
See section 4.1 of ref.\ \cite{QEDnf} for more discussion of this point.
In any case, the original calculations \cite{2brem,seq,dimreg}
of the $g{\to}ggg$ diagrams of figs.\ \ref{fig:crossed} and \ref{fig:seq}
discussed the UV divergence of each diagram and showed that they
indeed canceled.

The corresponding divergences of the virtual diagrams, however, will
not cancel.  Indeed, they must conspire to produce the known renormalization
of $\alphas$.
Ref.\ \cite{QEDnf} demonstrated how this worked out for large-$\Nf$ QED,
but the diagrammatics of renormalization of the QCD coupling is a little
more complicated.
We will also encounter a well-known annoyance of
Light Cone Perturbation Theory (LCPT): individual diagrams will contain
mixed UV-IR divergences that only cancel when the diagrams are summed
together.%
\footnote{
  For an example from calculations that are tangentially related to ours,
  see Beuf \cite{Beuf1,Beuf2} and
  H\"anninen, Lappi, and Paatelainen \cite{LaP,HLaP} on next-to-leading-order
  deep inelastic scattering (NLO DIS).
  For a description of the similarities and differences of our problem
  and theirs, see appendix B of ref.\ \cite{QEDnf}.
  For a very early result on obtaining the correct renormalization of
  the QCD coupling with LCPT in the context of vacuum diagrams, see
  ref.\ \cite{HarindranathZhang3}.
}


\subsubsection{UV and IR regulators}

We use dimensional regularization in $4{-}\eps$
spacetime dimensions for UV divergences.
However, we use the letter $d$ to refer to the number
of \textit{transverse spatial} dimensions
\begin {equation}
  d \equiv d_\perp = 2-\eps .
\end {equation}

For infrared divergences, we introduce a hard lower cut-off
$(p^+)_{\rm min}$ on light-cone momentum components $p^+$.
Hard momentum cut-offs complicate gauge invariance, but this is a
fairly standard procedure in LCPT, since LCPT is formulated specifically in
light-cone gauge $A^+{=}0$.
Note that $p^+$ is invariant under any
{\it residual}\/ gauge transformation that preserves light-cone gauge.
It would of course be nicer to use a more generally gauge invariant choice
of infrared regulator, but that would lead to more
complicated calculations.%
\footnote{
  \label {foot:IRdimreg}
  In particular, one might imagine using dimensional regularization for
  the infrared as well as the ultraviolet.  Unfortunately, the
  dimensionally-regulated expansions in $\eps$ that we currently have
  available \cite{dimreg,QEDnf} for the types of diagrams we need
  all made use of the fact that dimensional regularization was
  {\it only} needed for the ultraviolet.
}

We will write our IR cut-off on longitudinal momenta $p^+$ as
\begin {equation}
  (p^+)_{\rm min} = P^+ \delta
\label {eq:deltadef}
\end {equation}
where $P^+$ is the longitudinal momentum of the initial particle
in the double-splitting process and $\delta$ is an arbitrarily tiny
positive number.%
\footnote{
  A technicality concerning orders of limits: One should take the
  UV regulator $\eps \to 0$ before taking the IR regulator
  $\delta \to 0$.  Taking $\delta \to 0$ first would be
  equivalent to using dimensional
  regularization for the IR as well as the UV, which is
  currently problematic for the reason given in
  footnote \ref{foot:IRdimreg}.
}
For consistency of IR regularization of the theory,
this constraint must be applied to all particles in the process.
For instance, in a $g{\to}ggg$ process where $P^+$ splits into
daughters with longitudinal momenta $x P^+$, $y P^+$, and $z P_+$, we
require that the longitudinal momentum fractions $x$, $y$, and $z$
all exceed $\delta$. (This automatically guarantees
that internal particle lines in $g{\to}ggg$ diagrams also have
$p^+ > P^+ \delta$.) 
In a virtual correction to
$g \to gg$ where $P^+$ splits into $x P^+$ and $(1{-}x)P^+$,
we must have $x$ and $1{-}x$ greater than $\delta$, but we must
also impose that the momentum fractions of internal virtual lines
are greater than $\delta$ as well.  We'll see explicit examples below.
With this notation, the annoying mixed UV-IR divergences of LCPT are
proportional to $\eps^{-1} \ln\delta$, which is the product of
a logarithmic UV divergence $\eps^{-1}$ and a logarithmic IR divergence
$\ln\delta$.


\subsubsection{Results for UV (including mixed UV-IR) divergences}

We can read off the results for $1/\eps$ divergences from the
complete results given in appendix \ref{app:summary}.  However,
we will take the opportunity to be a little more concrete here
in the main text by stepping through the calculation for one
of the diagrams, but focusing on just the UV-divergent ($1/\eps$)
terms.  Then we'll put the diagrams together to see the
cancellation of mixed UV-IR divergences and the appearance of
the QCD beta function coefficient $\beta_0$.

Consider the first NLO $g{\to}gg$
diagram ($yx\bar xy$) in fig.\ \ref{fig:relateI},
which shows that diagram related by back-end transformation
to the $g{\to}ggg$ diagram $xy\bar y\bar x$.
The $1/\eps$ piece of the latter can be taken from ref.\ \cite{dimreg}
and is [see appendix \ref{app:details} of the current paper
for more detail]
\begin {align}
   \left[ \frac{d\Gamma}{dx\,dy} \right]_{xy\bar y\bar x} \approx{} &
   \frac{\CA^2 \alphas^2}{8\pi^2\epsilon}
     \bigl[ (i\Omega \sgn M)_{-1,x,1-x} + (i\Omega \sgn M)_{-(1-y),x,z} \bigr]
\nonumber\\ &\quad\times
     x y z^2 (1{-}x)(1{-}y)
       \left[
         (\alpha + \beta)(1{-}x)(1{-}y)
         + (\alpha + \gamma) x y
       \right] ,
\label {eq:xyyxdiv}
\end {align}
where
\begin {equation}
   \Omega_{x_1,x_2,x_3} \equiv 
   \sqrt{ \frac{-i \qhatA}{2 E}
          \left(\frac{1}{x_1}+\frac{1}{x_2}+\frac{1}{x_3}\right) } ,
   \qquad
   M_{x_1,x_2,x_3} \equiv - x_1 x_2 x_3 E ,
\label {eq:OmMdefs}
\end {equation}
and $(\alpha,\beta,\gamma)$ are functions of $x$ and $y$ that represent
various combinations of the helicity-dependent DGLAP splitting functions
associated with the vertices in the diagram.%
\footnote{
  Details
  of the definition of $(\alpha,\beta,\gamma)$
  in terms of DGLAP splitting functions are
  given in sections 4.5 and 4.6 of ref.\ \cite{2brem}.
  In order to make those definitions work with front-end transformations,
  one must additionally include absolute value signs as discussed after
  eq.\ (\ref{eq:abc}) of the current paper.
}
In this section we use $\approx$ to indicate that we are only
keeping $1/\eps$ terms.
Back-end transforming the above expression and swapping $x{\leftrightarrow}y$,
as indicated in fig.\ \ref{fig:relateI}, gives the corresponding result
for the virtual diagram $yx\bar xy$:
\begin {subequations}
\label {eq:crossedXpieces}
\begin {align}
   2\Re\left[ \frac{d\Gamma}{dx} \right]_{yx\bar x y}
   \approx&
   - \frac{\CA^2 \alphas^2}{4\pi^2\epsilon}
     \int_\delta^{1-x-\delta} dy \>
     \Re(i\Omega_{-1,y,1-y} + i\Omega_{-(1-x),y,z})
     x y z (1{-}x)(1{-}y)
\nonumber\\ &\qquad\times
       \left[
         (\alpha{+}\beta) z(1{-}x)(1{-}y)
         + (\alpha{+}\gamma) x y z
       \right] ,
\end {align}
where we have taken $2\Re(\cdots)$ to include the conjugate diagram
as well.

Doing similar calculations for the other crossed Class I diagrams
(the top line of fig.\ \ref{fig:relateI}), by using $g{\to}ggg$ results
for $x\bar y y\bar x$ and $x\bar y\bar x y$ from ref.\ \cite{dimreg}
and then transforming as in fig.\ \ref{fig:relateI}, gives
\begin {align}
   2\Re\left[ \frac{d\Gamma}{dx} \right]_{y\bar x xy}
   \approx&
   - \frac{\CA^2 \alphas^2}{4\pi^2\epsilon}
     \int_\delta^{1-x-\delta} dy \>
     \Re(i\Omega_{-1,y,1-y} + i\Omega_{-(1-x),y,z})
     x y z (1{-}x)(1{-}y)
\nonumber\\ &\qquad\times
       \left[
         - (\alpha + \beta) z(1{-}x)(1{-}y)
         + (\beta + \gamma) x y (1{-}x)(1{-}y)
       \right] ,
\\
   2\Re\left[ \frac{d\Gamma}{dx} \right]_{\bar x yxy}
   \approx&
   - \frac{\CA^2 \alphas^2}{4\pi^2\epsilon}
     \int_\delta^{1-x-\delta} dy \>
     \Re(i\Omega_{-1,x,1-x} + i\Omega_{-(1-x),y,z})
     x y z (1{-}x)(1{-}y)
\nonumber\\ &\qquad\times
       \left[
         - (\alpha + \gamma) x y z
         - (\beta + \gamma) x y (1{-}x)(1{-}y)
       \right] ,
\\
   2\Re\left[ \frac{d\Gamma}{dx} \right]_{yxy\bar x}
   \approx&
   - \frac{\CA^2 \alphas^2}{4\pi^2\epsilon}
     \int_\delta^{1-x-\delta} dy \>
     \Re(i\Omega_{-1,y,1-y} + i\Omega_{-1,x,1-x})
     x y z (1{-}x)(1{-}y)
\nonumber\\ &\qquad\times
       \left[
         - (\alpha + \gamma) x y z
         - (\beta + \gamma) x y (1{-}x)(1{-}y)
       \right] .
\label {eq:divyxyx}
\end {align}
\end {subequations}
Eqs.\ (\ref{eq:crossedXpieces}) sum to
\begin {align}
   \left[ \frac{d\Gamma}{dx} \right]_{\virtI~\rm crossed}
   \approx&
   \frac{\CA^2 \alphas^2}{2\pi^2\epsilon}
     \Re(i\Omega_{-1,x,1-x})
     \int_\delta^{1-x-\delta} dy \>
     x^2 y^2 z (1{-}x)(1{-}y)
\nonumber\\ &\qquad\times
       \left[
         (\alpha + \gamma) z
         + (\beta + \gamma) (1{-}x)(1{-}y)
       \right] .
\label{eq:crossedXdiv0}
\end {align}
Since we are focused here just on the $1/\eps$ pieces above, the integral
may be done using the explicit $d{=}2$ expressions (\ref{eq:abc})
for $(\alpha,\beta,\gamma)$.
But the combination $(\alpha + \gamma) z + (\beta + \gamma) (1{-}x)(1{-}y)$
appearing in (\ref{eq:crossedXdiv0}) turns out to be dimension-independent
in any case!  (See appendix \ref{app:abcdim}.)

Remember that for the crossed virtual diagrams, like all the Class I diagrams of
fig.\ \ref{fig:virtI}, taking $x \to 1{-}x$ generates other distinct
diagrams that need to be included as well.
So, do the $y$ integral in (\ref{eq:crossedXdiv0}),
combine the result with $x\to 1{-}x$
[as in (\ref{eq:dGnetNLOdef}) or (\ref{eq:VVR})],
and take the small-$\delta$ limit.  This gives
\begin {equation}
   \biggl(\left[ \frac{d\Gamma}{dx} \right]_{\virtI~\rm crossed}\biggr)
   + (x\to 1{-}x)
   \approx
   \left[ \frac{d\Gamma}{dx} \right]^{\rm LO}
   \frac{\CA\alphas}{\pi\epsilon}
   \bigl[
      - \tfrac{11}{3} + 2\ln\bigl(x(1{-}x)\bigr) - 6\ln\delta
   \bigr] ,
\label {eq:divcrossed}
\end {equation}
where $[d\Gamma/dx]^{\rm LO}$ is the leading-order single splitting
result%
\footnote{
   The QCD version of the leading-order rate
   goes back to BDMPS \cite{BDMPS12,BDMPS3}
   and Zakharov \cite{Zakharov}.
   For a discussion of how the QED version in our notation matches up
   with the original QED result of Migdal \cite{Migdal}, see
   appendix C.4 of ref.\ \cite{QEDnf}.
}
\begin {equation}
   \left[ \frac{d \Gamma}{dx} \right]^{\rm LO}
   = \frac{\alphas}{\pi} \, P(x) \, \Re(i\Omega_{-1,x,1-x})
     + O(\eps)
\label {eq:LO0}
\end {equation}
and $P(x)$ is the DGLAP $g{\to}gg$ splitting function.
A non-trivial feature of (\ref{eq:divcrossed}) is that the $y$ integration
in (\ref{eq:crossedXdiv0}),
combined with the addition of $x \to 1{-}x$, gave a result proportional to
the $P(x)$ in (\ref{eq:LO0}).  This is what will later make possible
the absorption of $1/\eps$ divergences
by renormalizing the $\alphas$ in the leading-order result.
For the time being, however,
note the unwanted mixed UV-IR divergence $\eps^{-1}\ln\delta$ in
(\ref{eq:divcrossed}).

Now turn to the sequential virtual diagrams.
The sum $2\Re[xy\bar x\bar y + x\bar x y\bar y + x\bar x \bar y y]$
of {\it non}-virtual sequential $g{\to}ggg$ diagrams shown in
fig.\ \ref{fig:seq} (together with their conjugates) represents
the sum of all time orderings of a tree-level process and so does
not give any net $1/\eps$ divergence.%
\footnote{
  This is shown explicitly by summing the individually divergent
  time-order diagrams in eq.\ (5.20) of \cite{dimreg}.
}
So there will also be no divergence in its back-end transformation,
which fig.\ \ref{fig:relateI} shows is equivalent to the
sum $2\Re[xy\bar x y + x\bar x y y + x\bar x \bar y \bar y]$
of three Class I sequential virtual diagrams.
Nor will there be any divergence to its front-end transformation
followed by the swap $x\leftrightarrow y$,
corresponding by fig.\ \ref{fig:relateII} to the
sum $2\Re[\bar y x \bar y \bar x + \bar y \bar y x \bar x + y y x \bar x]$
of three Class II sequential diagrams.  So none of these groups of
diagrams generate a divergence.

What remains of figs.\ \ref{fig:virtI} and \ref{fig:virtII}
is the Class I virtual diagram $xyy\bar x$ and the Class II
virtual diagram $x\bar y\bar y\bar x$, which are related to
each other via fig.\ \ref{fig:relateFund}.
As mentioned earlier, the result for $2\Re[xyy\bar x]$ can be converted from
the known result \cite{QEDnf} for the similar QED diagram of
fig.\ \ref{fig:QEDfund}.
The UV-divergent $1/\eps$ piece of that QED result was%
\footnote{
  \label{foot:QEDnfF42}
  This can be obtained by expanding eq.\ (F.42) of ref.\ \cite{QEDnf}
  in $\eps$ and replacing $\yfrake$ there by its definition
  $\yfrake \equiv \ye/(1-\xe)$.
  There was an overall sign error in eq.\ (F.42) of the original
  published version of ref.\ \cite{QEDnf}, which is treated
  correctly in the version above.
}
\begin {equation}
  2\Re \left[ \frac{d\Gamma}{d\xe} \right]_{xyy\bar x}
  \approx
  - 
  \frac{\Nf \alphaqed^2}{\pi^2 \eps} \,
  P_{e\to e}(\xe) \, \Re(i \Omega^{\rm QED} \sgn M)_{-1,\xe,1-\xe}
  \int_0^{1-x} \frac{d\ye}{1-\xe} \> P_{\gamma\to e}\bigl(\frac{\ye}{1-\xe}\bigr) .
\label{eq:QEDdiv}
\end{equation}
The translation from a QED diagram to a QCD diagram is explained
in our appendix \ref{app:methodFund} and gives
\begin {align}
  2\Re\left[ \frac{d\Gamma}{dx} \right]_{xyy\bar x}
  &\approx
  -
  \frac{\alphas^2}{2 \pi^2 \eps} \,
  P(x) \, \Re(i \Omega \sgn M)_{-1,x,1-x}
  \int_\delta^{1-x-\delta} \frac{dy}{1-x} \> P\bigl(\frac{y}{1-x}\bigr)
\nonumber\\
  &=
  - 
  \left[ \frac{d\Gamma}{dx} \right]^{\rm LO}
  \frac{\alphas}{2 \pi \eps} \,
  \int_\delta^{1-x-\delta} \frac{dy}{1{-}x} \> P\bigl(\frac{y}{1{-}x}\bigr) .
\end {align}
Our IR cut-off $\delta$ must now be included with the integration limits
because, unlike QED, LPM splitting rates are (non-integrably)
infrared divergent in QCD.
The $\sgn M$ factors are included above because, even though
$M_{-1,x,1-x}$ is positive for the $xyy\bar x$ diagram, this more general form
is consistent with the front-end transformation we are about to perform.

Since $xyy\bar x$ above is a Class I diagram, we need to also add in the
other diagram that is generated by $x \to 1{-}x$.
Finally, the transformation of fig.\ \ref{fig:relateFund} gives the
remaining (Class II) diagram $x\bar y\bar y\bar x$.%
\footnote{
  As discussed after eq.\ (\ref{eq:Pgg}), one must include an absolute
  value sign in the definition of $P(x)$ in order to make it work
  with front-end transformations using our conventions.
}
The sum of all three is
\begin {align}
  \left[ \frac{d\Gamma}{dx} \right]_{\rm other\,virt}
  &\approx
  - 
  \left[ \frac{d\Gamma}{dx} \right]^{\rm LO}
  \frac{\alphas}{2 \pi \eps} \,
     \left[
       \int_\delta^{1-x-\delta}\!\!
          \frac{dy}{1{-}x} \> P\bigl(\frac{y}{1{-}x}\bigr)
       + \int_\delta^{x-\delta}\!
          \frac{dy}{x} \> P\bigl(\frac{y}{x}\bigr)
       + \int_\delta^{1-\delta}\!\!
          dy \> P(y)
     \right]
\nonumber\\
   &\approx
     \left[ \frac{d\Gamma}{dx} \right]^{\rm LO}
     \frac{\CA\alphas}{\pi\epsilon}
     \bigl[
       \tfrac{11}{2} - 2\ln\bigl(x(1{-}x)\bigr) + 6\ln\delta
     \bigr] .
\label {eq:divother}
\end {align}

Adding (\ref{eq:divcrossed}) and (\ref{eq:divother}) gives
the total UV divergence from virtual corrections to single
splitting:
\begin {equation}
   \left[ \Delta \frac{d\Gamma}{dx} \right]^{\rm NLO}_{g\to gg}
   \approx
     -
     \left[ \frac{d\Gamma}{dx} \right]^{\rm LO}
     \frac{\beta_0\alphas}{\epsilon}
\label {eq:divNLO}
\end {equation}
with
\begin {equation}
   \beta_0 = -\frac{11\CA}{6\pi} \,.
\label {eq:beta0}
\end {equation}
The $\beta_0$ above is the same coefficient that appears in the
one-loop beta function for $\alphas = g^2/4\pi$:
\begin {equation}
   \frac{d\alphas}{d(\ln\mu)}
   = - \frac{( 11 \CA - 2 \Nf)}{6\pi} \,
       \alphas^2 ,
\label {eq:RNG}
\end {equation}
where $\Nf$ is the number of quark flavors.
The $\Nf$ term does not appear in (\ref{eq:beta0})
because we have not included quarks in our calculations,
consistent with our choice to work in the large-$\Nc$ limit
(for $\Nf$ fixed).

Note that the UV-IR mixed divergences have canceled between
(\ref{eq:divcrossed}) and (\ref{eq:divother}), as well as
the $\ln\bigl( x(1-x) \bigr)$ terms.  These cancellations had to occur in
order for the total divergence of the virtual diagrams to be
absorbed by usual QCD renormalization, as we'll now see.%
\footnote{
  There is something sloppy one might have tried in the preceding
  calculations that would have failed to produce the correct UV divergences,
  which we mention here as a caution to others because
  we unthinkingly tried it on our first attempt at
  this calculation.
  Suppose that we had set $\delta$ to zero in all the integration
  limits so that each IR-divergent integral we've done was
  divergent and ill-defined.  Then suppose that in each integral
  we scaled the integration variable $y$ so that each integral
  was now from 0 to 1, e.g.
  $\int_0^{1-x} dy \> f(y) \to (1{-}x) \int_0^1 dy \> f\bigl((1{-}x)y\bigr)$
  and similarly for $x \to 1{-}x$.  Now that the integration limits are
  the same, one could add together all the integrands for
  all the diagrams.  The combined integral would be convergent but
  does not give the correct result (\ref{eq:divNLO}).
  That's because one can get any incorrect answer by manipulating
  sums of ill-defined integrals.  To properly regularize a theory,
  one must first independently define the cut-off on the theory
  (in this case the
  IR cutoff on longitudinal momenta) and only then
  add up all diagrams calculated with that cut-off.
}


\subsubsection{Renormalization}

Following ref.\ \cite{QEDnf},%
\footnote{
  Specifically section 4.3.4 and footnote 26 of that reference.
  Our $\beta_0$ here corresponds to $2\Nf\alphaqed/3\pi$ in QED.
}
we find it simplest to implement
renormalization in this calculation by imagining that all diagrams have
been calculated using the bare (unrenormalized) coupling and then
rewriting $(\alphas)_{\rm bare}$ in terms of $(\alphas)_{\rm ren}$.
For the $\MSbar$-renormalization scheme, that's
\begin {equation}
   \alphas^{\rm bare} =
   \alphas^{\rm ren}
     + \frac{\beta_0}{2} (\alphas^{\rm ren})^2
       \Bigl( \frac{2}{\eps} - \gammaE + \ln(4\pi) \Bigr)
     + O(\alphas^3) .
\label {eq:renorm}
\end {equation}
When expressed in terms of renormalized $\alphas$, the $1/\eps$
divergences should then cancel in the combination%
\footnote{
   Though the $[\Delta\,d\Gamma/dx]^{\rm LO+NLO}_{g\to gg}$
   defined in (\ref{eq:ratebare})
   is UV finite, it is power-law IR divergent.
   Only in combination of
   the $g{\to}gg$ rates with $g{\to}ggg$ rates, such as
   (\ref{eq:dGnet}), are power-law IR divergences eliminated,
   leaving double-log IR divergences.
}
\begin {equation}
   \left[ \Delta \frac{d\Gamma}{dx} \right]^{\rm LO+NLO}_{g\to gg}
   \equiv
   \left[ \frac{d\Gamma}{dx} \right]^{\rm LO,bare}
   +
   \left[ \Delta \frac{d\Gamma}{dx} \right]^{\rm NLO,bare}_{g\to gg}
\label {eq:ratebare}
\end {equation}
through order $\alphas^2$.
Since the leading-order $[d\Gamma/dx]^{\rm bare}$ is proportional
to $\alphas^{\rm bare}$, (\ref{eq:renorm}) gives
\begin {equation}
   \left[ \frac{d\Gamma}{dx} \right]^{\rm LO,bare}
   =
   \left[ \frac{d\Gamma}{dx} \right]^{\rm LO,ren}
   +
   \frac{\beta_0 \alphas^{\rm ren}}{2}
      \left[ \frac{d\Gamma}{dx} \right]^{\rm LO,ren}_{d=2-\eps}
      \Bigl( \frac{2}{\eps} - \gammaE + \ln(4\pi) \Bigr)
   + O(\alphas^3) .
\label {eq:ren1}
\end {equation}
Note that, because it is multiplied by $2/\eps$, we will need to
use a $d{=}2{-}\eps$ formula for
$[ d\Gamma/dx ]^{\rm LO}$ in the last term above, as
indicated by the subscript.
We can now use (\ref{eq:ren1}) to
regroup terms in (\ref{eq:ratebare}) to write the
LO+NLO $g{\to}gg$ rate in terms of $\MSbar$ renormalized quantities
as
\begin {equation}
   \left[ \Delta \frac{d\Gamma}{dx} \right]^{\rm LO+NLO}_{g\to gg}
   =
   \left[ \frac{d\Gamma}{dx} \right]^{\rm LO,ren}
   +
   \left[ \Delta \frac{d\Gamma}{dx} \right]^{\rm NLO,ren}_{g\to gg}
\label {eq:rateren}
\end {equation}
with
\begin {equation}
   \left[ \Delta \frac{d\Gamma}{dx} \right]^{\rm NLO,ren}_{g\to gg}
   =
   \left[ \Delta \frac{d\Gamma}{dx} \right]^{\rm NLO,bare}_{g\to gg}
   +
   \frac{\beta_0 \alphas^{\rm ren}}{2}
      \left[ \frac{d\Gamma}{dx} \right]^{\rm LO,ren}_{d=2-\eps}
      \Bigl( \frac{2}{\eps} - \gammaE + \ln(4\pi) \Bigr) .
\label {eq:NLOren0}
\end {equation}
One can see from (\ref{eq:divNLO}) that the $1/\eps$ poles indeed
cancel in this renormalized $[ \Delta d\Gamma/dx ]^{\rm NLO}$.

There are many equivalent ways to introduce the $\MSbar$ renormalization
scale into the renormalization procedure outlined above.
Following ref.\ \cite{QEDnf},%
\footnote{
  See in particular the discussion of eq.\ (F.31)  of ref.\ \cite{QEDnf}.
}
we will introduce it by
writing the dimensionful bare $g^2/4\pi$
in $4{-}\eps$ spacetime dimensions as $\mu^\eps \alphas^{\rm bare}$,
where $\alphas^{\rm bare}$ is the usual dimensionless coupling for
$4$ spacetime dimensions.  As a result, every power of $\alphas$ in
our unrenormalized calculations comes with a power of $\mu^\eps$ which,
if multiplied by a $1/\eps$ UV divergence and expanded in $\eps$, will
generate the correct
logarithms $\ln\mu$ of the renormalization scale in our results,
as we detail next.


\subsubsection{Organization of Renormalized Results}
\label {sec:OrganizeRenorm}

Formulas for the NLO $g{\to}gg$ rate are given in appendix \ref{app:NLOsummary}.
Because of the fact that multiple diagrams contribute to cancellation of
$1/\eps$ poles in ways that are not particularly simple diagram by diagram,
we have organized our renormalized
result for $[d\Gamma/dx]^{\rm NLO, ren}_{g\to gg}$
slightly differently than the QED case of
ref.\ \cite{QEDnf}, in a way that we will explain here.

Also, we would like to write renormalized formulas in appendix
\ref{app:NLOsummary} in a way that makes transparent the dependence
on explicit renormalization scale logarithms $\ln\mu$.
The running (\ref{eq:RNG}) of $\alphas$, plus the fact that the
leading-order rate is proportional to $\alphas$, implies that
the renormalized NLO rate must have explicit $\mu$ dependence
\begin {equation}
  \left[ \Delta \frac{d\Gamma}{dx} \right]^{\rm NLO,ren}_{g\to gg}
  =
  -\left[ \frac{d\Gamma}{dx} \right]^{\rm LO} \beta_0 \alphas \ln\mu
  + \cdots
\label {eq:lnmu1}
\end {equation}
in order to cancel the implicit $\mu$ dependence
$d\alphas/d(\ln\mu)=\beta_0\alphas^2$ of $\alphas(\mu)$
from the LO rate.
In contrast, the NLO bare rate $[\Delta\,d\Gamma/dx]^{\rm NLO, bare}_{g\to gg}$
is proportional to
$(\mu^\eps\alphas)^2$, and so its divergence
(\ref{eq:divNLO}) generates
\begin {equation}
   \left[ \Delta \frac{d\Gamma}{dx} \right]^{\rm NLO,bare}_{g\to gg}
   =
     -
     \mu^{2\eps} \left[ \frac{d\Gamma}{dx} \right]^{\rm LO}_{d=2}
     \frac{\beta_0\alphas}{\epsilon}
     + \cdots
   =
     -
     \left[ \frac{d\Gamma}{dx} \right]^{\rm LO}_{d=2}
     \frac{\beta_0\alphas}{\epsilon}
     -2\left[ \frac{d\Gamma}{dx} \right]^{\rm LO} \beta_0 \alphas \ln\mu
     + \cdots .
\label {eq:lnmu2}
\end {equation}
The difference between the $\ln\mu$ terms of
(\ref{eq:lnmu1}) and (\ref{eq:lnmu2}) is made up by the last
term of the renormalization (\ref{eq:NLOren0}), as we'll now
make explicit while also keeping track of all $O(\eps^0)$
pieces of the conversion.

To start, we need the $d{=}2{-}\eps$ dimensional result for the leading-order
single splitting process, which appears in (\ref{eq:NLOren0}).
We'll find it convenient to write this as
\begin {subequations}
\label {eq:LOd}
\begin {equation}
   \left[ \frac{d \Gamma}{dx} \right]^{\rm LO}_{d=2-\eps}
   = 2\Re \left[ \frac{d \Gamma}{dx} \right]_{\substack{x\bar x\hfill\\d=2-\eps}} ,
\end {equation}
where complex-valued $[d\Gamma/dx]_{x \bar x}$ is the result for the
$x\bar x$ diagram of fig.\ \ref{fig:LO}:%
\footnote{
  Specifically, eqs.\ (3.1), (3.2) and (3.7) of ref.\ \cite{dimreg} give
  (\ref{eq:LOd}) above, except one needs to include the factor
  of $\mu^\eps$ discussed previously.
  See also the QED version in
  eq.\ (F.44) of ref.\ \cite{QEDnf}.
}
\begin {align}
  \left[ \frac{d\Gamma}{dx} \right]_{\substack{x\bar x\hfill\\d=2-\eps}}
  &=
  - \frac{\mu^\eps\alphas d}{8\pi} \,
    P(x) \,
    \Beta(\tfrac12{+}\tfrac{d}{4},-\tfrac{d}{4}) \,
    \Bigl( \frac{2\pi}{M_0\Omega_0} \Bigr)^{\eps/2}
    i \Omega_0
\nonumber\\
  &= \frac{\alphas}{2\pi} \, P(x) \,
    i\Omega_0
    \left[ 1 + \frac{\eps}{2} \ln\Bigl( \frac{\pi\mu^2}{M_0 \Omega_0}\Bigr)
           + O(\eps^2) \right]
  .
\end {align}
\end {subequations}
Here $\Beta(x,y) \equiv \Gamma(x)\,\Gamma(y)/\Gamma(x{+}y)$ is the
Euler Beta function;
we use the short-hand notations $\Omega_0$ and $M_0$ for
\begin {equation}
  \Omega_0 \equiv \Omega_{-1,x,1-x}
   = \sqrt{ \frac{-i\qhatA}{2 E}
            \left( -1 + \frac{1}{x} + \frac{1}{1-x} \right) }
   = \sqrt{ \frac{-i (1-x+x^2) \qhatA}{2x(1-x)E} } \,,
\label {eq:Om0def}
\end {equation}
\begin {equation}
   M_0 \equiv M_{-1,x,1-x} = x(1{-}x)E \,;
\end {equation}
and the DGLAP $g\to gg$ splitting function $P(x)$, given by (\ref{eq:Pgg}),
is independent of dimension (see appendix \ref{app:abcdim}).
Using (\ref{eq:LOd}), we rewrite the renormalized rate (\ref{eq:NLOren0})
as
\begin {equation}
   \left[ \Delta \frac{d\Gamma}{dx} \right]^{\rm NLO,ren}_{g\to gg}
   =
   \left[ \frac{d\Gamma}{dx} \right]_\renlog
   +
   \left[ \Delta \frac{d\Gamma}{dx} \right]^{\NLObar}_{g\to gg}
\label {eq:NLOren}
\end {equation}
with
\begin {equation}
   \left[ \frac{d\Gamma}{dx} \right]_\renlog
   \equiv
   - \beta_0\alphas
       \Re\left(
         \left[ \frac{d\Gamma}{dx} \right]_{\substack{x\bar x\hfill\\d=2}}
         \left[
            \ln \Bigl( \frac{\mu^2}{\Omega_0 E} \Bigr)
            + \ln\Bigl( \frac{x(1{-}x)}{4} \Bigr)
            + \gammaE
         \right]
       \right)
\label {eq:renlog}
\end {equation}
and
\begin {equation}
   \left[ \Delta \frac{d\Gamma}{dx} \right]^{\NLObar}_{g\to gg}
   \equiv
   \left[ \Delta \frac{d\Gamma}{dx} \right]^{\rm NLO,bare}_{g\to gg}
   +
     2\beta_0 \alphas \Re\left(
         \left[ \frac{d\Gamma}{dx} \right]_{\substack{x\bar x\hfill\\d=2}}
         \Bigl[
           \frac{1}{\eps} + \ln\Bigl(\frac{\pi\mu^2}{\Omega_0 E}\Bigr)
         \Bigr]
      \right) .
\label {eq:NLObar}
\end {equation}
The first term $[d\Gamma/dx]_{\renlog}$
of (\ref{eq:NLOren}) contains the correct
explicit $\ln\mu$ dependence of (\ref{eq:lnmu1}).  The
second term $[d\Gamma/dx]^{\NLObar}$ has,
by virtue of (\ref{eq:lnmu2}),
no net divergence $1/\eps$ and no net explicit dependence on
$\ln\mu$.  In appendix \ref{app:NLOsummary}, we implement this
combination (\ref{eq:NLObar})
by grouping all $1/\eps$ pieces of our unrenormalized
calculations into the form
\begin {equation}
  \sigbare \biggl(
     \frac{1}{\eps} + \ln\Bigl(\frac{\pi\mu^2}{\Omega_0 E}\Bigr)
  \biggr) .
\end {equation}
Setting $\sigbare{=}1$ displays unrenormalized formulas for
$[d\Gamma/dx]^{\rm NLO}$.  Setting
$\sigbare{=}0$ instead implements the combination
$[d\Gamma/dx]^{\NLObar}$ of (\ref{eq:NLObar}) once all diagrams
are summed over.
In this way, appendix \ref{app:NLOsummary} simultaneously presents
both bare and renormalized expressions for NLO $g\to gg$.

For later convenience, we find it useful to also define
\begin {equation}
   \left[ \frac{d\Gamma}{dx} \right]^{\LObar}_{g\to gg}
   \equiv
   \left[ \frac{d\Gamma}{dx} \right]^{\rm LO,ren}_{g\to gg}
   + \left[ \frac{d\Gamma}{dx} \right]_\renlog
\label {eq:dGLObar}
\end {equation}
so that we can rewrite (\ref{eq:rateren}) as
\begin {equation}
   \left[ \Delta \frac{d\Gamma}{dx} \right]^{\rm LO+NLO}_{g\to gg}
   =
   \left[ \frac{d\Gamma}{dx} \right]^{\LObar}
   +
   \left[ \Delta \frac{d\Gamma}{dx} \right]^{\NLObar}_{g\to gg} .
\label {eq:raterenbar}
\end {equation}
This is the meaning behind the notation we used back in
(\ref{eq:dGnet}).  The notation is convenient because, for our
final renormalized $g{\to}gg$ results listed in appendix
\ref{app:summary}, the notation distinguishes the
parts $[\Delta\,d\Gamma/dx]^{\NLObar}$ of our results that are expressed
in terms of $y$ integrals,%
\footnote{
  Specifically, $[\Delta\,d\Gamma/dx]^{\NLObar}$ is given by
  (\ref{eq:dGammaNLObar}) and the formulas following it with
  $\sigma_{\rm bare}=0$.
}
like in (\ref{eq:dGnetNLO1}),
from the parts $[d\Gamma/dx]^{\LObar}$ above that are not.


\section{IR divergences in Energy Loss Calculations}
\label {sec:IR}

We now discuss in detail how the IR behavior of
various measures of the development of
in-medium high-energy QCD parton showers depends only on the
combination
\begin {equation}
   \V(x,y) + \tfrac12 \R(x,y)
   \approx
   -\frac{\CA\alphas}{8\pi}
   \left[ \frac{d\Gamma}{dx} \right]^{\rm LO} \frac{\ln y}{y}
\label {eq:VRlimit2}
\end {equation}
of virtual and real diagrams introduced in (\ref{eq:VRlimit}),
for which power-law IR divergences cancel.
In this section, $\approx$ indicates an equality that is valid at
leading-log order.


\subsection {General shower evolution}

We start by looking generally at the evolution of the distribution
of partons in such a shower.  This will generalize, to NLO,
similar methods that
have been applied by Blaizot et al.\ at leading order
\cite{Nevolve1,Nevolve2}.%
\footnote{
  See also earlier leading-order work by Jeon and Moore \cite{JeonMoore},
  which avoided the $\hat q$ approximation and treated
  the quark-gluon plasma as weakly coupled.
}

In what follows, let $E_0$ be the energy of the initial parton that
starts the entire shower.  We will let $\zeta E_0$ refer to the energy
of some parton in the shower as the shower develops, and we will refer
to the distribution of shower partons in $\zeta$ at time $t$ as
$N(\zeta,E_0,t)$.  Formally, the total number of partons remaining in
the shower at time $t$ is then $\int_0^1 d\zeta \> N(\zeta,E_0,t)$, but
this particular integral is IR divergent, not least because some fraction of
the energy of the shower will have come to a stop in the medium
($\zeta{=}0$) and thermalized by time $t$.
However,
one may also
use $N(\zeta,E_0,t)$ to calculate IR-safe characteristics of the
shower, including $N(\zeta,E_0,t)$ itself for fixed $\zeta > 0$.%
\footnote{
  See the leading-order analysis of $N(\zeta,E_0,t)$ in
  refs.\ \cite{Nevolve1,Nevolve2}.  (Be aware that their analytic
  results depend on approximating $[d\Gamma/dx]^{\rm LO}$ by something
  more tractable.)
  For a next-to-leading-order example, see the related
  discussion of charge stopping distance and other moments of the
  charge stopping distribution for large-$\Nf$ QED in appendix C of
  ref.\ \cite{qedNfstop}.
}


\subsubsection {Basic Evolution Equation}

The basic evolution equation to start with is
(see appendix \ref{app:details} for some more detail)%
\footnote{
  This equation is only meant to be valid for particle energies
  $\zeta E$ large compared to the temperature $T$ of the plasma.
  In the high-energy and infinite-medium limit that we are working
  in, the evolution of particles in the shower whose energy has degraded
  to $\sim T$ has a negligible (i.e.\ suppressed by a power of $T/E$)
  effect on questions about in-medium shower development
  and calculations of where the shower deposits its energy into the
  plasma.  See, for example, the discussions in refs.\ \cite{stop}
  and \cite{Nevolve1}.  For discussion of some of the theory issues
  that would be involved in going beyond this high-energy approximation
  for single-splitting processes,
  see, as two examples, refs.\ \cite{JeonMoore} and \cite{Ghiglieri}.
}
\begin {equation}
  \frac{\partial}{\partial t} N(\zeta,E_0,t)
  =
  - \Gamma(\zeta E_0) \, N(\zeta,E_0,t)
  + \int_\zeta^1 \frac{dx}{x} \> 
    \left[ \frac{d\Gamma}{dx} \bigl(x,\tfrac{\zeta E_0}{x}\bigr) \right]_{\rm net}
    N\bigl( \tfrac{\zeta}{x}, E_0, t \bigr) ,
\label {eq:Nevolve0}
\end {equation}
where
\begin {equation}
   \left[ \frac{d\Gamma}{dx} (x,E) \right]_{\rm net}
\end {equation}
refers to the net rate (\ref{eq:dGnet}) to produce one daughter of energy
$xE$ (plus any other daughters) via single splitting or overlapping
double splitting from a parton of energy $E$.
The total splitting rate $\Gamma$ in the loss term is
\begin {equation}
   \Gamma(E)
   =
   \frac{1}{2!} \int_0^1 dx 
   \left\{
      \left[ \frac{d\Gamma}{dx} \right]^{\LObar}
      +
      \left[ \Delta \frac{d\Gamma}{dx} \right]^{\NLObar}_{g \to gg}
   \right\}
   +
   \frac{1}{3!} \int_0^1 dx \int_0^{1-x} dy
       \left[ \Delta \frac{d\Gamma}{dx\,dy} \right]_{g \to ggg}
   ,
\label {eq:Gtot}
\end {equation}
where the $1/2!$ and $1/3!$ are the final-state identical particle factors for
$g \to gg$ and $g \to ggg$.
The first and second terms in (\ref{eq:Nevolve0}) are respectively
loss and gain terms for $N(\zeta,E_0,t)$.
The gain term corresponds to the rate for any higher-energy particle
in the shower (energy $\zeta E_0/x$)
to split and produce a daughter whose energy is $\zeta E_0$.
To keep formulas simple here and throughout this discussion,
we will not explicitly write the IR cut-off $\delta$ in integration
limits.

By comparing (\ref{eq:Gtot}) to (\ref{eq:dGnet}), note that
\begin {equation}
   \Gamma(E) \not=
   \int_0^1 dx \> \left[ \frac{d\Gamma}{dx} (x,E) \right]_{\rm net}
\end {equation}
because of the different combinatoric factors involved in how
$[d\Gamma/dx]_{\rm net}$ is defined.  This is related to the fact
that (\ref{eq:Nevolve0}) should not conserve the total number of
partons: each $g \to gg$ should add a parton, and each
$g \to ggg$ should add two partons.%
\footnote{
  One way to see this clearly is to
  over-simplify the problem by {\it pretending} that splitting
  rates did not depend on energy $E$, then integrate both
  sides of (\ref{eq:Nevolve0}) over $\zeta$, and rewrite
  $\int_0^1 d\zeta \int_\zeta^1 dx/x = \int_0^1 dx \int_0^1 d\bar\zeta$
  with $\bar\zeta \equiv \zeta/x$.  Formally, this would give
  $\partial{\cal N}/\partial t =
    +(\Gamma_{g\to gg} + 2 \, \Delta\Gamma_{g\to ggg}) {\cal N}$,
  where ${\cal N}$ is the total number of partons in the shower and
  $\Gamma_{g \to gg} \equiv \Gamma^{\rm LO} + \Delta\Gamma^{\rm NLO}_{g\to gg}$.
  From the coefficients $+1$ and $+2$ of $\Gamma_{g\to gg}$ and
  $\Delta\Gamma_{g\to ggg}$ in this expression, one can see explicitly the
  number of partons added by each type of process.
}

The various pieces that go into the calculation of the right-hand side
of the evolution equation (\ref{eq:Nevolve0}) have various power-law
IR divergences which cancel in the combination of all the terms.
We now focus on identifying those divergences and showing how to
reorganize (\ref{eq:Nevolve0}) into an equivalent
form where power-law IR divergences
are eliminated from the integrals that must be done.


\subsubsection {$x\to 0$ or $1$ divergences at leading order}

To start, let's ignore $\NLObar$ corrections for a moment
and look at the leading-order version of (\ref{eq:Nevolve0}):
\begin {equation}
  \frac{\partial}{\partial t} N(\zeta,E_0,t)
  \simeq
  - \Gamma^{\LObar}(\zeta E_0) \, N(\zeta,E_0,t)
  + \int_\zeta^1 \frac{dx}{x} \> 
    \left[ \frac{d\Gamma}{dx} \bigl(x,\tfrac{\zeta E_0}{x}\bigr)
       \right]^{\LObar}
    N\bigl( \tfrac{\zeta}{x}, E_0, t \bigr)
\label {eq:NevolveLO}
\end {equation}
with
\begin {equation}
   \Gamma^{\LObar}(E)
   =
   \frac{1}{2!} \int_0^1 dx 
      \left[ \frac{d\Gamma}{dx} \right]^{\LObar} .
\label {eq:GLO}
\end {equation}
The leading-order rate $[d\Gamma/dx]^{\rm LO}$ diverges as
\begin {equation}
   \left[ \frac{d\Gamma}{dx} \right]^{\rm LO}
   \sim
   \frac{1}{[x(1{-}x)]^{3/2}}
   \qquad
   \mbox{as $x\to 0$ or $1$}
\label {eq:dGLOlims}
\end {equation}
[see eq.\ (\ref{eq:LO0}) with $\eps{=}0$].  Up to logarithmic factors,
this divergence is the same for $[d\Gamma/dx]^{\LObar}$ (\ref{eq:dGLObar})
as well. 
This means that the integral (\ref{eq:GLO}) that gives the total
rate $\Gamma^{\LObar}$ generates power-law IR divergences from both
the $x\to 0$ and $x\to 1$ parts of the integration region.
In contrast, the integral for the gain term in (\ref{eq:NevolveLO})
runs from $\zeta{>}0$ to $1$ and so only generates a
divergence from the $x\to 1$ behavior.
That means that we cannot get rid of the IR divergences simply
by directly combining the integrands.  However, if we first use
the identical final-particle symmetry $x \leftrightarrow 1{-}x$ of
$[d\Gamma/dx]^{\LObar}$ to rewrite (\ref{eq:GLO}) as
\begin {equation}
   \Gamma^{\LObar}(E)
   =
   \int_{1/2}^1 dx 
      \left[ \frac{d\Gamma}{dx} \right]^{\LObar} ,
\end {equation}
then we can combine the loss and gain terms in (\ref{eq:NevolveLO})
into
\begin {multline}
  \frac{\partial}{\partial t} N(\zeta,E_0,t)
  \simeq
  \int_0^1 dx
  \biggl\{
    -
    \left[ \frac{d\Gamma}{dx} \bigl(x,\zeta E_0\bigr) \right]^{\LObar}
    N\bigl( \zeta, E_0, t \bigr) \,
    \theta(x > \tfrac12 )
\\
    +
    \left[ \frac{d\Gamma}{dx} \bigl(x,\tfrac{\zeta E_0}{x}\bigr) \right]^{\LObar}
    N\bigl( \tfrac{\zeta}{x}, E_0, t \bigr) \,
    \frac{\theta(x > \zeta)}{x}
  \biggr\} .
\label {eq:NevolveLO2}
\end {multline}
Similar to (\ref{eq:dGnetNLO}),
we have implemented the actual limits of integration here using step
functions $\theta(\cdots)$ so that we may combine the integrands.
Because of the $\theta$ functions, the integrand has no support
for $x \to 0$ and so no divergence associated with $x \to 0$.
Because we have combined the integrands, however, one can see
that the integrand behaves like $1/(1{-}x)^{1/2}$ instead of
$1/(1{-}x)^{3/2}$ (\ref{eq:dGLOlims}) as $x \to 1$ because
of cancellation in that limit between the loss and gain contributions.
So the form (\ref{eq:NevolveLO2}) has the advantage that the integral
is completely convergent, and there are no IR divergences in this
equation for any given $\zeta > 0$.


\subsubsection {$y\to 0$ divergences at NLO}

As discussed in section \ref{sec:introIR}, $g{\to}ggg$ and NLO $g{\to}gg$
processes generate power-law IR divergences as the energy of the
softest real or virtual gluon (whose longitudinal momentum fraction we
often arrange to correspond to the letter $y$) goes to zero.
We have already discussed how
those power-law IR divergences cancel in the combination
$[\Delta\,d\Gamma/dx]_{\rm net}^{\NLObar}$ (\ref{eq:dGnetNLO}),
which is the combination
that appears in the NLO contribution to the gain term in
the evolution equation (\ref{eq:Nevolve0}).
But the loss term involves a different combination $\Gamma$
(\ref{eq:Gtot})
of real and virtual diagrams, and so we must
check that a similar cancellation occurs there.

\begin {figure}[t]
\begin {center}
  \includegraphics[scale=0.7]{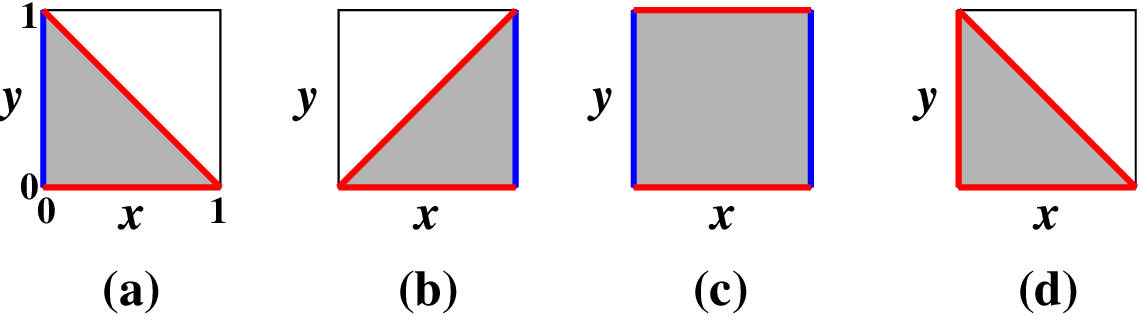}
  \caption{
     \label{fig:Gxylims}
     The integration regions (shaded) for the various terms in (\ref{eq:Gtot2})
     corresponding to
     (a) Class I virtual diagrams,
     (b) their $x \to 1{-}x$ cousins,
     (c) Class II virtual diagrams, and
     (d) $g \to ggg$ diagrams.
     The colored lines correspond to limits of the integration regions,
     which generate IR divergences.
     See the caption of fig.\ \ref{fig:Gxylims2} for the distinction
     between the red vs.\ blue lines here.
  }
\end {center}
\end {figure}

Recalling that our NLO $g \to gg$ diagrams consist of our Class I diagrams
(fig.\ \ref{fig:virtI}), their $x \to 1{-}x$ cousins, and
our class II diagrams (fig.\ \ref{fig:virtII}), the NLO contribution
to the total rate
(\ref{eq:Gtot}) is, in more detail,
\begin {multline}
   \Delta\Gamma^{\NLObar}(E)
   =
   \frac{1}{2!} \int_0^1 dx
   \biggl\{
     \biggl(
       \int_0^{1-x} dy \, \left[ \Delta \frac{d\Gamma}{dx\,dy} \right]_\virtI
     \biggr)
     + (x \to 1{-}x)
     + \int_0^1 dy \, \left[ \Delta \frac{d\Gamma}{dx\,dy} \right]_\virtII
   \biggr\}
\\
   +
   \frac1{3!}
   \int_0^1 dx
   \int_0^{1{-}x} dy \>
      \left[ \Delta \frac{d\Gamma}{dx\,dy} \right]_{g\to ggg} .
\label {eq:Gtot2}
\end {multline}
[Compare and contrast to (\ref{eq:dGnetNLOdef}).]
Fig.\ \ref{fig:Gxylims} shows the various integration regions corresponding
to the different terms above and the limits of integration
producing IR divergences (which is all of them).

\begin {figure}[t]
\begin {center}
  \includegraphics[scale=1.5]{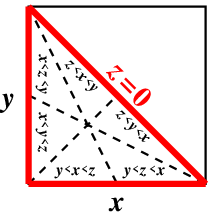}
  \caption{
     \label{fig:xyzRegions}
     Equivalent integration regions for $g \to ggg$ corresponding
     to permutations of the daughters $(x,y,z)$.
     The common vertex of these regions is at
     $(x,y,z) = (\tfrac13,\tfrac13,\tfrac13)$.
  }
\end {center}
\end {figure}

\begin {figure}[t]
\begin {center}
  \includegraphics[scale=0.7]{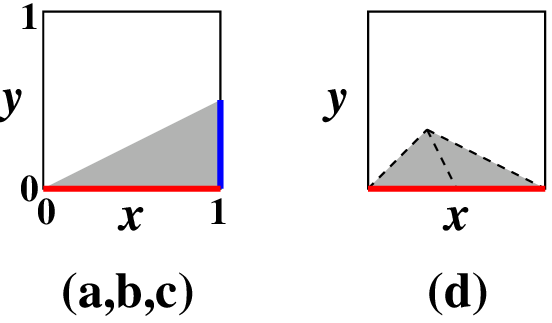}
  \caption{
     \label{fig:Gxylims2}
     The integration regions in (\ref{eq:Gtot4}).
     The labels (a,b,c) and (d) correspond to the original origin
     of these terms in fig.\ \ref{fig:Gxylims}.
     Colored lines again denote limits of integration associated
     with IR divergences.  Power-law divergences associated with
     the red lines above cancel each other in (\ref{eq:Gtot4}).
     Blue line divergences only cancel when loss and gain
     terms are combined in (\ref{eq:SevolveNLO}).  The origins of the
     red vs.\ blue divergences here are depicted by the red vs.
     blue lines in fig.\ \ref{fig:Gxylims}.
  }
\end {center}
\end {figure}

We will now align the location of the IR divergences so that we can
eventually combine the different integrals and eliminate
power-law divergences.
First, note by change $x \to 1{-}x$ of integration variables,
the ``$(x \to 1{-}x)$'' term in (\ref{eq:Gtot2}) gives the same
result as the ``$\virtI$'' term.  Second, simultaneously use
the $x \to 1{-}x$ and $y \to 1{-}y$ symmetries of Class II diagrams
to divide the integration region of fig.\ \ref{fig:Gxylims}c in half
diagonally, giving
\begin {multline}
   \Delta\Gamma^{\NLObar}(E)
   =
   \int_0^1 dx \int_0^{1-x} dy \>
   \biggl\{
     \left[ \Delta \frac{d\Gamma}{dx\,dy} \right]_\virtI
     + \left[ \Delta \frac{d\Gamma}{dx\,dy} \right]_\virtII
   \biggr\}
\\
   +
   \frac1{3!}
   \int_0^1 dx
   \int_0^{1{-}x} dy \>
      \left[ \Delta \frac{d\Gamma}{dx\,dy} \right]_{g\to ggg} .
\label {eq:Gtot3}
\end {multline}
For the NLO $g{\to}gg$ contributions, we now divide the integration
region into (i) $0 < y < (1{-}x)/2$ and (ii) $(1{-}x)/2 < y < 1{-}x$
and change integration variables $y \to z = 1{-}x{-}y$ in the latter,
similar to the manipulations used earlier to obtain (\ref{eq:dGnetNLO}).
For the $g{\to}ggg$ contributions, note that permutation symmetry for
the three final daughters $(x,y,z)$ implies the integral over each of
the six regions shown in fig.\ \ref{fig:xyzRegions} is the same.
We can therefore replace the integral over all six regions by three
times the integral over the bottom two, depicted by the shaded
region of fig.\ \ref{fig:Gxylims2}d.  [We will see later the
advantage of integrating over these two regions instead of reducing
the integral to just one region.]
Eq.\ (\ref{eq:Gtot3}) can
then be written as
\begin {multline}
   \Delta\Gamma^{\NLObar}(E)
   =
   \int_0^1 dx \int_0^{1/2} dy
   \biggl\{
     \V(x,y) \, \theta(y<\tfrac{1-x}{2})
     + \tfrac12 \R(x,y) \, \theta(y<x) \, \theta(y<\tfrac{1-x}{2})
   \biggr\} ,
\end {multline}
with $v$ and $r$ defined as in (\ref{eq:VRdef}).
We will find it convenient to change
integration variable
$x \to 1{-}x$ in the first term and rewrite the equation as
\begin {multline}
   \Delta\Gamma^{\NLObar}(E)
   =
   \int_0^1 dx \int_0^{1/2} dy
   \biggl\{
     \V(1{-}x,y) \, \theta(y<\tfrac{x}{2})
     + \tfrac12 \R(x,y) \, \theta(y<x) \, \theta(y<\tfrac{1-x}{2})
   \biggr\} .
\label {eq:Gtot4}
\end {multline}
The integration regions corresponding to the two terms in
(\ref{eq:Gtot4}) are shown in fig.\ \ref{fig:Gxylims2}, where
the only IR divergences correspond to $y{\to}0$ or $x{\to}1$.

The rationale for the last change was to convert
$x{\to}0$ divergences into $x{\to}1$ divergences (the blue line
in fig.\ \ref{fig:Gxylims2}), which we will later see then
cancel similar $x{\to}1$ divergences in the gain term of the
evolution equation.  For the moment, however, we focus
only on the $y{\to}0$ divergences of (\ref{eq:Gtot4}), depicted
by the red lines in fig.\ \ref{fig:Gxylims2}.
In the limit $y\to 0$ (for fixed $x$), the integrand in
(\ref{eq:Gtot4}) approaches
\begin {equation}
  \V(1{-}x,y) + \tfrac12 \R(x,y)
  = \V(1{-}x,y) + \tfrac12 \R(z,y)
  \simeq \V(1{-}x,y) + \tfrac12 \R(1{-}x,y) ,
\label {eq:VRlimitG}
\end {equation}
where the first equality follows because $g \to ggg$ is symmetric under
permutations of $(x,y,z)$.
The right-hand side of (\ref{eq:VRlimitG})
is the same combination as (\ref{eq:VRlimit}) but with
$x \to 1{-}x$.  In fig.\ \ref{fig:dbllogCheck}, we verified
numerically that $y^{-3/2}$ divergences (which generate power-law IR divergences
when integrated) indeed cancel in this combination, leaving behind
the double-log divergence shown in (\ref{eq:VRlimit}) [which happens
to be symmetric under $x \to 1{-}x$].  Interested readers can find
non-numerical information on how the $y^{-3/2}$ divergences cancel in
appendix \ref{app:IRcancel}.

One can now see why we did not replace the integral of $r(x,y)$ over
the two sub-regions shown in fig.\ \ref{fig:Gxylims2} by, for example,
twice the integral of just the left-hand sub-region $(x < y < z)$.
If we had done the latter, there would be no $r$ term for
$x > 1/2$ and so nothing would cancel the $y^{-3/2}$ divergence
of $v(1{-}x,y)$ for $x > 1/2$.  We had to be
careful how we organized things to achieve our goal that the $y$ integral in
(\ref{eq:Gtot4}) not generate a power-law IR divergence for
any value of $x$.

Next, we turn to our final goal for this section
of showing that the integrals
in the evolution equation for
$N(\zeta,E_0,t)$ can be arranged to directly avoid
power-law IR divergences for the entire integration over {\it both}\/
$x$ and $y$.


\subsubsection {$x\to 0$ or $1$ divergences at NLO}
\label {sec:xNLOdivs}

By using (\ref{eq:dGnetNLO}), (\ref{eq:NevolveLO2}), and (\ref{eq:Gtot4})
in the shower evolution equation (\ref{eq:Nevolve0}), we can now
combine integrals to avoid all power-law divergences:
\begin {subequations}
\label {eq:NevolveRV}
\begin {equation}
  \frac{\partial}{\partial t} N(\zeta,E_0,t)
  =
  {\cal S}^{\LObar} + {\cal S}^{\NLObar}
\end {equation}
where
\begin {multline}
  {\cal S}^{\LObar}
  =
  \int_0^1 dx
  \biggl\{
    -
    \left[ \frac{d\Gamma}{dx} \bigl(x,\zeta E_0\bigr) \right]^{\LObar}
    N\bigl( \zeta, E_0, t \bigr) \,
    \theta(x > \tfrac12 )
\\
    +
    \left[ \frac{d\Gamma}{dx} \bigl(x,\tfrac{\zeta E_0}{x}\bigr) \right]^{\LObar}
    N\bigl( \tfrac{\zeta}{x}, E_0, t \bigr) \,
    \frac{\theta(x > \zeta)}{x}
  \biggr\}
\end {multline}
and
\begin {multline}
  {\cal S}^{\NLObar}
  =
  \int_0^1 dx \int_0^{1/2} dy
  \biggl\{
     - \Bigl[
         \V(1{-}x,y) \, \theta(y<\tfrac{x}{2})
         + \tfrac12 \R(x,y) \, \theta(y<x) \, \theta(y<\tfrac{1-x}{2})
       \Bigr]
       N\bigl( \zeta, E_0, t \bigr)
\\
     + \Bigl[
         \V(x,y) \, \theta(y<\tfrac{1-x}{2})
         + \V(1{-}x,y) \, \theta(y<\tfrac{x}{2})
         + \R(x,y) \, \theta(y<\tfrac{1-x}{2})
       \Bigr]
       N\bigl( \tfrac{\zeta}{x}, E_0, t \bigr) \,
       \frac{\theta(x > \zeta)}{x}
  \biggr\} .
\label {eq:SevolveNLO}
\end {multline}
\end {subequations}
We've previously seen that the LO piece ${\cal S}^{\LObar}$ is free of
divergences.  And we've seen that the loss and gain terms of the
NLO piece ${\cal S}^{\NLObar}$ are each free of power-law divergences
associated with $y\to 0$ (with fixed $x$).  Now consider divergences
of ${\cal S}^{\NLObar}$ associated with the behavior of $x$.
The integrand in (\ref{eq:SevolveNLO}) has no
support as $x \to 0$ (fixed $y$).  And for $x\to 1$ (fixed $y$),
there is a cancellation between the loss and gain terms.
So there is no divergence of ${\cal S}^{\NLObar}$
associated with $x \to 0$ or $x \to 1$.%
\footnote{
  This statement relies on the observation that the various
  NLO $g \to gg$ differential rates making up $v(x,y)$ diverge
  no faster than $s^{-3/2}$ as some parton with longitudinal momentum
  fraction $s$ becomes soft, e.g.\ $(1-x)^{-3/2}$ as $x \to 1$.
  The cancellation between the gain and loss terms
  in (\ref{eq:SevolveNLO}) reduces
  that by one power, to $(1-x)^{-1/2}$, which is an integrable
  singularity and so generates no divergence for (\ref{eq:SevolveNLO}).
}

In summary, the only IR divergences coming from ${\cal S}^{\NLObar}$
are the uncanceled double-log divergences associated with
$y \to 0$.


\subsection {Absorbing double logs into \boldmath$\hat q$
  and comparison with known results}

Refs.\ \cite{Blaizot,Iancu,Wu} have previously performed leading-log
calculations of overlap corrections and shown that the double-log
IR divergences can be absorbed into the medium parameter $\hat q$.
We will now verify that the double-log piece of our results produces
the same modification \cite{Wu0} of $\hat q$.


\subsubsection {Double-log correction for $[d\Gamma/dx]_{\rm net}$}
\label {sec:qhateff}

Let's start with the relatively simple situation of the
$[d\Gamma/dx]_{\rm net}$ introduced in section \ref{sec:introIR}.
From the discussion of (\ref{eq:dGnetNLO}) through (\ref{eq:VRlimit}),
the double-log divergence of the NLO contribution to
$[d\Gamma/dx]_{\rm net}$ is given by%
\footnote{
  Note that (\ref{eq:renlog}) has no IR double log contribution, so the
  distinction between (LO,NLO) and $(\LObar,\NLObar)$ can be ignored
  for this discussion.
}
\begin {align}
  \left[ \frac{d\Gamma}{dx} \right]_{\rm net}^{\rm NLO}
  \approx
  - \frac{\CA\alphas}{4\pi} \left[ \frac{d\Gamma}{dx} \right]^{\rm LO}
    \int_\delta^{1/2} dy \> \frac{\ln y}{y}
  \approx
  \frac{\CA\alphas}{8\pi} \ln^2\delta
 ,
\label {eq:dGnetNLODblLog}
\end {align}
where we have re-introduced our sharp IR cut-off $\delta$.
Combining (\ref{eq:dGnetNLODblLog}) with
$[d\Gamma/dx]_{\rm net} = [d\Gamma/dx]^{\rm LO} + [d\Gamma/dx]_{\rm net}^{\rm NLO}$
gives
\begin {equation}
  \left[ \frac{d\Gamma}{dx} \right]_{\rm net} \simeq
  \left[
     1 + \frac{\CA\alphas}{8\pi} \ln^2\delta
  \right]
  \left[ \frac{d\Gamma}{dx} \right]^{\rm LO}
  .
\label {eq:dGLOeff}
\end {equation}
Since $[d\Gamma/dx]^{\rm LO} \propto \sqrt{\hat q/E}$
[see (\ref{eq:OmMdefs}) and (\ref{eq:LO0})],
the double-log correction above can be
absorbed at this order by replacing $\hat q$ by
\begin {equation}
  \hat q_{\rm eff} = 
  \left[
     1 + \frac{\CA\alphas}{4\pi} \ln^2\delta 
  \right]
  \hat q .
\label {eq:qhateffdelta}
\end {equation}

The corresponding
leading-log modification of $\hat q$ from earlier literature
\cite{Wu0,Blaizot,Iancu,Wu} is usually
expressed in the final form
\begin {equation}
  \hat q_{\rm eff}(L) = 
  \left[
     1 + \frac{\CA\alphas}{2\pi} \ln^2\Bigl( \frac{L}{\tau_0} \Bigr)
  \right]
  \hat q ,
\label {eq:qhateffStd}
\end {equation}
where $L$ is the thickness of the medium and $\tau_0$ is taken to be
of order the mean free path for elastic scattering in the medium.
In order to compare (\ref{eq:qhateffdelta}) and (\ref{eq:qhateffStd}),
we need to translate.

First, for simplicity, we have been working in the infinite-medium
approximation, which assumes that
the size of the medium is large compared to all relevant formation lengths.
Eq.\ (\ref{eq:qhateffStd}) instead focuses on the
phenomenologically often-relevant case where the width $L$ of the medium
is $\lesssim$ the formation time $t_{\rm form}(x)$ associated with the harder
splitting $x$.  One may convert at leading-log level
by considering the boundary case where
\begin {equation}
   L \sim t_{\rm form}(x) .
\label {eq:Lsim}
\end {equation}
Parametric substitutions like this inside the arguments of logarithms
are adequate for a leading-log analysis.

\begin {figure}[t]
\begin {center}
  \begin{picture}(450,150)(0,0)
    \put(0,25){\includegraphics[scale=0.5]{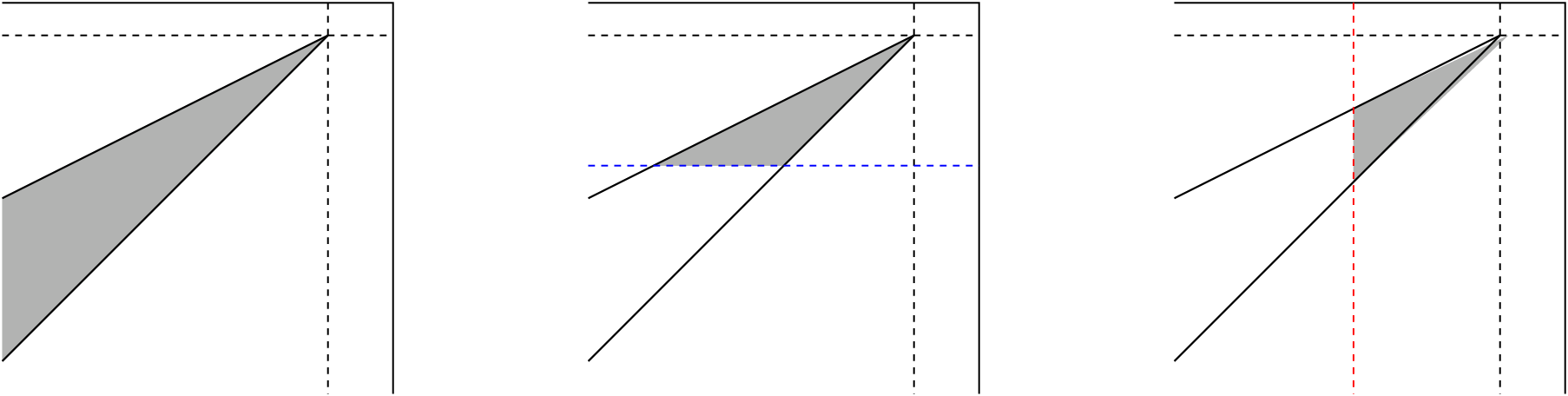}}
    \put(0,115){$\scriptstyle{\Delta t\,\sim\,t_{\rm form}(x)}$}
    \put(83,46){\rotatebox{-90}{$\scriptstyle{y\,\sim\,x}$}}
    \put(50,140){$\ln y$}
    \put(113,90){\rotatebox{-90}{$\ln\Delta t$}}
    \put(50,0){(a)}
    \put(163,115){$\scriptstyle{\Delta t\,\sim\,t_{\rm form}(x)}$}
    \put(246,46){\rotatebox{-90}{$\scriptstyle{y\,\sim\,x}$}}
    \put(213,140){$\ln y$}
    \put(276,90){\rotatebox{-90}{$\ln\Delta t$}}
    \put(213,0){(b)}
    \put(178,80){$\color{blue}\scriptstyle{\Delta t\,\sim\,\tau_0}$}
    \put(326,115){$\scriptstyle{\Delta t\,\sim\,t_{\rm form}(x)}$}
    \put(409,46){\rotatebox{-90}{$\scriptstyle{y\,\sim\,x}$}}
    \put(376,140){$\ln y$}
    \put(439,90){\rotatebox{-90}{$\ln\Delta t$}}
    \put(376,0){(c)}
    \put(365,46){\rotatebox{-90}{$\color{red}\scriptstyle{y\,\sim\,\delta}$}}
%
  \end{picture}
  \caption{
     \label {fig:region}
     The region of integration (\ref{eq:region})
     giving rise to a double log in the
     $\hat q$ approximation with (a) no cut off, (b) the cut off
     $\Delta t \sim \tau_0$ used in earlier literature, and (c)
     the IR regulator $y \sim \delta$ used in our calculations.
     See text for discussion.
  }
\end {center}
\end {figure}
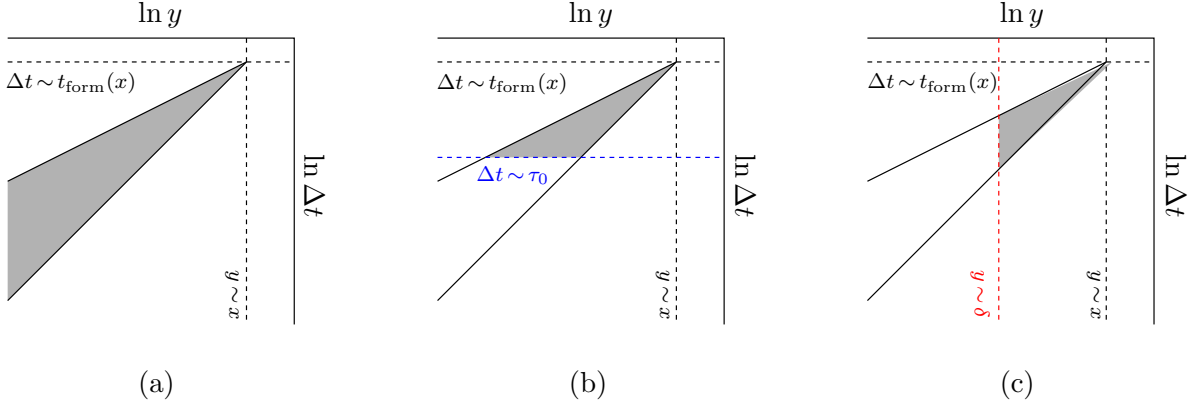

What remains is to translate between the use of two different types
of cut-offs
in (\ref{eq:qhateffdelta}) and (\ref{eq:qhateffStd}): $\delta$ and
$\tau_0$.  To understand the effect of the cut-offs, it is useful
to review where double logs come from in the $\hat q$ approximation,
at first ignoring the cut-offs altogether.  Parametrically, the IR
double log arises from an integral of the form
\begin {equation}
   \iint \frac{dy}{y} \> \frac{d(\Delta t)}{\Delta t}
\label {eq:Dlog0}
\end {equation}
over the integration region shown in fig.\ \ref{fig:region}a, given by%
\footnote{
   Using (\ref{eq:Lsim}) and $t_{\rm form}(\xi) \sim \sqrt{\xi E/\hat q}$
   for small $\xi$,
   (\ref{eq:tregion}) can be put in the form
   $y \sqrt{E/x\hat q} \ll \Delta t \ll \sqrt{y E/\hat q}$ presented
   in eq.\ (9.3) of ref.\ \cite{2brem} for
   $y \ll x \le z$.  The equivalence, in turn, with notation used in some
   of the original work on double logs in the NLO LPM effect is
   discussed in appendix F.1 of ref.\ \cite{2brem}.
}
\begin {subequations}
\label {eq:region}
\begin {equation}
   \frac{yE}{\hat q L}
   \ll \Delta t
   \ll t_{\rm form}(y) \,.
\label {eq:tregion}
\end {equation}
Using
$t_{\rm form}(y) \sim \sqrt{yE/\hat q}$ for small $y$,
these inequalities can
be equivalently expressed as a range on $y$:
\begin {equation}
   \frac{\hat q (\Delta t)^2}{E}
   \ll y
   \ll \frac{\hat q L\,\Delta t}{E}
   \,.
\label {eq:yregion}
\end {equation}
\end {subequations}

Now consider two different ways to evaluate the double logarithm
(\ref{eq:Dlog0}).  The first method is to add a lower cut-off
$\tau_0$ on $\Delta t$, as in fig.\ \ref{fig:region}b.
Using (\ref{eq:yregion}), that's
\begin {equation}
   \approx
   \int_{\tau_0}^L \frac{d(\Delta t)}{\Delta t}
   \int_{\hat q(\Delta t)^2/E}^{\hat q L\,\Delta t/E} \frac{dy}{y}
   =
   \int_{\tau_0}^L \frac{d(\Delta t)}{\Delta t} \,
   \ln\Bigl( \frac{L}{\Delta t} \Bigr)
   =
   \tfrac12 \ln^2\Bigl( \frac{L}{\tau_0} \Bigr) .
\label {eq:Dlogtau}
\end {equation}
Alternatively, adding a lower cut-off $\delta$ on $y$ as in
fig.\ \ref{fig:region}c, using (\ref{eq:Lsim}), and assuming
$x \le \frac12$ so that parametrically
$t_{\rm form}(x) \sim \sqrt{x E/\hat q}$,
the double log (\ref{eq:Dlog0}) is regulated to
\begin {equation}
   \approx
   \int_\delta^x \frac{dy}{y}
   \int_{y E/\hat q L}^{t_{\rm form}(y)} \frac{d(\Delta t)}{\Delta t}
   =
   \int_\delta^x \frac{dy}{y}
   \ln \Bigl( \frac{\hat q L \, t_{\rm form}(y)}{y E} \Bigr)
   \approx
   \int_\delta^x \frac{dy}{y}
   \ln \Bigl( \sqrt{\frac{x}{y}} \Bigr)
   =
   \tfrac14 \ln^2 \Bigl( \frac{\delta}{x} \Bigr) .
\label {eq:Dlogdelta}
\end {equation}
When we extract just the double log dependence $\ln^2\delta$
on the parameter $\delta$, there is no difference (for fixed $x$) at leading-log
order between $\ln^2( \delta/x )$ and $\ln^2\delta$.
At that level, comparison of (\ref{eq:Dlogtau}) and (\ref{eq:Dlogdelta})
gives the leading-log translation
\begin {equation}
  \ln^2\Bigl( \frac{L}{\tau_0} \Bigr)
  \longrightarrow
  \tfrac12 \ln^2\delta
\label {eq:DlogTranslate}
\end {equation}
between IR-regularization with $\tau_0$ and $\delta$.
Applied to the standard double log result (\ref{eq:qhateffStd}),
this translation exactly reproduces the double log behavior
(\ref{eq:qhateffdelta}) of our own results.

We will return to the $x$ dependence of (\ref{eq:Dlogdelta})
when we later examine sub-leading single-log corrections in section
\ref{sec:SingleLog}.

Our $\delta$ is simply a formal IR regulator.
In contrast, there is a plausible physical reason for using
the elastic mean free path $\tau_0$ as an IR regulator at the
double log level: The $\hat q$ approximation used throughout our
discussion and earlier literature
is a multiple-scattering approximation that requires
long time periods compared to the mean free time between collisions.
However, beyond leading-log order, the use of a
$\tau_0$ cut-off would be problematic for full NLO calculations.
In our calculations, a $\tau_0$ cut-off would interfere with
the correct UV-renormalization of $\alphas$,
which comes from $\Delta t \to 0$
(and small enough time scales that even $\hat q$-approximation propagators
faithfully reproduce vacuum propagators).  So in this paper
we have just chosen the formal IR regulator, $\delta$, that seemed most
convenient for our calculations.

In order to use IR-regulated results for NLO splitting rates, one
must either compute quantities that are IR-safe in the $\hat q$
approximation or else make an appropriate matching calculation for
soft emission that takes into account how the QCD LPM effect turns off
for formation lengths $\lesssim \tau_0$.


\subsubsection {Physics scales:
  What if you wanted to take $\delta$ more seriously?}
\label {sec:scales}

\begin {figure}[t]
\begin {center}
  \begin{picture}(180,200)(0,0)
    \put(60,75){\includegraphics[scale=0.5]{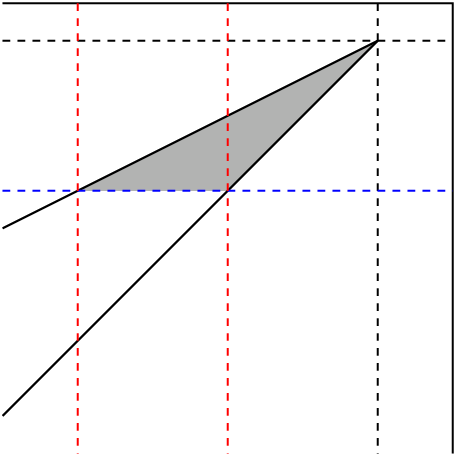}}
    \put(0,137){$\color{blue}{\scriptstyle{
       \Delta t\,\sim\,\tau_0 \sim 1/g^2T
     }}$}
    \put(75,75){\rotatebox{-90}{$\color{red}{
       \scriptstyle{yE\,\sim\,\hat q\tau_0^2\,\sim\,T}
     }$}}
    \put(111,75){\rotatebox{-90}{$\color{red}{
       \scriptstyle{yE\,\sim\,\hat q L \tau_0^{}\,\sim\,\sqrt{x E T}}
     }$}}
    \put(110,190){$\ln y$}
    \put(173,140){\rotatebox{-90}{$\ln\Delta t$}}
  \end{picture}
  \caption{
     \label {fig:region2}
     Parametric scales associated with various features of
     fig.\ \ref{fig:region}b.  The expressions in terms of
     $\tau_0$, $\hat q$ and $L$ match those
     of the original work \cite{Wu0} on double log corrections
     to $\hat q$.
  }
\end {center}
\end {figure}
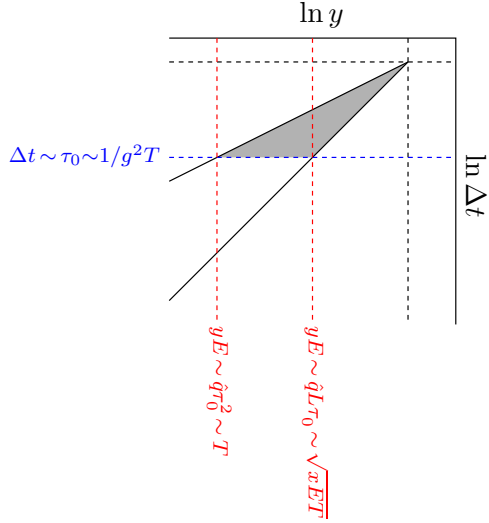

Though we are simply taking $\delta$ as a formal IR cut-off for
calculations involving the $\hat q$ approximation, we should mention
what the physics scales are where our $\hat q$-based analysis would
break down if one used our results for calculations that were
sensitive to the value of $\delta$.  The situation is complicated
because there are potentially two scales to consider,
indicated in
fig.\ \ref{fig:region2}.  We have given parametric formulas
for those scales for the case of a weakly-coupled quark-gluon
plasma.  One may translate to a strongly-coupled 
quark-gluon plasma, in both the figure and the discussion below,
simply by erasing the factors of $g$.

Parametrically, the mean free time between (small-angle)
elastic collisions with the medium is $\tau_0 \sim 1/g^2 T$,
and $\hat q$ is $\sim g^4 T^3$.  Using the
limits (\ref{eq:yregion}) on $y$, as well as (\ref{eq:Lsim})
and $t_{\rm form(x)} \sim \sqrt{xE/\hat q}$,
one then finds
for $\Delta t \sim \tau_0$ the corresponding
soft gluon energies $yE$ indicated in the figure.

Our formalism breaks down for $yE$ smaller
than the lower limit $yE \sim T$ because gluons of energy $T$
cannot be treated as high-energy compared to the plasma.
Note that if one correspondingly chose $\delta \sim T/E$
without also constraining $\Delta t$, then the resulting double
log region would be larger than has been conventionally assumed
in the literature.  In contrast, if one chose
$\delta \sim \sqrt{xT/E}$, corresponding to the other red line
in fig.\ \ref{fig:region2}, then one would guarantee that
$\Delta t \gtrsim \tau_0$ but the resulting double log region
would be smaller than the one used in the literature.
There is no choice of $\delta$ alone that corresponds to the
traditional shaded region of fig.\ \ref{fig:region2}.


\subsubsection {Double-log correction for shower evolution equation}

The gain term of the shower evolution equation (\ref{eq:Nevolve0})
depends only on the combination $[d\Gamma/dx]_{\rm net}$ of rates,
and so the same redefinition (\ref{eq:qhateffdelta}) will absorb
the double logarithmic divergence.  One expects that this must
also work for the loss term in (\ref{eq:Nevolve0}), which depends
on the combination $\Gamma$, but we should make sure.  Since we found that
only $y \to 0$ ultimately contributes to the double logarithm
in our later version (\ref{eq:NevolveRV}) of the  evolution equation,
we can focus on the $y{\to}0$ behavior of the NLO loss term
for fixed $x$, which corresponds to the $y{\to}0$ behavior of the
integrand of (\ref{eq:Gtot4}) for $\Delta\Gamma^{\NLObar}$.
Using (\ref{eq:VRlimitG}) and
(\ref{eq:VRlimit}), the
double log generated by the $y$ integration in (\ref{eq:Gtot4})
is
\begin {equation}
   \Delta\Gamma^{\rm NLO}
   \approx
   - \frac{\CA\alphas}{8\pi} 
   \int_0^1 dx \left[ \frac{d\Gamma}{dx} \right]^{\rm LO} 
   \int_\delta^{1/2} dy \> \frac{\ln y}{y}
   \approx
   \frac{\CA\alphas}{16\pi} 
   \int_0^1 dx \left[ \frac{d\Gamma}{dx} \right]^{\rm LO} 
   \ln^2\delta .
\end {equation}
When combined with the leading-order rate $\Gamma^{\rm LO}$ given
by (\ref{eq:GLO}), we have
\begin {equation}
   \Gamma
   \approx
   \frac{1}{2!}
   \int_0^1 dx \>
   \left[
     1 + \frac{\CA\alphas}{8\pi} \ln^2\delta
   \right]
   \left[ \frac{d\Gamma}{dx} \right]^{\rm LO} ,
\end {equation}
which indeed involves the same correction to $[d\Gamma/dx]^{\rm LO}$,
and so to $\hat q$, as (\ref{eq:dGLOeff}).


\subsection {Why not talk about \boldmath$dE/dL$?}

In the literature, it is common to discuss energy loss per unit
length ($dE/dL$) for a high-energy particle.  This makes sense
only if one can unambiguously identify the original particle after
a process that has degraded its energy.
For many applications of the LPM effect, the energy loss occurs by
radiation that is soft compared to the initial particle energy $E$,
and so one can identify the particle afterwards as the only one that
still has very high energy.  In this paper, however, we have been
focused on the case of a very thick medium (thick compared to
formation lengths).  In that case, hard bremsstrahlung is an
important aspect of energy loss.  If the two daughters of a splitting
have comparable energies, it becomes more difficult to say which
is the successor of the original.  For a double-splitting process
beginning with a quark, one can unambiguously (for large $\Nc$) choose
to follow the original quark.  But, for processes that begin with
$g{\to}gg$, the distinction is less clear.

\begin {figure}[t]
\begin {center}
  \includegraphics[scale=0.7]{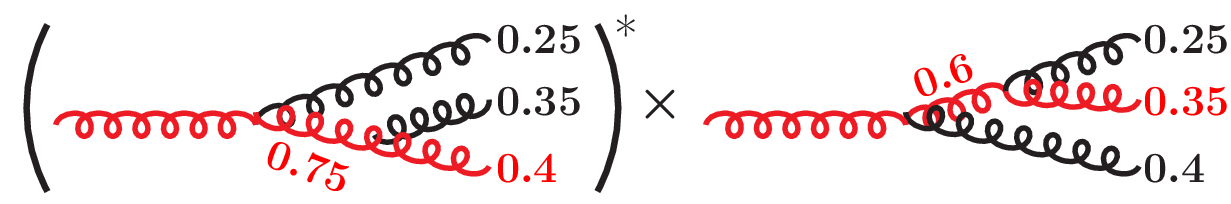}
  \caption{
     \label{fig:ambiguityGlue}
     An example of an interference between two different amplitudes
     for double splitting $g \to gg \to ggg$.  The numbers show the
     energy fractions of gluons relative to the first gluon
     that initiated the double-splitting process.  The red follows
     the highest-energy daughter of each individual $g{\to}gg$
     process
  }
\end {center}
\end {figure}

One possibility might be to formally define $dE/dL$ for
$g{\to gg}$ processes by always following after each splitting
the daughter gluon that has the highest energy of the two daughters.
Unfortunately, this procedure is ill-defined when analyzing
the effect of overlapping formation times on successive splittings.
Consider the interference shown in fig.\ \ref{fig:ambiguityGlue}
of two different amplitudes for double splitting $g \to gg \to ggg$.
For each amplitude, the red gluon line shows which gluon we would
follow by choosing the highest-energy daughter of each individual
$g{\to}gg$ splitting.  The two amplitudes do not agree on which
of the final three gluons is the successor of the original gluon.
That's not a problem if the individual splittings are well
enough separated
that the interference can be ignored, i.e.\ if formation lengths
for the individual splittings do not overlap.  But since we are
interested specifically in calculating such interference,
we have no natural way of defining which gluon to follow.
This is why we have avoided $dE/dL$ and
focused on more general measures of shower evolution.

The above argument generalizes to $g \to ggg$ points made in
ref.\ \cite{qedNfstop} about
$e \to \gamma e \to \bar e e e$, $q \to gq \to \bar q q q$
and $q\ \to gq \to ggq$.
However, in those cases, ref.\ \cite{qedNfstop} noted that
$dE/dL$ was nonetheless well-defined in the large $\Nf$ or
$\Nc$ limits respectively.  In contrast, the $g{\to}ggg$ interference
shown in fig.\ \ref{fig:ambiguityGlue} is unsuppressed in
the large-$\Nc$ limit.


\subsection {Similar power-law IR cancellations}

LPM splitting rates and overlap corrections scale with energy like
$\sqrt{\hat q/E}$, up to logarithms.
For situations where rates are proportional to a power $E^{-\nu}$
of energy,
ref.\ \cite{qedNfstop} discusses how to derive relatively
simple formulas for the stopping distance of a shower, and more
generally formulas for various moments of the distribution of where
the energy of the shower is deposited.  Those formulas can
also be adapted to the case where the rates also have
single-logarithmic
dependence $E^{-\nu} \ln E$.  This is adequate for
analyzing stopping distances for QED showers \cite{qedNfstop},
but the application to QCD, which has double logs, is unclear.
But even for QCD, one can use those stopping
length formulas as yet
another context in which to explore the cancellation of power-law
IR divergences.
See appendix \ref{app:lstop} for that analysis.


\section{IR single logarithms}
\label {sec:SingleLog}

\subsection{Numerics}

\begin {figure}[t]
\begin {center}
  \includegraphics[scale=0.55]{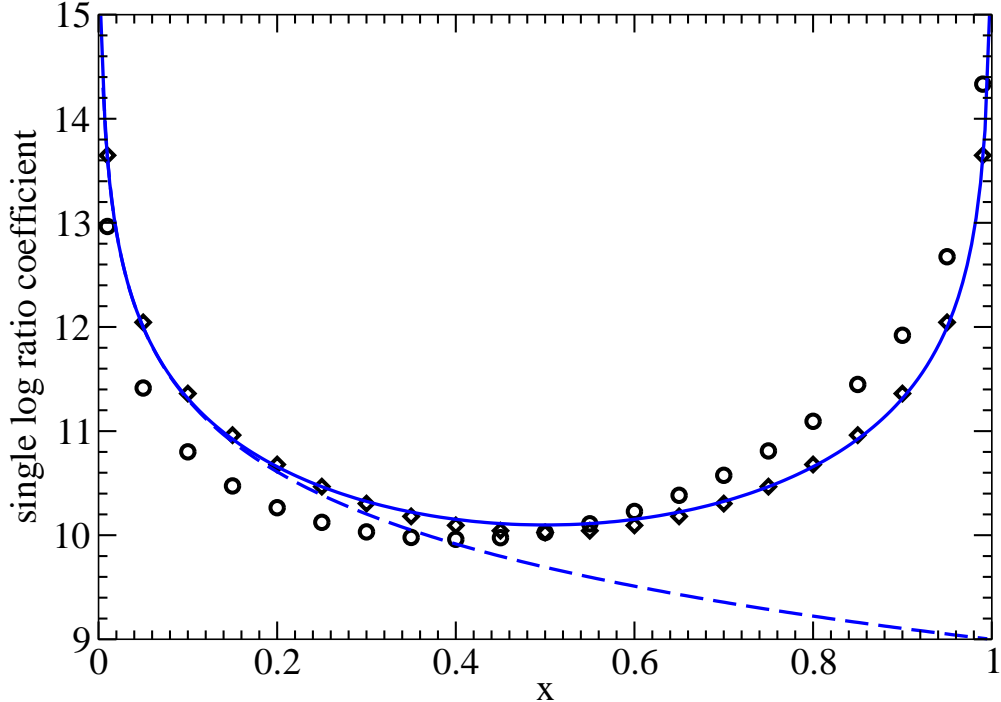}
  \caption{
    \label {fig:SingleLogs}
    [See {\it note added} at end of the main text.]
    The single-log coefficients $s$ (circles) and $\bar s$ (diamonds)
    as a function of $x$.  The left-most and right-most data points are
    for $x{=}0.01$ and $x{=}0.99$, while all other data points
    are evenly spaced at
    $x{=}0.05$, $0.10$, $0.15$, ..., $0.85$, $0.90$, $0.95$.
    For comparison,
    the dashed blue curve shows the anticipated small-$x$ behavior
    (\ref{eq:sbarsmallx}) with constant $c$ fit by (\ref{eq:c}),
    and the solid blue curve shows the
    educated guess (\ref{eq:sbarform}) for the full $x$ dependence.
  }
\end {center}
\end {figure}

In (\ref{eq:VRlimit}) and section \ref{sec:qhateff}, we extracted the
known IR double logarithm from the slope of a straight-line fit
to the small-$y$ behavior of our full numerical results when plotted
as
\begin {equation}
  \frac{
     \V(x,y) + \tfrac12 \R(x,y)
     \vphantom{\Big|}
  }{
     \frac{\CA\alphas}{8\pi}
     \left[ \frac{d\Gamma}{dx} \right]^{\rm LO} \frac{1}{y}
     \vphantom{\Big|}
  }
\label {eq:VRratio2}
\end {equation}
vs.\ $\ln y$, as in fig.\ \ref{fig:dbllogCheck}.
The sub-leading single-log behavior can
be similarly found, for each value of $x$,
from the {\it intercept}\/ of that straight-line fit.
Specifically, refine (\ref{eq:VRlimit}) to include single-log
effects by writing
\begin {equation}
   \V(x,y) + \tfrac12 \R(x,y)
   \simeq
   -\frac{\CA\alphas}{8\pi}
   \left[ \frac{d\Gamma}{dx} \right]^{\rm LO} \frac{\bigl(\ln y + s(x)\bigr)}{y}
   \,.
\label {eq:VRwithSingle}
\end {equation}
Here, the $y^{-1} \ln y$ term generates the known double-log behavior
$\propto \ln^2\delta$ after integration over $y$, and the new
$s(x)\,y^{-1}$ term allows for additional single-log behavior
$\propto \ln\delta$.
Then the combination (\ref{eq:VRratio2}) behaves at small $y$ like
\begin {equation}
  \frac{
     \V(x,y) + \tfrac12 \R(x,y)
     \vphantom{\Big|}
  }{
     \frac{\CA\alphas}{8\pi}
     \left[ \frac{d\Gamma}{dx} \right]^{\rm LO} \frac{1}{y}
     \vphantom{\Big|}
  }
  \simeq
  -\bigl( \ln y + s(x) \bigr) .
\label {eq:VRratioSingle}
\end {equation}
The right-hand side represents the straight line fit of
fig.\ \ref{fig:dbllogCheck}, and the intercept of that fit at $\ln y = 0$
gives $-s(x)$.
Our numerical results for $s(x)$ are shown by circles in
fig.\ \ref{fig:SingleLogs}.
Note that $s(x)$ is not symmetric under $x \to 1{-}x$.
That's because we defined $v(x,y)$ in (\ref{eq:Vdef})
to contain Class I virtual diagrams but not their
$x \to 1{-}x$ cousins.

We do not have anything interesting to say about the precise shape of
$s(x)$ itself.  But we can get to something interesting if we note that our
original discussion (\ref{eq:VRlimit}) of the small-$y$ behavior of
$v(x,y)+\tfrac12 r(x,y)$ was in the context of $[d\Gamma/dx]_{\rm net}$,
where $v(x,y)+\tfrac12 r(x,y)$ appeared in the $x \leftrightarrow 1{-}x$
symmetric combination
\begin {equation}
   \bigl[ \V(x,y) + \tfrac12 \R(x,y) \bigr] + \bigl[ x \to 1{-}x \bigr]
\label {eq:VR2}
\end {equation}
of (\ref{eq:VR}).
For this combination, the single log piece corresponds to twice
the average
\begin {equation}
  \bar s(x) \equiv \frac{s(x)+s(1{-}x)}{2}
\end {equation}
of $s(x)$ over $x \leftrightarrow 1{-}x$.
This $\bar s(x)$ is depicted by the diamonds in
fig.\ \ref{fig:SingleLogs}.
And even though we currently have only numerical results for $\bar s(x)$,
we will be able to make some interesting observations about its form
by comparing our numerics to an educated guess that we will discuss in
a moment.

$[d\Gamma/dx]_{\rm net}$, and thus $\bar s(x)$, also appears in
our other discussions of IR behavior, such as the gain term
in the evolution equation (\ref{eq:Nevolve0}) for the gluon
distribution $N(\zeta,E_0,t)$.
The loss term of that equation depends on the total rate
$\Gamma$, which treats the two identical daughters of $g \to gg$
processes $x$ and $1{-}x$ on an equal footing.%
\footnote{
  As was true for $[d\Gamma/dx\,dy]_{\rm net}$, the $r(x,y)$ contribution
  representing $g \to ggg$ is symmetric in
  $x \leftrightarrow z \equiv 1{-}x{-}y$ rather than $x\leftrightarrow 1{-}x$,
  but the difference is unimportant in the $y{\to}0$ limit we are using
  to extract IR divergences.  More specifically, the
  difference between $r(x,y) = r(1{-}x{-}y,y)$ and $r(1{-}x,y)$ is
  parametrically smaller as $y{\to}0$ than the $1/y$ terms responsible
  for the single-log IR divergence under discussion.
}
So $\bar s(x)$ is the relevant function for single log divergences,
regardless of the fact that we found it convenient to rewrite
$\Gamma$ in (\ref{eq:NevolveRV}) in a way
that obscured the $x \leftrightarrow 1{-}x$ symmetry of $g{\to}gg$
so that we could make more explicit the cancellation of
power-law IR divergences.%
\footnote{
  If desired, one could achieve both goals by replacing the integrand in
  (\ref{eq:NevolveRV}) by its average
  over $x \leftrightarrow 1{-}x$.
}


\subsection {Educated guess for form of \boldmath$\bar s(x)$}

Let's now return to the issue of $x$ dependence in the translation
of the standard double log result $\ln^2(L/\tau_0)$ in (\ref{eq:Dlogtau})
to the $\ln^2\delta$ of our calculations in (\ref{eq:Dlogdelta}).
Previously, when we compared the two, we ignored the $x$ dependence
of the $\ln^2(\delta/x)$ in (\ref{eq:Dlogdelta}).  Now keeping track of
that $x$ dependence, the translation (\ref{eq:DlogTranslate}) becomes
\begin {equation}
  \ln^2\Bigl( \frac{L}{\tau_0} \Bigr)
  \longrightarrow
  \tfrac12 \ln^2\Bigl( \frac{\delta}{x} \Bigr) .
\label {eq:DlogTranslatex}
\end {equation}
Here we assume $x < 1{-}x$, and
the arguments of the double logarithms are only {\it parametric} estimates.
Rewrite the right-hand side
of (\ref{eq:DlogTranslatex}) as $\ln^2 \Delta$ with
$\Delta \sim \delta/x$.  For $x \ll 1$, this parametric relation
suggests that $\Delta \simeq \#\delta/x$ for some proportionality
constant $\#$.
So (\ref{eq:DlogTranslatex}) suggests
that a more precise substitution for $x \ll 1$ would be
\begin {equation}
  \ln^2\Bigl( \frac{L}{\tau_0} \Bigr)
  \longrightarrow
  \tfrac12 \ln^2\Bigl( \# \frac{\delta}{x} \Bigr)
  =
  \tfrac12 \ln^2\delta
     + \left[ \ln\Bigl(\frac{1}{x} \Bigr) + \ln\# \right] \ln\delta
     + \mbox{(IR convergent)}
  .
\label {eq:DlogTranslatex2}
\end {equation}
Eq.\ (\ref{eq:DlogTranslatex2}) contains information about the
small-$x$ dependence
of the coefficient of the sub-leading, single IR-logarithm $\ln\delta$.

In a moment, we will attempt to generalize to a guess of the
behavior for all values of $x$,
but first let's see how (\ref{eq:DlogTranslatex})
compares to our numerics.  Consider the logarithms arising
from a symmetrized $\bar s$ version of (\ref{eq:VRwithSingle}), whose
integral over $y$ would be proportional to
\begin {equation}
  - \int_\delta dy \> \frac{\bigl( \ln y + \bar s(x) \bigr)}{y}
  = \tfrac12 \ln^2\delta + \bar s(x) \ln \delta
    + (\mbox{IR convergent}) .
\label {eq:sbarlogint}
\end {equation}
Comparison of (\ref{eq:DlogTranslatex2}) with (\ref{eq:sbarlogint})
suggests that
\begin {equation}
  \bar s(x) \simeq \ln\Bigl(\frac{1}{x} \Bigr) + c
  \qquad (y \ll x \ll 1) ,
\label {eq:sbarsmallx}
\end {equation}
where $c=\ln\#$ is a constant that is not determined by this argument
and must be fit to our numerics.  The dashed blue curve in
fig.\ \ref{fig:SingleLogs} shows (\ref{eq:sbarsmallx}) with%
\astfootnote{
  See {\it note added} at end of main text.
}
\begin {equation}
   c = 9.0
\label {eq:c}
\end {equation}
on the graph of our full numerical results.
The form (\ref{eq:sbarsmallx}) works well for small $x$.

To make an educated guess for the full $x$ dependence of
$\bar s(x)$, we need to replace (\ref{eq:sbarsmallx}) by
something symmetric in $x \leftrightarrow 1{-}x$.
The formation time $t_{\rm form}(x)$, related to the harmonic
oscillator frequency $\Omega_0$ of (\ref{eq:Om0def}) by
\begin {equation}
  \frac{1}{[t_{\rm form}(x)]^2} \sim |\Omega_0|^2
  =
  \left|
    \frac{-i \qhatA}{2E} \Bigl( -1 + \frac{1}{x} + \frac{1}{1-x} \Bigr)
  \right| ,
\label {eq:Om0sqr}
\end {equation}
is symmetric in $x \leftrightarrow 1{-}x$ and plays a major role in
the LPM effect.  So, even though our arguments about double logs
have only been parametric, let us see what happens if we guess
that the $1/x$ in
(\ref{eq:sbarsmallx}) is arising from the small $x$ behavior of
(\ref{eq:Om0sqr}), and so we replace (\ref{eq:sbarsmallx}) by
\begin {equation}
  \bar s(x) =
  \ln \Bigl( -1 + \frac{1}{x} + \frac{1}{1-x} \Bigr) + c .
\label {eq:sbarform}
\end {equation}
This guess is shown by the solid blue curve in fig.\ \ref{fig:SingleLogs}.


\subsection {How well does the educated guess work?}

\begin {figure}[t]
\begin {center}
  \includegraphics[scale=0.55]{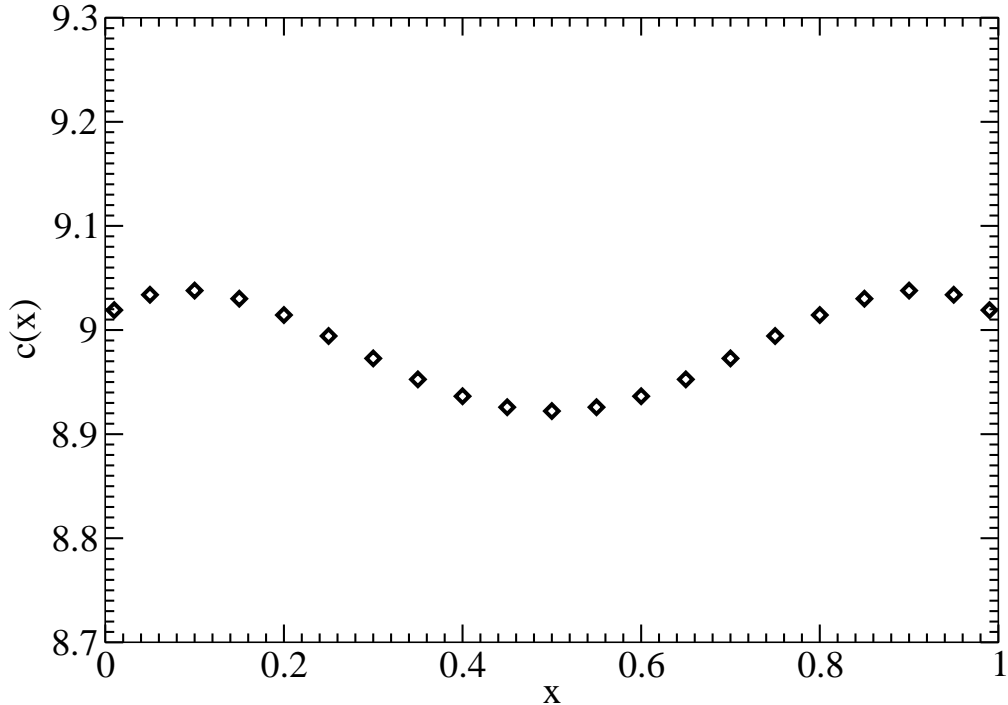}
  \caption{
    \label {fig:c}
    [See {\it note added} at end of the main text.]
    Extraction via (\ref{eq:cx})
    of the $x$-dependence of the ``constant'' $c$
    in the form (\ref{eq:sbarform}) for $\bar s(x)$.
  }
\end {center}
\end {figure}

As the figure shows, (\ref{eq:sbarform}) captures the $x$ dependence
of the single log coefficient $\bar s(x)$
very well.  However, it is not quite perfect.
To see the discrepancies, one may use (\ref{eq:VRwithSingle})
together with (\ref{eq:sbarform}) to extract from our numerical
results for $v(x,y)+\tfrac12r(x,y)$ the best choice $c(x)$
of $c$ for each {\it individual}\/ value of $x$:
\begin {equation}
  c(x) \equiv
  \lim_{y\to 0}
  \left\{
     \frac{ \frac12\bigl(
               [\V(x,y) + \tfrac12 \R(x,y)] + [x \leftrightarrow 1{-}x]
             \bigr) }
          {  -\frac{\CA\alphas}{8\pi}
             \left[ \frac{d\Gamma}{dx} \right]^{\rm LO} \frac{1}{y}
             \vphantom{\Big|} }
  -
  \left[
     \ln y + \ln \Bigl( -1 + \frac{1}{x} + \frac{1}{1-x} \Bigr) 
  \right]
  \right\} .
\label {eq:cx}
\end {equation}
If the guess (\ref{eq:sbarform}) for the form of $\bar s(x)$ were
exactly right, then $c(x)$ would be an $x$-independent constant.
But fig.\ \ref{fig:c} shows a small variation of our
$c(x)$ with $x$.  Our educated guess is
a good approximation but appears not to be the entire
story for understanding IR single logs.
The variation of $c(x)$ in fig.\ \ref{fig:c} is the reason that
we have not bothered to determine the small-$x$ value of $c$
in (\ref{eq:sbarsmallx}) to better precision than (\ref{eq:c}).

We should note that the value of $c$ will be IR-regularization
scheme dependent.
If we had regulated the IR with a smooth cut-off at $p^+ \sim P^+ \delta$
instead of a hard cut off, a different value of $c$ would be needed
to keep the physics the same on the right-hand side of
(\ref{eq:sbarlogint}) with the different meaning of $\delta$.


\section{Theorist Error}
\label {sec:error}

The results presented in Appendix \ref{app:summary} for overlap
effects on double splitting calculations represent the culmination
of a very long series of calculations
\cite{2brem,seq,dimreg,QEDnf} that required addressing many subtle
technical issues as well as many involved arguments computing
expansions in $\eps$ for novel dimensionally-regulated quantities.
In the absence of calculations by an independent group using
independent methods,
a natural worry must be whether somewhere our group
might have made a mistake that would noticeably affect our final
results.  We refer to this possibility as ``theorist error,'' in
contrast to ``theoretical error'' estimates of
uncertainty arising from the approximations used.

Though we cannot absolutely guarantee the absence of theorist error,
we think it useful to
list a number of cross-checks and features of our calculations.
Some of these check our treatment of technical subtleties of
the calculation.

{\it 1.}
  The power-law IR divergences computed for real and virtual diagrams
  in the $\hat q$ approximation cancel
  each other, as discussed in this paper.
  Sub-leading IR divergences, which do not cancel,
  correctly reproduce the IR double log \cite{Wu0} known from previous,
  independent calculations \cite{Blaizot,Iancu,Wu}
  that analyzed overlap effects in leading-log approximation.

{\it 2.}
  Our calculation generates the correct $1/\eps$ UV divergences for the known
  renormalization of $\alphas$.
  This includes the cancellation of mixed UV-IR divergences,
  which is one of the subtleties of Light-Cone Perturbation Theory.

{\it 3.}
  In the soft limit $y \ll x \ll 1$ of $g \to ggg$,
  crossed \cite{2brem} and sequential \cite{seq} diagrams give
  contributions to $\Delta\Gamma/dx\,dy$ that behave like
  $\ln(x/y)/x y^{3/2}$.  But the logarithmic enhancement of
  these $1/x y^{3/2}$ contributions {\it cancels}\/ when
  all $g {\to} ggg$ processes are
  added together, reassuringly consistent with the
  Gunion-Bertsch picture presented in appendix B of ref.\ \cite{seq}.
  When our formalism is applied instead to large-$\Nf$ QED \cite{QEDnf},
  the analogous logarithm does not cancel.  In that case,
  its coefficient reassuringly matches what one would expect from
  DGLAP-like arguments, as explained in section 2.2.3 of ref.\ \cite{QEDnf}.

{\it 4.}
  One of the technical subtleties of our methods has to do with identifying the
  correct branch to take for logarithms $\ln C$ of complex or negative numbers,
  which may arise in dimensional regularization, for example, from
  the expansion of a $C^\eps$.  See section 4.6 and appendix H of
  ref.\ \cite{dimreg}, as well as appendix H.1 of ref.\ \cite{QEDnf}, for
  examples where the determination of the appropriate branch requires
  care.  Making a mistake of $\pm 2\pi i$ in the
  evaluation of a logarithm would generally have a significant effect
  on our results.  But we do have some consistency checks on such
  ``$\pi$ terms'' that result from the logarithm of the phases of
  complex numbers in our calculation.  One check is illustrated by
  appendix \ref{app:IRcancel}, where $\pi$ terms associated
  with individual diagrams must all cancel as one part of the
  cancellation of IR power-law divergences.
  A different, somewhat indirect cancellation test of
  $\pi$ terms generated by dimensional regularization is given
  in appendix D of ref.\ \cite{dimreg}.

{\it 5.}
  Here is another test of an $O(\eps^0)$ term in the expansion of
  dimensional regularization of a UV-divergent diagram.
  Recall that both $g{\to}ggg$ and NLO $g{\to}gg$ processes have
  power-law IR divergences of the form
  $\int_\delta dy/y^{3/2} \sim \delta^{-1/2}$, where the power law $y^{-3/2}$
  matches a physical argument given in section I.D of ref.\ \cite{seq}.
  In the calculation of divergent diagrams, the UV-sensitive piece
  of the calculation is isolated
  into what are called ``pole'' pieces in refs.\ \cite{2brem,seq,dimreg,QEDnf}
  and in appendix \ref{app:summary}.
  These pole pieces are evaluated analytically with
  dimensional regularization and
  yield $1/\eps$ divergences plus finite $O(\eps^0)$
  contributions.  The remaining UV-insensitive contributions to the
  diagrams are evaluated with numerical integration.  For
  some of the crossed virtual diagrams (top line of fig.\ \ref{fig:virtI}),
  both the $O(\eps^0)$ pole piece and the UV-insensitive numerical
  integral%
\footnote{
    In formulas, the pole piece of the crossed virtual diagrams
    corresponds to eq.\ (\ref{eq:ApoleIc}) for $A^{\rm pole}_\virtIc$.
    whereas the UV-insensitive piece is the integral shown in
    (\ref{eq:AvirtIc}).  For more details on exactly how the pole
    piece is defined, see appendix \ref{app:method}.
}
  turn out to have spurious IR divergences that are {\it more} IR divergent
  than the power-law divergences we have discussed.  However,
  they also turn out to exactly cancel each other.  For example,
  in appendix \ref{app:Dxi}, we show how the integral
  associated with $2\Re(x y \bar y\bar x)$ has an unwanted
  $\int dy/y^2 \sim \delta^{-1}$
  divergence from $y{\to}0$ that is canceled by the $O(\eps^0)$ piece of the
  UV-divergent pole term.%
\footnote{
  This is unrelated (as far as we know) to a different class of cases,
  where individual diagrams have unwanted
  IR divergences that are only canceled by similar divergences of
  another diagram.  See the two pairs of $\int dz/z^{5/2}$ divergences
  in Table \ref{tab:limits} in appendix \ref{app:IRcancel}.
}


\section{Conclusion}
\label {sec:conclusion}

The results of this paper (combined with those of earlier papers)
are the complete formulas in appendix
\ref{app:summary} for the effects of overlapping formation times
associated with the various $g{\to}ggg$ and $g{\to}gg$ processes of
figs.\ \ref{fig:crossed}--\ref{fig:virtII}.  But there are still
missing pieces we need before we can answer the qualitative question
which motivates this work:
Are overlap effects small enough that an
in-medium shower can be treated as a collection of individual
high-energy partons, assuming one first absorbs potentially large
double logarithms into the effective value of $\hat q$?

First, for a complete calculation, we will also need processes involving
longitudinal gluon exchange and direct 4-gluon vertices, such as
in fig.\ \ref{fig:later}.  The methods for computing those diagrams
are known, and so it should only take an investment of care and time
to include them.

More importantly, our results as given are double-log IR divergent.
The known double-log IR divergence can easily be subtracted away from our
results and absorbed into the effective value of $\hat q$ reviewed
in section \ref{sec:qhateff}.  However, this potentially leaves behind
a sub-leading {\it single}-log IR divergence.  We've seen from numerics
that much of those single-log divergences can also be absorbed into
$\hat q_{\rm eff}$ by accounting for the $x$ dependence of the
natural choice of scale for the double-log contribution to
$\hat q_{\rm eff}$, but there remains a smaller part of the single-log
IR divergences that is not yet understood.
In order to make progress and understand the structure of the
single logarithms, we hope in the future to extract analytic (as opposed to
numerical) results for them from our full diagrammatic results.
We have also not yet determined whether diagrams involving longitudinal
gluon exchange, which have so far been left out, contribute to
IR single logarithms.

It would be extremely helpful, both conceptually and as a check of our
own work, if someone can figure out a way to directly and independently
compute the sub-leading single-log IR divergences without going through
the entire complicated and drawn-out process that we have used to
compute our full results.


\section*{\it Note added}

After this paper was published,
we found an error in eq.\ (\ref{eq:wrong})
for ${\cal A}_{\rm seq}^{\rm pole}(x,y)$,
which gave incorrect $i\pi$
terms when the arguments $x$ or $y$ are negative and so
generated an incorrect result when front-end transformed
for evaluation of type II virtual sequential diagrams.
The correct version is derived in appendix A of ref.\ \cite{logs}.
Here, we have chosen to leave (\ref{eq:wrong}) and
figs.\ \ref{fig:SingleLogs} and \ref{fig:c} as in the original
publication but will describe the changes.
The correct formula for ${\cal A}_{\rm seq}^{\rm pole}(x,y)$ that works
with front-end transformations is quoted in a footnote below
(\ref{eq:wrong}).
The single-log coefficients $s(x)$, $\bar s(x)$,
and $c(x)$ plotted in
figs.\ \ref{fig:SingleLogs} and \ref{fig:c}
are changed to the numbers shown minus $4\pi$.
(See, for example, fig.\ 5 of ref.\ \cite{logs}.)
Correspondingly, the choice $c = 9.0$ in (\ref{eq:c}) should instead be
$c = 9.0-4\pi$.


\begin{acknowledgments}

We are very grateful to Risto Paatelainen for valuable conversations
concerning cancellation of divergences in Light Cone Perturbation Theory.
We also thank Yacine Mehtar-Tani for several discussions over
the years concerning double-log corrections.
This work was supported, in part, by the U.S. Department
of Energy under Grant No.~DE-SC0007984 (Arnold and Gorda)
and the National Natural
Science Foundation of China under
Grant Nos.~11935007, 11221504 and 11890714 (Iqbal).

\end{acknowledgments}

\appendix

\section {Summary of Rate Formulas}
\label {app:summary}

In this appendix, we collect final results for the elements
contributing to the leading-order $g{\to}gg$ rate,
its NLO corrections, and the $g{\to}ggg$ rate:
\begin {equation}
   \left[ \frac{d\Gamma}{dx} \right]^{\rm LO} , 
   \qquad
   \left[ \Delta \frac{d\Gamma}{dx} \right]_{g \to gg}^{\rm NLO} ,
   \qquad
   \left[ \Delta \frac{d\Gamma}{dx\,dy} \right]_{g \to ggg} .
\end {equation}
Throughout this appendix, we define
\begin {equation}
  z \equiv 1{-}x{-}y
\end {equation}
as in the main text.

We remind readers that in this paper we have not included diagrams
involving 4-gluon vertices or instantaneous interactions via
longitudinal gauge boson exchange,
such as the examples of fig.\ \ref{fig:later}.


\subsection{Leading-order splitting rate}

\subsubsection{$d{=}2$ transverse spatial dimensions}

In our notation, the leading-order $g{\to}gg$ rate is
\begin {equation}
   \left[ \frac{d \Gamma}{dx} \right]^{\rm LO}
   = \frac{\alphas}{\pi} \, P(x) \, \Re(i\Omega_0)
\label {eq:LO}
\end {equation}
with
\begin {equation}
   \Omega_0
   = \sqrt{ \frac{-i\qhatA}{2 E}
            \left( -1 + \frac{1}{x} + \frac{1}{1-x} \right) }
   = \sqrt{ \frac{-i (1-x+x^2) \qhatA}{2x(1-x)E} }
\label {eq:Om0}
\end {equation}
and the $g{\to}g$ DGLAP splitting function
\begin {equation}
  P(x)
  = \CA \left| \frac{1 + x^4 + (1-x)^4}{x(1-x)} \right|
  = \CA \left| \frac{2(1 - x + x^2)^2}{x(1-x)} \right| .
\label {eq:Pgg}
\end {equation}
Here and throughout this paper, our $P(x)$ is just the function above
and does ${\it not}$ include the pieces of the usual DGLAP splitting function
used to include the effect of virtual diagrams.
In particular, the $1/(1{-}x)$ above is just the ordinary function
$1/(1{-}x)$ and not the
distribution $1/(1{-}x)_+$, and our $P(x)$ above does not contain a
$\delta$-function term $\delta(1{-}x)$.
When we need to deal with virtual diagrams in this paper, we will do so
explicitly.

The absolute value signs in (\ref{eq:Pgg}) may seem redundant since
the absolute value is taken of a quantity that is manifestly positive
for $0 < x < 1$.
They are included so that our definition of $P(x)$ works with front-end
transformations, for the same reasons described after (\ref{eq:abc}) below.

For the sake of later formulas for virtual corrections, it will
be helpful to also express the above result in terms
of the $x\bar x$ diagram of fig.\ \ref{fig:LO} as
\begin {equation}
   \left[ \frac{d \Gamma}{dx} \right]^{\rm LO}
   = 2\Re \left[ \frac{d \Gamma}{dx} \right]_{x\bar x}
\label {eq:xxLO2}
\end {equation}
with
\begin {equation}
  \left[ \frac{d \Gamma}{dx} \right]_{x\bar x} = 
  \frac{\alphas}{2\pi} \, P(x) \, i\Omega_0 .
\end {equation}


\subsubsection{$d{=}2{-}\eps$ transverse spatial dimensions}

Equations (\ref{eq:LO}) and (\ref{eq:xxLO2}) are
all we need for the leading-order result for
renormalized calculations.  However, for comparison with intermediate,
unrenormalized results, the $d{=}2{-}\eps$ version is given in
(\ref{eq:LOd}).


\subsubsection{$\LObar$ rate}

For some applications, we have found it convenient to group together
the LO rate with the NLO renormalization logarithm as
\begin {equation}
   \left[ \Delta \frac{d\Gamma}{dx} \right]^{\LObar}
   \equiv
   \left[ \Delta \frac{d\Gamma}{dx} \right]^{\rm LO}
   + \left[ \frac{d\Gamma}{dx} \right]_\renlog ,
\label {eq:dGLObar2}
\end {equation}
where $[d\Gamma/dx]_\renlog$ is given by (\ref{eq:renlog2}).


\subsection{\boldmath$g\to ggg$ rate}
\label {app:gggSummary}

For the diagrams considered in this paper, we have
\begin {equation}
   \left[ \Delta \frac{d\Gamma}{dx\,dy} \right]_{g{\to}ggg}
   =
   \left[ \frac{d\Gamma}{dx\,dy} \right]_{\rm crossed}
   +
   \left[ \Delta \frac{d\Gamma}{dx\,dy} \right]_{\rm seq} ,
\end {equation}
where the first term represents the crossed diagrams of
fig.\ \ref{fig:crossed}
and the second term the sequential diagrams of fig.\ \ref{fig:seq}.
A summary of the formulas for these rates
appears in appendix A of
ref.\ \cite{4point}.  We will also present them here (i) for
convenient reference in this paper, especially since many of
the new formulas we need are related, (ii) because some
minor modifications are needed
[to (\ref{eq:abc}) and (\ref{eq:abcbar}) below]
to make the formulas work in a simple way with front-end transformations,
(iii) because we've rewritten some old formulas [such as (\ref{eq:Apole})]
in a way that makes clearer their relation to some new formulas
[such as (\ref{eq:ApoleIc})],
and (iv) to include some notational definitions [such as (\ref{eq:Om3})
and (\ref{eq:OmEx})] that were omitted from the summary in ref.\ \cite{4point}.


\subsubsection{Crossed Diagrams}
\label {app:crossed}

Here we collect the result for the crossed diagrams \cite{2brem}
as corrected by ref.\ \cite{dimreg}.  A brief summary of the
interpretation of each piece below can be found in section VIII of
ref.\ \cite{2brem}.

\begin {equation}
   \left[ \frac{d\Gamma}{dx\>dy} \right]_{\rm crossed}
   =
   A(x,y) + A(z,y) + A(x,z)
\label {eq:summary1}
\end {equation}
\begin {align}
   A(x,y) &=
   A^{\rm pole}(x,y)
   + \int_0^\infty d(\Delta t) \> 
     2 \Re \bigl[ B(x,y,\Delta t) + B(y,x,\Delta t) \bigr]
\label {eq:summaryA}
\end {align}
\begin {align}
   B(x,y,\Delta t) &=
       C(\{\hat x_i\},\alpha,\beta,\gamma,\Delta t)
       + C(\{x'_i\},\beta,\alpha,\gamma,\Delta t)
       + C(\{\tilde x_i\},\gamma,\alpha,\beta,\Delta t)
\nonumber\\
   &=
       C({-}1,y,z,x,\alpha,\beta,\gamma,\Delta t)
       + C\bigl({-}(1{-}y),{-}y,1{-}x,x,\beta,\alpha,\gamma,\Delta t\bigr)
\nonumber\\ &\qquad\qquad
       + C\bigl({-}y,{-}(1{-}y),x,1{-}x,\gamma,\alpha,\beta,\Delta t\bigr)
\label {eq:summaryB}
\end {align}
\begin {equation}
   C = D - \lim_{\hat q\to 0} D
\label {eq:summaryC}
\end {equation}
\begin {align}
   D(x_1,&x_2,x_3,x_4,\alpha,\beta,\gamma,\Delta t) =
\nonumber\\ &
   \frac{\CA^2 \alphas^2 M_\ix M_\fx}{32\pi^4 E^2} \, 
   ({-}x_1 x_2 x_3 x_4)
   \Omega_+\Omega_- \csc(\Omega_+\Delta t) \csc(\Omega_-\Delta t)
\nonumber\\ &\times
   \Bigl\{
     (\beta Y_\bx Y_\Ax + \alpha \Ybar_{\bx\Ax} Y_{\bx\Ax}) I_0
     + (\alpha+\beta+2\gamma) Z_{\bx\Ax} I_1
\nonumber\\ &\quad
     + \bigl[
         (\alpha+\gamma) Y_\bx Y_\Ax
         + (\beta+\gamma) \Ybar_{\bx\Ax} Y_{\bx\Ax}
        \bigr] I_2
     - (\alpha+\beta+\gamma)
       (\Ybar_{\bx\Ax} Y_\Ax I_3 + Y_\bx Y_{\bx\Ax} I_4)
   \Bigl\}
\label {eq:summaryD}
\end {align}
\begin {align}
   A^{\rm pole}(x,y) &=
    \frac{\CA^2 \alphas^2}{8\pi^2} \,
    x y z (1{-}x) (1{-}y)
    \Re \biggl(
       -i
       \bigl( \Omega_{-1,1-x,x} + \Omega_{-(1-y),z,x}
        + \Omega_{-1,1-y,y} + \Omega_{-(1-x),z,y} \bigr)
\nonumber\\ & \qquad~ \qquad \times
       \Bigl\{
       \bigl(
         (\alpha{+}\beta) z(1{-}x)(1{-}y)
         + (\alpha{+}\gamma) xyz
       \bigr)
       \left[ \ln \left( \tfrac{z}{(1{-}x)(1{-}y)} \right) - i\pi \right]
\nonumber\\ & \qquad\qquad\qquad
       + 2(\alpha{+}\beta{+}\gamma) x y z
   \Bigr\}
   \biggr)
\label {eq:Apole}
\end {align}
\begin {subequations}
\label {eq:I}
\begin {align}
   I_0 &=
   \frac{4\pi^2}{(X_\bx X_\Ax - X_{\bx\Ax}^2)}
\displaybreak[0]\\
   I_1 &=
   - \frac{2\pi^2}{X_{\bx\Ax}}
   \ln\left( 1 - \frac{X_{\bx\Ax}^2}{X_\bx X_\Ax} \right)
\displaybreak[0]\\
   I_2 &=
   \frac{2\pi^2}{X_{\bx\Ax}^2}
     \ln\left( 1 - \frac{X_{\bx\Ax}^2}{X_\bx X_\Ax} \right)
   + \frac{4\pi^2}{(X_\bx X_\Ax - X_{\bx\Ax}^2)}
\displaybreak[0]\\
   I_3 &=
   \frac{4\pi^2 X_{\bx\Ax}}{X_\Ax(X_\bx X_\Ax - X_{\bx\Ax}^2)}
\displaybreak[0]\\
   I_4 &=
   \frac{4\pi^2 X_{\bx\Ax}}{X_\bx(X_\bx X_\Ax - X_{\bx\Ax}^2)}
\end {align}
\end {subequations}
\begin {subequations}
\label {eq:XYZdef}
\begin {align}
   \begin{pmatrix} X_\bx & Y_\bx \\ Y_\bx & Z_\bx \end{pmatrix}
   &\equiv
   \begin{pmatrix} |M_\ix|\Omega_\ix & 0 \\ 0 & 0 \end{pmatrix}
     - i a_\bx^{-1\top} \uOmega \cot(\uOmega\,\Delta t)\, a_\bx^{-1}
\\
   \begin{pmatrix} X_\Ax & Y_\Ax \\ Y_\Ax & Z_\Ax \end{pmatrix}
   &\equiv
   \begin{pmatrix} |M_\fx|\Omega_\fx & 0 \\ 0 & 0 \end{pmatrix}
     - i a_\Ax^{-1\top} \uOmega \cot(\uOmega\,\Delta t)\, a_\Ax^{-1}
\\
   \begin{pmatrix} X_{\bx\Ax} & Y_{\bx\Ax} \\ \Ybar_{\bx\Ax} & Z_{\bx\Ax} \end{pmatrix}
   &\equiv
   - i a_\bx^{-1\top} \uOmega \csc(\uOmega\,\Delta t) \, a_\Ax^{-1}
\end {align}
\end {subequations}
\begin {equation}
   \uOmega \equiv \begin{pmatrix} \Omega_+ & \\ & \Omega_- \end{pmatrix}
\label {eq:uOmega}
\end {equation}
\begin {subequations}
\label {eq:OMif}
\begin {equation}
   M_\ix = x_1 x_4 (x_1{+}x_4) E ,
   \qquad
   M_\fx = x_3 x_4 (x_3{+}x_4) E
\end {equation}
\begin {equation}
   \Omega_\ix
   = \sqrt{ 
     -\frac{i \hat q_{\rm A}}{2E}
     \left( \frac{1}{x_1} 
            + \frac{1}{x_4} - \frac{1}{x_1{+}x_4} \right)
   } ,
   \qquad
   \Omega_\fx
   = \sqrt{ 
     -\frac{i \hat q_{\rm A}}{2E}
     \left( \frac{1}{x_3} + \frac{1}{x_4}
            - \frac{1}{x_3 + x_4}
     \right)
   }
\label {eq:Omif}
\end {equation}
\end {subequations}
\begin {equation}
   \Omega_{\xi_1,\xi_2,\xi_3}
   = \sqrt{ 
     -\frac{i \hat q_{\rm A}}{2E}
     \left( \frac{1}{\xi_1} + \frac{1}{\xi_2} + \frac{1}{\xi_3} \right)
   }
\label {eq:Om3}
\end {equation}
\begin {equation}
   a_\ybx =
   \begin{pmatrix} C^+_{34} & C^-_{34} \\ C^+_{12} & C^-_{12} \end{pmatrix}
\label {eq:af}
\end {equation}
\begin {equation}
   a_\yx =
   \frac{1}{(x_1+x_4)}
   \begin{pmatrix}
       -x_3 & -x_2 \\
       \phantom{-}x_4 &  \phantom{-}x_1
   \end {pmatrix}
   a_\ybx
\label {eq:ai}
\end {equation}
\begin {align}
   \begin{pmatrix} \alpha \\ \beta \\ \gamma \end{pmatrix}
   =
   \phantom{+}
   & \begin{pmatrix} - \\ + \\ + \end{pmatrix} \Biggl[
       \left| \frac{x}{y^3z(1{-}x)^3(1{-}y)^3} \right|
       + \left| \frac{y}{x^3z(1{-}x)^3(1{-}y)^3} \right|
\nonumber\displaybreak[3]\\ & \qquad\qquad
       + \left| \frac{1{-}x}{x^3y^3z(1{-}y)^3} \right|
       + \left| \frac{1{-}y}{x^3y^3z(1{-}x)^3} \right|
   \Biggr]
\nonumber\\
   + & \begin{pmatrix} + \\ - \\ + \end{pmatrix} \Biggl[
       \left| \frac{x}{y^3z^3(1{-}x)(1{-}y)} \right|
       + \left| \frac{y}{x^3z^3(1{-}x)(1{-}y)} \right|
\nonumber\displaybreak[3]\\ & \qquad\qquad
       + \left| \frac{z}{x^3y^3(1{-}x)(1{-}y)} \right|
       + \left| \frac{1}{x^3y^3z^3(1{-}x)(1{-}y)} \right|
   \Biggr]
\nonumber\\
   + & \begin{pmatrix} + \\ + \\ - \end{pmatrix} \Biggl[
        \left| \frac{1{-}x}{xyz^3(1{-}y)^3} \right|
        + \left| \frac{1{-}y}{xyz^3(1{-}x)^3} \right|
\nonumber\displaybreak[3]\\ & \qquad\qquad
        + \left| \frac{z}{xy(1{-}x)^3(1{-}y)^3} \right|
        + \left| \frac{1}{xyz^3(1{-}x)^3(1{-}y)^3} \right|
   \Biggr]
\label {eq:abc}
\end {align}
Note that the $(\alpha,\beta,\gamma)$ used in
the definition (\ref{eq:summaryB}) of $B$ are implicitly
functions (\ref{eq:abcbar})
of the {\it arguments} $x$ and $y$ of $B(x,y,\Delta t)$
[with $z \equiv 1{-}x{-}y$].  This is important in formulas
such as (\ref{eq:summaryA}),
where in some terms those local arguments are replaced by
other variables.

Eq.\ (\ref{eq:abc}) gives $(\alpha,\beta,\gamma)$ for $d{=}2$.
However, as explained in appendix \ref{app:abcdim} (which gives
the more general formulas for $d{=}2{-}\eps$), the $d{=}2$ formulas for
$(\alpha,\beta,\gamma)$ are all that is needed here in
appendix \ref{app:summary}.

The absolute value signs in (\ref{eq:abc}) may seem unnecessary since
$g \to ggg$ processes have parton longitudinal momentum fractions
$x$, $y$, $z$, $1{-}x$, $1{-}y$ all
positive.  The advantage
of including an absolute value sign around every such parton momentum
fraction [which is equivalent to the use of absolute value signs in
(\ref{eq:abc})] is that they make front-end transformations
like (\ref{eq:frontend}) work in a simple way, despite the fact that
the front-end transformation replaces $x$ by a negative number.%
\footnote{
  They are the QCD version of the absolute value signs used in
  eq.\ (A22) of ref.\ \cite{QEDnf}, which are discussed in
  footnote 38 of ref.\ \cite{QEDnf}.  One could alternatively dispense
  with the absolute value signs in the QCD case (\ref{eq:abc})
  above by noting that
  negating $x$ in that formula would, without absolute value signs,
  simply introduces a common overall minus sign in the values of
  ($\alpha$,$\beta$,$\gamma$), which could be accounted for by modifying
  the sign of the front-end transformation formula (\ref{eq:frontend}).
  We've chosen to
  introduce the absolute value signs, however, so that our overall
  sign convention
  for front-end transformations will be the same as it was in the QED
  case of ref.\ \cite{QEDnf}.
}

The $\hat q\to 0$ limit for the vacuum piece in (\ref{eq:summaryC})
corresponds to taking all $\Omega$'s to zero and so making the
replacements
\begin {equation}
   \Omega_\ix \to 0,
   \qquad
   \Omega_\fx \to 0,
   \qquad
   \uOmega \cot(\uOmega\,\Delta t) \to (\Delta t)^{-1} \underline{\openone} ,
   \qquad
   \uOmega \csc(\uOmega\,\Delta t) \to (\Delta t)^{-1} \underline{\openone} ,
\end {equation}
\begin {equation}
  \Omega_\pm \csc(\Omega_\pm \Delta t) \to (\Delta t)^{-1} ,
\end {equation}
where $\underline{\openone}$ is the identity matrix.
For numerical evaluation, one must take care that the above
takes $X_{\yx\ybx} \to 0$ and so
\begin {equation}
  I_1 \to 0,
  \qquad
  I_2 \to \frac{2\pi^2}{X_\yx X_\ybx} \,.
\end {equation}


\subsubsection{4-particle frequencies and normal modes}
\label {app:modes}

Here we collect formulas for the large-$\Nc$ frequencies and normal modes
associated with 4-particle propagation (section V.B of ref.\ \cite{2brem}).

\begin {equation}
  \Omega_\pm =
  \left[ - \frac{i\hat q_{\rm A}}{4E} \left(
    \frac1{x_1} + \frac1{x_2} + \frac1{x_3} + \frac1{x_4} \pm \sqrt\Delta
  \right) \right]^{1/2}
\label {eq:Omegapm}
\end {equation}
\begin {equation}
  \Delta =
   \frac1{x_1^2} + \frac1{x_2^2} + \frac1{x_3^2} + \frac1{x_4^2}
   + \frac{(x_3{+}x_4)^2+(x_1{+}x_4)^2}{x_1 x_2 x_3 x_4}
\end {equation}
\begin {subequations}
\label {eq:Cmodes}
\begin {align}
   C^\pm_{34}
   &=
   \frac{x_2}{x_3+x_4} \sqrt{\frac{x_1 x_3}{2 N_\pm E}}
   \left[
      \frac1{x_3}-\frac1{x_1}+\frac1{x_4}+\frac{x_1}{x_3 x_2} \pm \sqrt\Delta
   \right]
\\
   C^\pm_{12}
   &=
   - \frac{x_4}{x_1+x_2} \sqrt{\frac{x_1 x_3}{2 N_\pm E}}
   \left[
      \frac1{x_1}-\frac1{x_3}+\frac1{x_2}+\frac{x_3}{x_1 x_4} \pm \sqrt\Delta
   \right]
\end {align}
\end {subequations}
\begin {equation}
   N_\pm \equiv
   - x_1 x_2 x_3 x_4 (x_1+x_3)\Delta
   \pm (x_1 x_4 + x_2 x_3)(x_1 x_2 + x_3 x_4) \sqrt\Delta
\label {eq:Npm}
\end {equation}


\subsubsection{Sequential Diagrams}
\label {app:seqsummary}

Here we collect the result for the sequential diagrams \cite{seq}.  A
brief summary of the interpretation of each piece below can be found
in section III of ref.\ \cite{seq}.  Symbols such as $\Omega_\pm$ or
$a_\yx$, which
are written in the exact same notation as symbols defined above,
are given by their definitions above.

\begin {align}
   \left[ \Delta \frac{d\Gamma}{dx\>dy} \right]_{\rm sequential}
   = \quad
   & {\cal A}_\seq(x,y) + {\cal A}_\seq(z,y) + {\cal A}_\seq(x,z)
\nonumber\\
   + ~ &
   {\cal A}_\seq(y,x) + {\cal A}_\seq(y,z) + {\cal A}_\seq(z,x)
\label {eq:dGammaseq}
\end {align}
\begin {equation}
   {\cal A}_\seq(x,y)
   =
   {\cal A}^{\rm pole}_\seq(x,y)
   + \int_0^{\infty} d(\Delta t) \>
     \Bigl[
        2 \Re \bigl( B_{\rm seq}(x,y,\Delta t) \bigr)
        + F_{\rm seq}(x,y,\Delta t)
     \bigr]
\label {eq:Aseq}
\end {equation}
\begin {equation}
   B_\seq(x,y,\Delta t) =
       C_\seq(\{\hat x_i\},\bar\alpha,\bar\beta,\bar\gamma,\Delta t)
   =
       C_\seq({-}1,y,z,x,\bar\alpha,\bar\beta,\bar\gamma,\Delta t)
\label {eq:Bseq}
\end {equation}
\begin {equation}
   C_\seq = D_\seq - \lim_{\hat q\to 0} D_\seq
\end {equation}
\begin {align}
   D_\seq(x_1,&x_2,x_3,x_4,\bar\alpha,\bar\beta,\bar\gamma,\Delta t) =
\nonumber\\ &
   \frac{\CA^2 \alphas^2 M_\ix M_\fx^\seq}{32\pi^4 E^2} \, 
   ({-}x_1 x_2 x_3 x_4)
   \Omega_+\Omega_- \csc(\Omega_+\Delta t) \csc(\Omega_-\Delta t)
\nonumber\\ &\times
   \Bigl\{
     (\bar\beta Y_\yx^\seq Y_\xbx^\seq
        + \bar\alpha \Ybar_{\yx\xbx}^{\,\seq} Y_{\yx\xbx}^\seq) I_0^\seq
     + (\bar\alpha+\bar\beta+2\bar\gamma) Z_{\yx\xbx}^\seq I_1^\seq
\nonumber\\ &\quad
     + \bigl[
         (\bar\alpha+\bar\gamma) Y_\yx^\seq Y_\xbx^\seq
         + (\bar\beta+\bar\gamma) \Ybar_{\yx\xbx}^{\,\seq} Y_{\yx\xbx}^\seq
        \bigr] I_2^\seq
\nonumber\\ &\quad
     - (\bar\alpha+\bar\beta+\bar\gamma)
       (\Ybar_{\yx\xbx}^{\,\seq} Y_\xbx^\seq I_3^\seq + Y_\yx^\seq Y_{\yx\xbx}^\seq I_4^\seq)
   \Bigl\}
\label {eq:Dseq}
\end {align}
\begin {align}
   F_\seq(x,y,\Delta t) =
   \frac{\alphas^2 P(x) P(\yfrak)}{4\pi^2(1-x)}
   \Bigl[ &
      \Re\bigl(i(\Omega\sgn M)_{E,x}\bigr) \,
      \Re\bigl( \Delta t \, \Omega_{(1-x)E,\yfrak}^2
                \csc^2(\Omega_{(1-x)E,\yfrak} \, \Delta t) \bigr)
\nonumber\\
      + &
      \Re\bigl(i(\Omega\sgn M)_{(1-x)E,\yfrak}\bigr) \,
      \Re\bigl( \Delta t \, \Omega_{E,x}^2
                \csc^2(\Omega_{E,x} \, \Delta t) \bigr)
   \Bigr]
\label {eq:Fseq}
\end {align}
\begin {equation}
   {\cal A}_\seq^{\rm pole}(x,y)
   = - \frac{\alphas^2 \, P(x) \, P(\yfrak)}{4\pi^2(1-x)}
       \Re\bigl[
          \bigl(i (\Omega\sgn M)_{E,x} + i (\Omega\sgn M)_{(1-x)E,\yfrak}\bigr)
          \bigl(1 + \tfrac{i\pi}{2}\bigr)
       \bigr]
\label {eq:wrong}
\end {equation}
[Warning: See the {\it note added} at the end of main text,
discussing a correction to (\ref{eq:wrong}).%
\astfootnote{
  The correct formula \cite{logs} for (\ref{eq:wrong}) is
  \begin {multline}
   {\cal A}_{\rm seq}^{\rm pole}(x,y) =
   - \frac{\alphas^2 \, P(x) \, P(\yfrak)}{4\pi^2(1-x)}
   \Re\Bigl[
     i (\Omega\sgn M)_{E,x} (1+\tfrac{i\pi}{2} \sgn M_{(1-x)E,\yfrak})
  \\
     +
     i (\Omega\sgn M)_{(1-x)E,\yfrak} (1+\tfrac{i\pi}{2} \sgn M_{E,x})
   \Bigr]
  \nonumber
  \end {multline}
  or equivalently
\begin {equation*}
  {\cal A}_{\rm seq}^{\rm pole}(x,y) =
  - \frac{\alphas^2 \, P(x) \, P(\yfrak)}{4\pi^2(1-x)}
  \Re\bigl[
    i (\Omega\sgn M)_{E,x} + i (\Omega\sgn M)_{(1-x)E,\yfrak})
  \bigr]
  \bigl(1-\tfrac{\pi}{2} \sgn M_{E,x} \sgn M_{(1-x)E,\yfrak}\bigr)
  .
  \nonumber
  \end {equation*}
}%
]
\begin {equation}
  \yfrak \equiv \frac{y}{1-x}
\end {equation}
\begin {subequations}
\label {eq:Iseq}
\begin {align}
   I_0^\seq &=
   \frac{4\pi^2}{[X_\yx^\seq X_\xbx^\seq - (X_{\yx\xbx}^\seq)^2]}
\displaybreak[0]\\
   I_1^\seq &=
   - \frac{2\pi^2}{X_{\yx\xbx}^\seq}
   \ln\left( 1 - \frac{(X_{\yx\xbx}^\seq)^2}{X_\yx^\seq X_\xbx^\seq} \right)
\displaybreak[0]\\
   I_2^\seq &=
   \frac{2\pi^2}{(X_{\yx\xbx}^\seq)^2}
     \ln\left( 1 - \frac{(X_{\yx\xbx}^\seq)^2}{X_\yx^\seq X_\xbx^\seq} \right)
   + \frac{4\pi^2}{[X_\yx^\seq X_\xbx^\seq - (X_{\yx\xbx}^\seq)^2]}
\displaybreak[0]\\
   I_3^\seq &=
   \frac{4\pi^2 X_{\yx\xbx}^\seq}
        {X_\xbx^\seq[X_\yx^\seq X_\xbx^\seq - (X_{\yx\xbx}^\seq)^2]}
\displaybreak[0]\\
   I_4^\seq &=
   \frac{4\pi^2 X_{\yx\xbx}^\seq}
        {X_\yx^\seq[X_\yx^\seq X_\xbx^\seq - (X_{\yx\xbx}^\seq)^2]}
\end {align}
\end {subequations}
\begin {subequations}
\label {eq:XYZseqdef}
\begin {align}
   \begin{pmatrix} X_\yx^\seq & Y_\yx^\seq \\ Y_\yx^\seq & Z_\yx^\seq \end{pmatrix}
   &\equiv
   \begin{pmatrix} |M_\ix|\Omega_\ix & 0 \\ 0 & 0 \end{pmatrix}
     - i a_\yx^{-1\top} \uOmega \cot(\uOmega\,\Delta t)\, a_\yx^{-1}
   =
   \begin{pmatrix} X_\yx & Y_\yx \\ Y_\yx & Z_\yx \end{pmatrix} ,
\\
   \begin{pmatrix} X_\xbx^\seq & Y_\xbx^\seq \\ Y_\xbx^\seq & Z_\xbx^\seq \end{pmatrix}
   &\equiv
   \begin{pmatrix} |M_\fx^\seq|\Omega_\fx^\seq & 0 \\ 0 & 0 \end{pmatrix}
     - i (a_\xbx^\seq)^{-1\top} \uOmega \cot(\uOmega\,\Delta t)\, (a_\xbx^\seq)^{-1} ,
\\
   \begin{pmatrix} X_{\yx\xbx}^\seq & Y_{\yx\xbx}^\seq \\
                   \Ybar_{\yx\xbx}^\seq & Z_{\yx\xbx}^\seq \end{pmatrix}
   &\equiv
   - i a_\yx^{-1\top} \uOmega \csc(\uOmega\,\Delta t) \, (a_\xbx^\seq)^{-1}
\end {align}
\end {subequations}
\begin {equation}
   M_\fx^\seq = x_2 x_3 (x_2{+}x_3) E
\label {eq:Mfseq}
\end {equation}
\begin {equation}
   \Omega_\fx^\seq
   = \sqrt{ 
     -\frac{i \hat q_{\rm A}}{2E}
     \left( \frac{1}{x_2} + \frac{1}{x_3}
            - \frac{1}{x_2 + x_3}
     \right)
   }
\label {eq:Omfseq}
\end {equation}
\begin {equation}
   M_{E,x} = x(1{-}x)E
\end {equation}
\begin {equation}
   \Omega_{E,x}
   = \sqrt{ 
     -\frac{i \hat q_{\rm A}}{2E}
     \left( \frac{1}{x} + \frac{1}{1{-}x} - 1 \right)
   }
\label {eq:OmEx}
\end {equation}
\begin {equation}
   a_\xbx^\seq \equiv 
   \begin{pmatrix} 0 & 1 \\ 1 & 0 \end{pmatrix} a_\yx
\end {equation}
\begin {align}
   \begin{pmatrix}
      \bar\alpha \\ \bar\beta \\ \bar\gamma
   \end{pmatrix}_{x\bar y\bar x y}
   =
   \phantom{+}
   & \begin{pmatrix} - \\ + \\ + \end{pmatrix}
       \left| \frac{4}{xyz(1{-}x)^6} \right|
\nonumber\\
   + & \begin{pmatrix} + \\ - \\ + \end{pmatrix} \Biggl[
       \left| \frac{1}{x^3y^3z^3(1{-}x)^2} \right|
       + \left| \frac{z}{x^3y^3(1{-}x)^2} \right|
       + \left| \frac{x}{y^3z^3(1{-}x)^2} \right|
       + \left| \frac{y}{x^3z^3(1{-}x)^2} \right|
   \Biggr]
\nonumber\\
   + & \begin{pmatrix} + \\ + \\ - \end{pmatrix} \Biggl[
       \left| \frac{(1{-}x)^2}{x^3y^3z^3} \right|
       + \left| \frac{z}{x^3y^3(1{-}x)^6} \right|
       + \left| \frac{xz}{y^3(1{-}x)^6} \right|
\nonumber\\ & \qquad\qquad
       + \left| \frac{y}{x^3z^3(1{-}x)^6} \right|
       + \left| \frac{xy}{z^3(1{-}x)^6} \right|
   \Biggr]
\label {eq:abcbar}
\end {align}
A comment similar to the one following (\ref{eq:abc}) applies here to
the dependence of $(\bar\alpha,\bar\beta,\bar\gamma)$ on the arguments
$x$ and $y$ of ${\cal A}_\seq$.


\subsection{NLO \boldmath$g\to gg$ rate}
\label {app:NLOsummary}

With regard to renormalization, we are going to make our summary formulas
in this subsection do double duty by introducing a variable
\begin {equation}
  \sigren \equiv
  \begin {cases}
    1, & \mbox{for renormalized results}; \\
    0, & \mbox{for unrenormalized results} ,
  \end {cases}
\end {equation}
and its complement
\begin {equation}
   \sigbare \equiv 1 - \sigren .
\end {equation}

In the renormalized case ($\sigren{=}1, \sigbare{=}0$),
the $\alphas$ in the leading-order
splitting rate (\ref{eq:LO})
is MS-bar renormalized $\alphas$ with renormalization scale
$\mu$, and we have chosen to group all of the $\mu$-dependence of
the NLO diagrams into the term shown explicitly in (\ref{eq:renlog2})
below, as discussed in section \ref{sec:OrganizeRenorm}.
In the unrenormalized case ($\sigren{=}0,\sigbare{=}1$),
the $\alphas$ in the leading-order
splitting rate (\ref{eq:LO})
is instead the bare $\alphas$, and we show the $1/\eps$ and $\ln\mu$
dependence of the NLO diagrams individually for each diagram.

Along the lines discussed in section \ref{sec:OrganizeRenorm}, we write
\begin {equation}
   \left[ \Delta \frac{d\Gamma}{dx} \right]^{\rm NLO}_{g\to gg}
   =
   \left[ \Delta \frac{d\Gamma}{dx} \right]^{\NLObar}_{g\to gg}
   +
   \sigren\left[ \frac{d\Gamma}{dx} \right]_\renlog
\end {equation}
with
\begin {equation}
   \left[ \frac{d\Gamma}{dx} \right]_\renlog
   \equiv
   - \beta_0\alphas
       \Re\left(
         \left[ \frac{d\Gamma}{dx} \right]_{\substack{x\bar x\hfill\\d=2}}
         \left[
            \ln \Bigl( \frac{\mu^2}{\Omega_0 E} \Bigr)
            + \ln\Bigl( \frac{x(1{-}x)}{4} \Bigr)
            + \gammaE
         \right]
       \right) ,
\label {eq:renlog2}
\end {equation}
\begin {equation}
   \beta_0 = -\frac{11\,\CA}{6\pi} \,,
\end {equation}
and
\begin {align}
   \left[ \Delta \frac{d\Gamma}{dx} \right]^{\NLObar}_{g\to gg}
   &=
     \biggl(
       \left[ \Delta \frac{d\Gamma}{dx} \right]_\virtI
     \biggr)
     + (x \to 1{-}x)
   + \left[ \Delta \frac{d\Gamma}{dx} \right]_\virtII
\nonumber\\
   &=
     \biggl(
       \int_0^{1-x} dy \, \left[ \Delta \frac{d\Gamma}{dx\,dy} \right]_\virtI
     \biggr)
     + (x \to 1{-}x)
   +
   \int_0^1 dy \, \left[ \Delta \frac{d\Gamma}{dx\,dy} \right]_\virtII .
\label {eq:dGammaNLObar}
\end {align}

In what follows, we will further subdivide Class I diagrams
into what we call (Class Ic) crossed virtual diagrams, given by the
first row of fig.\ \ref{fig:relateI} plus conjugates;
(Class Is) back-end sequential
virtual diagrams, given by the remaining three diagrams of
fig.\ \ref{fig:relateI} plus conjugates;
and $2\Re(x y y \bar x$), given by the last
diagram of fig.\ \ref{fig:virtI} plus conjugate:
\begin {equation}
  \left[ \Delta \frac{d\Gamma}{dx\,dy} \right]_\virtI
  =
  \left[ \frac{d\Gamma}{dx\,dy} \right]_\virtIc
  +
  \left[ \Delta \frac{d\Gamma}{dx\,dy} \right]_\virtIs
  +
  2\Re \left[ \frac{d\Gamma}{dx\,dy} \right]_{xyy\bar x} .
\label {eq:virtIsummary}
\end {equation}
Similarly, we subdivide Class II diagrams into
(Class IIs) front-end sequential virtual diagrams, given by the three
diagrams of fig.\ \ref{fig:relateII} plus conjugates; and
$2\Re(x \bar y \bar y \bar x$), given by the last
diagram of fig.\ \ref{fig:virtII} plus conjugate:
\begin {equation}
  \left[ \Delta \frac{d\Gamma}{dx\,dy} \right]_\virtII
  =
  \left[ \Delta \frac{d\Gamma}{dx\,dy} \right]_\virtIIs
  +
  2\Re \left[ \frac{d\Gamma}{dx\,dy} \right]_{x\bar y\bar y\bar x} .
\end {equation}


\subsubsection {Crossed Virtual Diagrams}

\begin {equation}
  \left[ \Delta \frac{d\Gamma}{dx\, dy} \right]_\virtIc
  =
  A_\virtIc^{\rm pole}(x,y)
   + \int_0^\infty d(\Delta t) \>
   \bigl[
     2 \Re \bigl( B_\virtIc(x,y,\Delta t) \bigr)
     - ({\cal D}_2)_\virtIc(x,y,\Delta t)
   \bigr]
\label {eq:AvirtIc}
\end {equation}

\begin {align}
   B_\virtIc(x,y,\Delta t)
   &=
       - C({-}1,x,z,y,\alpha,\beta,\gamma,\Delta t)
       - C\bigl({-}(1{-}x),{-}x,1{-}y,y,\beta,\alpha,\gamma,\Delta t\bigr)
\nonumber\\ &\quad
       - C\bigl({-}y,{-}(1{-}y),x,1{-}x,\gamma,\alpha,\beta,\Delta t\bigr)
       - C\bigl(x,z,y,{-}1,\gamma,\beta,\alpha,\Delta t\bigr)
\label {eq:BvirtIc}
\end {align}
Above, $C$ and $(\alpha,\beta,\gamma)$ are the same as
(\ref{eq:summaryC}) and (\ref{eq:abc}) for $g{\to}ggg$ crossed
diagrams.

\begin {multline}
   ({\cal D}_2)_{\virtIc}(\Delta t) =
   \frac{\CA^2 \alphas^2}{4\pi^2} \, 
   \Re\bigl[ i \Omega_0^3\,\Delta t\,\csc^2(\Omega_0\,\Delta t) \bigr]
   x^2 y^2 z (1{-}x)(1{-}y)
\\ \times
   \bigl[
      (\alpha{+}\gamma)z + (\beta{+}\gamma)(1{-}x)(1{-}y)
   \bigr]
\label {eq:D2virtIc}
\end {multline}

\begin {align}
   A^{\rm pole}_\virtIc&(x,y) =
    \frac{\CA^2 \alphas^2}{8\pi^2} \,
    x y z (1{-}x) (1{-}y)
    \Re \biggl\{
\nonumber\\ & \quad
       i \Omega_{-1,y,1-y}
       \biggl[
       \Bigl(
         (\alpha{+}\beta) z(1{-}x)(1{-}y)
         - (\beta{+}\gamma) xy(1{-}x)(1{-}y)
       \Bigr)
       \left[ \ln\left( \tfrac{z}{(1{-}x)(1{-}y)} \right) - i\pi \right]
\nonumber\\ & \hspace{7em}
       + 2(\alpha{+}\beta{+}\gamma) xy(1{-}x)(1{-}y)
       \biggr]
\nonumber\\ & \quad
       + i \Omega_{-(1-x),y,z}
       \biggl[
       \Bigl(
         (\alpha{+}\beta) z(1{-}x)(1{-}y)
         + (\alpha{+}\gamma) xyz
       \Bigr)
       \left[ \ln\left( \tfrac{z}{(1{-}x)(1{-}y)} \right) - i\pi \right]
\nonumber\\ & \hspace{7em}
       + 2(\alpha{+}\beta{+}\gamma) xyz
       \biggr]
\nonumber\\ & \quad
       + i \Omega_{-1,x,1-x}
       \biggl[
       \Bigl(
         (\alpha{+}\gamma) xyz
         + (\beta{+}\gamma) xy(1{-}x)(1{-}y)
       \Bigr)
       \Bigl[
         4 \sigbare
            \Bigl(
               \tfrac{1}{\eps} + \ln\bigl(\tfrac{\pi\mu^2}{\Omega_0 E} \bigr)
            \Bigr)
\nonumber\\ & \hspace{10em}
         + 4 - \ln\bigl( x^2 y^2 z (1{-}x)(1{-}y) \bigr) - i\pi
       \Bigr]
\nonumber\\ & \hspace{7em}
       - 2(\alpha{-}\beta{+}\gamma) xyz
       - 2(-\alpha{+}\beta{+}\gamma) xy(1{-}x)(1{-}y)
       \biggr]
   \biggr\}
\label {eq:ApoleIc}
\end {align}
Note: The shorthand notation $\Omega_0$ (\ref{eq:Om0}) used above is
the same as the $\Omega_{-1,x,1-x}$ (\ref{eq:Om3})
also appearing above, but we have used
the latter to make explicit the similar structures of the three
terms in (\ref{eq:ApoleIc}).


\subsubsection {Sequential Virtual Diagrams}

\begin {align}
  \left[ \Delta \frac{d\Gamma}{dx\, dy} \right]_\virtIs
  &=
  - \tfrac12 \bigl[ {\cal A}_\seq(x,y) + {\cal A}_\seq(x,z) \bigr]
\label {eq:AseqBE}
\\
  \left[ \Delta \frac{d\Gamma}{dx\, dy} \right]_\virtIIs
  &=
  - \tfrac12 (1{-}y)^{-1/2}
  \bigl[
    {\cal A}_\seq\bigl(\tfrac{-y}{1-y},\tfrac{x}{1-y}\bigr)
    +
    {\cal A}_\seq\bigl(\tfrac{-y}{1-y},\tfrac{1-x}{1-y}\bigr)
  \bigr]
\label {eq:AseqFE}
\end {align}
Above, ${\cal A}_\seq$ is the same as (\ref{eq:Aseq}) for
sequential $g{\to}ggg$ diagrams.
See appendix \ref{sec:DeriveVirtSeq} for alternative ways
to write (\ref{eq:AseqFE}) and for comments concerning the
physical meaning of the $\Delta t$ integration variable of
(\ref{eq:Aseq}) in the context of (\ref{eq:AseqFE}).


\subsubsection {$2\Re(x y y \bar x)$}

\begin {equation}
  2\Re \left[ \frac{d\Gamma}{dx\, dy} \right]_{x y y\bar x}
  =
  \tfrac12 \bigl[ A_\new(x,y) + A_\new(x,z) \bigr]
\label {eq:dGfund}
\end {equation}
\begin {equation}
  A_\new(x,y) =
  A_\new^{\rm pole}(x,y) +
  \int_0^\infty d(\Delta t) \>
      2\Re\bigl[ B_\new(x,y,\Delta t) \bigr]
\label {eq:Anew}
\end {equation}

\begin {align}
   B_\new(x,y,\Delta t)
   =
   D_\new(-1,y,z,x,\bar\alpha,\bar\beta,\bar\gamma,\Delta t)
\end {align}
Above, $(\bar\alpha,\bar\beta,\bar\gamma)$ are the same as (\ref{eq:abcbar}).
\begin {align}
   D_\new(x_1,x_2,x_3,x_4,\bar\alpha,\bar\beta,\bar\gamma,&\Delta t) =
\nonumber\\
   - \frac{\CA^2 \alphas^2 M_\ix^2}{32\pi^4 E^2} \, 
   ({-}x_1 x_2 x_3 x_4)
   &
   \biggl\{
     \Omega_+\Omega_- \csc(\Omega_+\Delta t) \csc(\Omega_-\Delta t)
\nonumber\\ &\quad \times
     \Bigl[
       (\bar\beta Y_\yx^\new Y_\yx^\new
          + \bar\gamma \Ybar_{\yx\yx'}^{\,\new} Y_{\yx\yx'}^\new) I_0^\new
       + (2\bar\alpha{+}\bar\beta{+}\bar\gamma) Z_{\yx\yx'}^\new I_1^\new
\nonumber\\ &\quad\qquad
       + \bigl[
           (\bar\alpha{+}\bar\gamma) Y_\yx^\new Y_\yx^\new
           + (\bar\alpha{+}\bar\beta) \Ybar_{\yx\yx'}^{\,\new} Y_{\yx\yx'}^\new
          \bigr] I_2^\new
\nonumber\\ &\quad\qquad
       - (\bar\alpha{+}\bar\beta{+}\bar\gamma)
         (\Ybar_{\yx\yx'}^{\,\new} Y_\yx^\new I_3^\new
              + Y_\yx^\new Y_{\yx\yx'}^\new I_4^\new)
     \Bigl]
\nonumber\\ &
     - (2\bar\alpha{+}\bar\beta{+}\bar\gamma) \tfrac{x_2 x_3}{x_1 x_4}
         \, {\cal D}_2^{(\bbI)}(\Omega_\ix \sgn M_\ix, \Delta t)
   \biggr\}
\label {eq:Dnew}
\end {align}

\begin {equation}
   {\cal D}_2^{(\bbI)}(\Omega,\Delta t) =
   2\pi^2
   \left[
     \frac{\ln(2i\Omega \,\Delta t)}{(\Delta t)^2}
     - i\Omega^3\,\Delta t \csc^2(\Omega\, \Delta t)
   \right]
\label {eq:D2Isummary}
\end {equation}

\begin {equation}
   A^{\rm pole}_\new(x,y) =
    \frac{\alphas^2}{2\pi^2} \,
    \frac{P(x)\,P(\frac{y}{1-x})}{1-x} \,
    \Re \biggl\{
       (i \Omega_0 \sgn M_0)
       \Bigl[
         - \sigbare
            \Bigl(
               \tfrac{1}{\eps}
               + \ln\bigl(\tfrac{\pi\mu^2}{E\Omega_0\sgn M_0} \bigr)
            \Bigr)
         + \tfrac12 \ln(x y z)
       \Bigr]
    \biggr\}
\label {eq:Apolenew}
\end {equation}
Here the $I^{\rm new}_n$ are the same as the
$I^\seq_n$ of (\ref{eq:Iseq}) except that the $(X,Y,Z)^\seq$ there are
replaced by
\begin {subequations}
\begin {align}
   \begin{pmatrix} X_\yx^\new & Y_\yx^\new \\ Y_\yx^\new & Z_\yx^\new \end{pmatrix}
   =
   \begin{pmatrix} X_{\yx'}^\new & Y_{\yx'}^\new
                   \\ Y_{\yx'}^\new & Z_{\yx'}^\new \end{pmatrix}
   \equiv &
   \begin{pmatrix} |M_\ix|\Omega_\ix & 0 \\ 0 & 0 \end{pmatrix}
     - i a_\yx^{-1\top} \uOmega \cot(\uOmega\,\Delta t)\, a_\yx^{-1} ,
\\
   \begin{pmatrix} X_{\yx\yx'}^\new & Y_{\yx\yx'}^\new \\[2pt]
                   \Ybar_{\yx\yx'}^\new & Z_{\yx\yx'}^\new \end{pmatrix}
   \equiv &
     - i a_\yx^{-1\top} \uOmega \csc(\uOmega\,\Delta t)\, a_\yx^{-1} .
\end {align}
\end {subequations}
The $M$'s, $\Omega$'s and $a$'s are as in section
\ref{app:crossed} with $(x_1,x_2,x_3,x_4)$ set
to
$
  (\hat x_1,\hat x_2,\hat x_3,\hat x_4)=
  (-1,y,z,x)
$,
and $(\Omega,M)_0 \equiv (\Omega,M)_{-1,x,1-x}$.
The only reason that the
factors of $\sgn M$ in (\ref{eq:Dnew}) and
(\ref{eq:Apolenew}) are necessary
is to accommodate the transformation to $2\Re(x \bar y \bar y \bar x)$
below.


\subsubsection {$2\Re(x \bar y \bar y \bar x)$}

\begin {equation}
  2\Re \left[ \frac{d\Gamma}{dx\, dy} \right]_{x \bar y \bar y \bar x}
  =
  \tfrac12 \bigl[ \Abar_\new(x,y) + \Abar_\new(1{-}x,y) \bigr]
\end {equation}
\begin {equation}
  \Abar_\new(x,y) =
  \Abar_\new^{\rm pole}(x,y)
  + \int_0^\infty \frac{d(\widetilde{\Delta t})}{(1{-}x)^{1/2}} \>
      2\Re\bigl[
         B_\new(\tfrac{-x}{1-x},\tfrac{y}{1-x},\widetilde{\Delta t}) \bigr]
\label {eq:Abarnew}
\end {equation}

\begin {multline}
   \Abar^{\rm pole}_\new(x,y) =
    \frac{\alphas^2}{2\pi^2} \,
    P(x)\,P(y) \,
\\ \times
    \Re \biggl\{
       i \Omega_{-1,x,1-x}
       \Bigl[
         - \sigbare
            \Bigl(
               \tfrac{1}{\eps} + \ln\bigl(\tfrac{\pi\mu^2}{\Omega_0 E} \bigr)
            \Bigr)
         + \tfrac12 \ln\bigl( x y (1-x) (1-y) \bigr)
         + \tfrac{i\pi}{2}
       \Bigr]
    \biggr\}
\label {eq:Abarpolenew}
\end {multline}
See appendix \ref{sec:DeriveSEII} for alternative ways
to write (\ref{eq:Abarnew}) and for comments concerning the
physical meaning of the $\widetilde{\Delta t}$ integration variable.


\section {More details on some formulas}
\label {app:details}

\paragraph*{Eq.\ (\ref{eq:dGnetNLOdef}):}
As discussed in section \ref{sec:SymFactor}, our differential rates
do not contain final state symmetry factors, which must then be
included when integrating.
The factor of $1/2$ in the last term of (\ref{eq:dGnetNLOdef}) is
an identical final-state particle factor for two $(y,z)$ of the
three daughters $(x,y,z)$.  The factor is $1/2$ instead of $1/3!$
because the question that defines our $[d\Gamma/dx]_{\rm net}$
(what is the rate for $g\to ggg$ where any one of the daughters
has energy $xE$)
distinguishes the role of $x$, which is not integrated over, from
the roles of $y$ and $z$.

\paragraph*{Eq.\ (\ref{eq:dGnetNLO1}):}
To obtain this equation from (\ref{eq:dGnetNLOdef}),
split the integration region $0 < y < 1$ for $[\Delta\Gamma/dx\,dy]_\virtII$
into (i) $0 < y < 1{-}x$ and (ii) $1{-}x < y < 1$.  Then use the
symmetry of Class II diagrams under
$y \to 1{-}y$ (see fig.\ \ref{fig:virtII})
to change the latter to $0 < y < x$.

\paragraph*{Eq.\ (\ref{eq:dGnetNLO}):}
To obtain this equation from (\ref{eq:dGnetNLO1}),
split the integration region $0 < y < 1{-}x$ into
(i) $0 < y < (1{-}x)/2$ and (ii) $(1{-}x)/2 < y < 1{-}x$, and
then make the change of integration variable $y \rightarrow
z \equiv 1{-}x{-}y$ for the latter.
Finally, note that $[d\Gamma/dx\,dy]_{g\to ggg}$ is symmetric
under permutations of the three daughters, and so in particular under
$y \leftrightarrow z$.
By the way, given the constraints of the $\theta$ functions, any upper
limit $\ge 1/2$ could be used
for the integral signs in (\ref{eq:dGnetNLO}): we've chosen $1/2$
just because that is the largest $y$ for which the integrand can
be non-zero if one considers all possible values of $x$.

\paragraph*{Eq.\ (\ref{eq:xyyxdiv}):}
Eqs.\ (4.36--37) of ref.\ \cite{dimreg} give that the $1/\eps$ piece
of $xy\bar y\bar x$ is
\begin {multline}
   \left[ \frac{d\Gamma}{dx\,dy} \right]_{xy\bar y\bar x} \simeq
   \frac{\CA^2 \alphas^2}{8\pi^2\epsilon}
     \bigl[ (i\Omega_\ix \sgn M_\ix)^{d/2} + (i\Omega_\fx \sgn M_\fx)^{d/2} \bigr]
\\ \times
     \hat x_1^2 \hat x_2 \hat x_3^2 \hat x_4
     (\hat x_1+\hat x_4)^2 (\hat x_3+\hat x_4)^2
       \left[
         (\alpha + \beta)
         - \frac{(\alpha + \gamma) \hat x_2 \hat x_4}
                {(\hat x_1+\hat x_4)(\hat x_3+\hat x_4)}
       \right] ,
\label {eq:xyyxdivDetail}
\end {multline}
where $(\hat x_1,\hat x_2,\hat x_3,\hat x_4) \equiv (-1,y,z,x)$
and $(M,\Omega)_\ix = (M,\Omega)_{-1,x,1-x}$ and
$(M,\Omega)_\fx = (M,\Omega)_{-(1-y),x,z}$.
Then expand (\ref{eq:xyyxdivDetail}) in $\eps$ for $d=2{-}\eps$.

\paragraph*{Eq.\ (\ref{eq:Nevolve0}):}
The type of convolution integral
shown in the gain term is standard for any type of
splitting process.  But to make our
discussion self-contained, we note that its form can be understood
by initially writing the gain term as
\begin {equation}
  \int_\zeta^1 d\zeta' \int_0^1 dx \>
  \delta(\zeta - x\zeta') \,
  \left[ \frac{d\Gamma}{dx} (x,\zeta' E_0) \right]_{\rm net}
  N(\zeta', E_0, t) ,
\end {equation}
where $\zeta' E_0 > \zeta E_0$ is the energy of a particle in the shower
that decays into a daughter carrying fraction $x$ of the parent's
energy.  The $\delta$ function requires that the daughter's energy
$x\zeta' E_0$ match the energy $\zeta E_0$ we are looking for,
and all possibilities for $\zeta'$ and $x$ are integrated over.
Using the $\delta$ function to do the $\zeta'$ integral gives the
gain term in (\ref{eq:Nevolve0}).

\paragraph*{Eq.\ (\ref{eq:DxiIntegral}):}
The desired integral is convergent, but it will be useful to integrate
the two terms separately.  We must introduce a regulator
to split up the integration because the integral of each
term by itself is divergent.
So consider the more general convergent integral
\begin {equation}
  f(\eps) \equiv \int_0^\infty d\tau \> \tau^\eps
    \Bigl( \frac{1}{\tau^2} - \frac{1}{\sh^2\tau} \Bigr)
    \ln(a \tau)
\label {eq:f}
\end {equation}
and follow logic similar to dimensional regularization.
By scaling arguments, similar to dimensional regularization,
the integral of any power must be zero.
For example,
\begin {equation}
  \int_0^\infty  \frac{d\tau}{\tau^{2-\eps}}
  \longrightarrow 0 ,
\label {eq:f0a}
\end {equation}
and then differentiating this result with respect to $\eps$ gives
\begin {equation}
  \int_0^\infty  \frac{d\tau}{\tau^{2-\eps}} \, \ln\tau
  \longrightarrow 0 .
\label {eq:f1b}
\end {equation}
Writing $\ln(a\tau) = \ln a + \ln\tau$, (\ref{eq:f0a}) and (\ref{eq:f1b})
then give
\begin {equation}
  \int_0^\infty  \frac{d\tau}{\tau^{2-\eps}} \, \ln(a\tau)
  \longrightarrow 0 ,
\label {eq:f0b}
\end {equation}
and so the first term in (\ref{eq:f}) integrates to zero
with this regularization.
We are left with
\begin {equation}
  f(\eps) = - \int_0^\infty d\tau \> \frac{\tau^\eps \ln(a\tau)}{\sh^2\tau}
  \,.
\label {eq:f1}
\end {equation}
Consider $\eps > 1$ (for which this integral is convergent), and
then later analytically
continue to $\eps = 0$.  We can rewrite (\ref{eq:f1}) as
\begin {equation}
  f(\eps) = f_0(\eps) \ln a + \frac{d f_0(\eps)}{d\eps}
\label{eq:f2}
\end {equation}
with
\begin {equation}
  f_0(\eps) \equiv - \int_0^\infty d\tau \> \frac{\tau^\eps}{\sh^2\tau} \,.
\end {equation}
From eq.\ (3.527.1) of Gradshteyn and Ryzhik \cite{GR},%
\footnote{
  One may obtain this result by hand by integrating once by parts to
  turn $\tau^\eps/\sh^2(\tau)$ into $\eps \tau^{\eps-1}\cth\tau$.
  Then expand
  $\cth\tau = (e^\tau+e^{-\tau})/(e^{\tau}-e^{-\tau})
   = 1 + 2 e^{-2\tau} + 2 e^{-4\tau} + \cdots$
  and integrate term by term, treating the first term
  $\int_0^\infty \tau^{\eps-1}$ as zero (similar to dimensional regularization).
  [Alternatively, one could put in a large-$\tau$ cut-off $\tau_{\rm max}$,
  and then $\int_0^{\tau_{\rm max}} \tau^{\eps-1}$
  would cancel the previously-ignored boundary terms from
  the integration by parts in the limit $\tau_{\rm max}\to\infty$.]
} 
\begin {equation}
  f_0(\eps) = -2^{1-\eps} \, \Gamma(1+\eps) \, \zeta(\eps)
  = 1 + (\ln\pi - \gammaE)\eps + O(\eps^2) ,
\label {eq:somefint}
\end {equation}
where $\zeta$ is the Riemann zeta function.
Eq.\ (\ref{eq:f2}) then gives the desired result for our
original integral by taking the limit $\eps{=}0$.
To calm any doubts about this derivation, one may simply
check the answer numerically.

\paragraph*{Eq.\ (\ref{eq:GxiIntegral1}):}
Following the technique used above, consider the final integral in
(\ref{eq:GxiIntegral1}) as a special case of
\begin {equation}
  g(\eps) \equiv \int_0^\infty d\tau \> \tau^\eps
    \left[
      \frac{1}{\tau^2}
      \ln \Bigl(
        \frac{1-e^{-2\tau}}{2\tau}
      \Bigr)
      +
      \frac{\tau}{\sinh^2\tau}
    \right] .
\label {eq:g}
\end {equation}
Writing
$\ln\bigl((1-e^{-2\tau})/2\tau\bigr) = \ln\bigl(1-e^{-2\tau}) - \ln(2\tau$),
the integral of $\tau^{\eps-2} \ln(2\tau)$ vanishes as in (\ref{eq:f0b}).
Expanding $\ln(1-e^{-2\tau})$ in powers of $e^{-2\tau}$ and
integrating term by term,
\begin {equation}
  \int_0^\infty d\tau \> \tau^{\eps-2} \ln(1-e^{-2\tau})
  = -2^{1-\eps} \, \Gamma(\eps-1) \, \zeta(\eps)
  = -\frac{1}{\eps} - \ln\pi + \gammaE - 1 + O(\eps) .
\end {equation}
Adapting (\ref{eq:somefint}),
\begin {equation}
  \int_0^\infty d\tau \> \frac{\tau^{1+\eps}}{\sh^2\tau}
  = 2^{-\eps} \, \Gamma(2+\eps) \, \zeta(1+\eps)
  = \frac{1}{\eps} - \ln2 + 1 + O(\eps) .
\end {equation}
Putting everything together and then setting $\eps{=}0$
gives
\begin {equation}
  \int_0^\infty d\tau \>
    \left[
      \frac{1}{\tau^2}
      \ln \Bigl(
        \frac{1-e^{-2\tau}}{2\tau}
      \Bigr)
      +
      \frac{\tau}{\sinh^2\tau}
    \right]
  =
  - \bigl[ \ln(2\pi)-\gammaE \bigr] ,
\end {equation}
as used in (\ref{eq:GxiIntegral1}).


\section {\boldmath$(\alpha,\beta,\gamma)$ in \boldmath$d{=}2{-}\eps$
          dimensions}
\label {app:abcdim}

Eqs.\ (\ref{eq:abc}) and (\ref{eq:abcbar}) present $d{=}2$ results from
refs.\ \cite{2brem} and \cite{seq} for
$(\alpha,\beta,\gamma)$ and $(\bar\alpha,\bar\beta,\bar\gamma)$, which are
various combinations of helicity-dependent DGLAP splitting functions
that arise in calculations of $g\to ggg$ diagrams.
However, in this paper, we use these same quantities in the calculation
of virtual diagrams for $g \to gg$, which are UV-divergent.
So one might expect that when an $\alpha$ or $\beta$ or $\gamma$ is
multiplied by a divergent $1/\eps$, then we need to know the
$O(\eps)$ corrections to $(\alpha,\beta,\gamma)$ in order to calculate
the finite pieces of our $g{\to}gg$ virtual diagrams, similar
to what happens for QED in ref.\ \cite{QEDnf}.%
\footnote{
  Specifically, see the discussion at the end of appendix F.3 of
  ref.\ \cite{QEDnf}.
}
In this appendix, we present $d{=}2{-}\eps$ results for
$(\alpha,\beta,\gamma)$ and $(\bar\alpha,\bar\beta,\bar\gamma)$.
However, we will see that, in the final results of
appendix \ref{app:summary}, $(\alpha,\beta,\gamma)$
and $(\bar\alpha,\bar\beta,\bar\gamma)$ only appear
in combinations where the $O(\eps)$ pieces cancel, and so
the original $d{=}2$ results are all that are actually needed there.

The first important fact is that the helicity-{\it averaged}\/ $g{\to}gg$
DGLAP splitting function $P(x)$ given in (\ref{eq:Pgg}) does not
depend at all on dimension and so has no $O(\eps)$ correction.
[See, for example, eq.\ (17) of ref.\ \cite{Pdim}, which one may
verify independently.]
This lack of dependence on dimension is special to helicity-averaged
$g{\to}gg$ splitting.
Splittings involving quarks do depend on dimension,
but we do not consider those in the large-$\Nc$ limit of
gluon-initiated showers considered in this paper.

For the particular combinations of helicity-{\it dependent}\/
splitting functions that we need, we found it easiest to do the calculation
from scratch.  The helicity basis is unwieldy in general dimensions
since there are no longer simply two helicities $\pm$, and we find it simpler
to do the calculation in a basis of linear polarizations.  Other than that,
we will follow the same notation and normalization conventions and
derivations that were used for the $d{=}2$ case in sections 4.5 and 4.6 of
ref.\ \cite{2brem}.  (See also appendix C of ref.\ \cite{2brem}.)

Following ref.\ \cite{2brem},
we write our splitting vertex matrix elements in the form
\begin {equation}
  \langle \p_j,\p_k|\delta H|\p_i\rangle = g \bcalT_{i\to j k} \cdot\P_{jk}
  \equiv \frac{g T^{\rm color}_{i \to jk} \bcalP_{i\to jk}}{2 E^{3/2}} \cdot\P_{jk},
\label {eq:mat1}
\end {equation}
where
\begin {equation}
  \P_{jk} \equiv x_k\p_j - x_j\p_k
\end {equation}
and the $\p$'s represent transverse momentum, with $\p_i = \p_j+\p_k$.
The $\bcalT_{i\to jk}$ factor above implicitly depends on the polarization,
longitudinal momentum fractions,
and color states of the parent $i$ and daughters $j,k$.  The color
factor is the $T^{\rm color}_{i \to jk}$ above, which is $-i f^{abc}$ for
$g \to gg$.

One may then extract the splitting functions $\bcalP_{i\to jk}$ from
the corresponding matrix elements in the nearly-collinear limit, and
it's easiest to do this by temporarily choosing the axes so that
the parent has transverse momentum zero: $\p_i = 0$ above.
Then define $\q \equiv \p_j = -\p_k$.  One can calculate that the
matrix element ${\cal M}$
for the three-gluon interaction is given by
\begin {equation}
  i {\cal M}_{\rm rel} =
  2 g f^{abc} \left[
     \frac{q_K \delta_{IJ}}{\xi_k} - q_I \delta_{JK}
     + \frac{q_J \delta_{KI}}{\xi_j}
  \right]
  =
  2 g f^{abc} x_i \left[
     \frac{q_K \delta_{IJ}}{x_k} - \frac{q_I \delta_{JK}}{x_i}
     + \frac{q_J \delta_{KI}}{x_j}
  \right] .
\label {eq:Mrel}
\end {equation}
Here, capital roman letters $I,J,K$ run over $1,2,\cdots,d{=}2{-}\eps$ and
index a basis for the (linear) transverse polarization states
of the particles $i,j,k$.
$\xi_j = x_j/x_i$ and $\xi_k = x_k/x_i$ are the longitudinal momentum
fractions of
the two daughters relative to their immediate parent in
$g \to gg$.
The $x$'s are longitudinal momentum fractions
of the various particles in this one particular $g{\to}gg$ splitting relative
to the original particle that initiated the
entire double splitting process.  In the nearly-collinear limit
relevant to high-energy bremsstrahlung, the energies of the particles
in this $g{\to}gg$ splitting are then $(E_i,E_j,E_k) = (x_i,x_j,x_k)E$,
where $E$ is the energy of the original particle that initiated the
double-splitting process.
We bring this up in order to match conventions with the analysis in
ref.\ \cite{2brem}.  That analysis used non-relativistic normalization
of states, and so the desired matrix element is related to the
more conventional ${\cal M}_{\rm rel}$ above by
\begin {align}
  \langle \p_j,\p_k|\delta H|\p_i\rangle
  &= \frac{{\cal M}_{\rm rel}}{(2 E_i)^{1/2} (2 E_j)^{1/2} (2 E_k)^{1/2}}
  = \frac{{\cal M}_{\rm rel}}{(x_i x_j x_k)^{1/2} (2 E)^{3/2}}
\nonumber\\
  &=
  \frac{-ig f^{abc}}{(2 x_i)^{1/2}(x_j x_k E)^{3/2}}
  \left[
     x_i x_j q_K \delta_{IJ} - x_j x_k q_I \delta_{JK}
     + x_k x_i q_J \delta_{KI}
  \right] .
\label {eq:mat2}
\end {align}
With our temporary convention that $\p_i=0$, we have
$\P_{jk} = (x_k{+}x_j)\q = x_i\q$.  Then comparison of
(\ref{eq:mat1}) with (\ref{eq:mat2}) gives the components
of $\bcalP_{i\to jk}$ to be
\begin {equation}
  {\cal P}^n_{i\to jk} =
  \frac{\sqrt2}{(x_i x_j x_k)^{3/2}}
  \left[
     x_i x_j \delta_{IJ} \delta_{nK} - x_j x_k \delta_{JK} \delta_{nI}
     + x_k x_i \delta_{KI} \delta_{nJ}
  \right] .
\label {eq:calPn}
\end {equation}
In this appendix, we will assume that all the $(x_i,x_j,x_k)$ are
positive and will not bother with the absolute value signs that
were included in (\ref{eq:abc}) to be consistent with
front-end transformations.

We can now use (\ref{eq:calPn}) in the definition of the combinations
$(\alpha,\beta,\gamma)$ in eqs.\ (4.37--38) of ref.\ \cite{2brem}, which is
\begin {multline}
   \alpha(x,y) \, \delta^{\bar n n} \delta^{\bar m m}
     + \beta(x,y) \, \delta^{\bar n \bar m} \delta^{nm}
     + \gamma(x,y) \, \delta^{\bar n m} \delta^{n \bar m}
\\
   \equiv
   \frac{1}{d} \sum_{I_\ix}
   \sum_{I_\xx,I_\yx,I_\zx}
   \Bigl[
   \sum_{\bar I}
   {\cal P}^{\bar n}_{\bar I \to I_\zx,I_\xx}\bigl(1{-}y \to z,x\bigr) \,
   {\cal P}^{\bar m}_{I_\ix \to \bar I, I_\yx}\bigl(1 \to 1{-}y,y\bigr)
   \Bigr]^*
\\ \times
   \Bigl[
   \sum_I
   {\cal P}^n_{I \to I_\zx,I_\yx}\bigl(1{-}x \to z,y\bigr) \,
   {\cal P}^m_{I_\ix \to I,I_\xx}\bigl(1 \to 1{-}x,x\bigr)
   \Bigr] .
\label {eq:abcdef}
\end {multline}
[Here, we've indexed the possible linear polarization states using the
letter $I$, whereas in ref.\ \cite{2brem} the helicity basis was used,
indicated by the letter $h$ there.]  Plugging (\ref{eq:calPn}) into
the right-hand side of (\ref{eq:abcdef}) and doing all the sums over
polarization indices for $d$ transverse dimensions, we can then extract
from (\ref{eq:abcdef}) the results for $(\alpha,\beta,\gamma)$.
For $d=2$, the results are given in (\ref{eq:abc}) here and were
originally presented in ref.\ \cite{2brem}.  For general
$d$, we find
\begin {subequations}
\begin {align}
   \alpha &= \alpha_{d{=}2}
   - \frac{ 16(d{-}2) }{ d x^2 y^2 z (1{-}x)^2 (1{-}y)^2 } \,,
\\
   \beta &= \beta_{d{=}2}
   + \frac{ 16(d{-}2) }{ d x^2 y^2 z^2 (1{-}x) (1{-}y) } \,,
\\
   \gamma &= \gamma_{d{=}2}
   - \frac{ 16(d{-}2) }{ d x y z^2 (1{-}x)^2 (1{-}y)^2 } \,.
\end {align}
\end {subequations}
The $(d{-}2)/d$ terms above {\it cancel}\/ in the combination
\begin {equation}
  (\alpha + \gamma) z + (\beta + \gamma) (1{-}x)(1{-}y) ,
\end {equation}
which is the only combination that appears multiplying a
UV-divergent $1/\eps$ in our results summarized in appendix
\ref{app:summary} [see (\ref{eq:ApoleIc})].  For that reason,
there is no problem with just using the $d{=}2$ values
(\ref{eq:abc}) in appendix \ref{app:summary}.

A similar procedure determines $(\bar\alpha,\bar\beta,\bar\gamma)$,
which are defined by eqs.\ (E.2,E.3) of ref.\ \cite{seq} as
\begin {multline}
   \bar\alpha(x,y) \, \delta^{\bar n n} \delta^{\bar m m}
     + \bar\beta(x,y) \, \delta^{\bar n \bar m} \delta^{nm}
     + \bar\gamma(x,y) \, \delta^{\bar n m} \delta^{n \bar m}
\\
   \equiv
   \frac{1}{d} \sum_{I_\ix}
   \sum_{I_\xx,I_\yx,I_\zx}
   \Bigl[
   \sum_{\bar I}
   {\cal P}^{\bar n}_{\bar I \to I_\zx,I_\yx}\bigl(1{-}x \to z,y\bigr) \,
   {\cal P}^{\bar m}_{I_\ix \to \bar I, I_\xx}\bigl(1 \to 1{-}x,x\bigr)
   \Bigr]^*
\\ \times
   \Bigl[
   \sum_I
   {\cal P}^n_{I \to I_\zx,I_\yx}\bigl(1{-}x \to z,y\bigr) \,
   {\cal P}^m_{I_\ix \to I,I_\xx}\bigl(1 \to 1{-}x,x\bigr)
   \Bigr] .
\label {eq:abcbardef}
\end {multline}
This gives
\begin {subequations}
\label {eq:abcbardim}
\begin {align}
  \bar\alpha &=
    \bar\alpha_{d{=}2}
    \,,
\\
  \bar\beta &=
    \frac{4 d}{x y z (1{-}x)^6}
    - \bar\gamma
\\
  \bar\gamma &=
    \frac{8(x-yz)}{x^2y^2z^2(1{-}x)^4}
    + \frac{32}{d x^2y^2z^2(1{-}x)^2} \,,
\end {align}
\end {subequations}
One can check that these results satisfy the QCD version of the identity
of eq. (F32) of ref.\ \cite{QEDnf}:%
\footnote{
  This identity was first given in the earlier paper ref.\ \cite{dimreg}
  eq.\ (5.17) but had to be corrected as discussed in ref.\ \cite{QEDnf}
  appendix F.3.
}
\begin {equation}
   \bar\alpha + \tfrac1{d} \bar\beta + \tfrac1{d} \bar\gamma
   =
   \frac{ P^{(d)}(x) \, P^{(d)}\bigl(\frac{y}{1{-}x}\bigr) }
        { \CA^2 x^2 y^2 z^2 (1{-}x)^3} \,,
\label {eq:PPcheck}
\end {equation}
remembering that for the case of $g\to gg$, the polarization-averaged
splitting functions $P^{(d)}\kern-1pt(x)$ do not in fact depend on dimension $d$.

In our summary of results in appendix \ref{app:summary},
$(\bar\alpha,\bar\beta,\bar\gamma)$ either appear in formulas where
there are no UV-divergent $1/\eps$ factors, or else only appear
implicitly in $d$-independent combinations like
$P(x)\,P(\cdots)$
in (\ref{eq:Apolenew}) and (\ref{eq:Abarpolenew}).
So the general-$d$ formulas (\ref{eq:abcbardim}) are not necessary
for our results.


\section {Details on transforming previous work to
          NLO \boldmath$g{\to}gg$ diagrams}
\label {app:method}

In this appendix, we give more detail about computing NLO $g{\to}gg$
diagrams.  Since many of those diagrams are transformations of
$g{\to}ggg$ diagrams, we start with the latter.


\subsection {Prelude: \boldmath$g{\to}ggg$ Crossed Diagrams}

Though previous work \cite{2brem,dimreg} has calculated
$g{\to}ggg$ processes with dimensional regularization,
those calculations were
complete only for {\it sums} of crossed diagrams for which
UV divergences $1/\eps$ canceled (as they must for
tree-level processes).  The transformations to
virtual crossed diagrams in fig.\ \ref{fig:relateI} do not
involve such UV-canceling collections of $g{\to}ggg$ diagrams,
and so we now need complete results for {\it individual}\/
$g{\to}ggg$ crossed diagrams.
Consistently combining calculations of UV divergences with
finite numerical integrals requires going slightly beyond
what was done in ref.\ \cite{dimreg}, and here we will organize
the calculation using the methods developed in ref.\ \cite{QEDnf}.

In our calculations, UV divergences arise as $\Delta t{\to}0$
divergences of single integrals $\int_0^\infty d(\Delta t) \> F(\Delta t)$
of some function $F(\Delta t)$.
The full integrals are complicated enough that we do not know how to
do them analytically.
As explained in section 4.3.2 of ref.\ \cite{QEDnf}, our
method for isolating the UV divergences and combining them with
numerical integration is to rewrite
\begin {multline}
   \int_0^\infty d(\Delta t) \> F_d(\Delta t)
   =
   \lim_{\mbox{\small``$\scriptstyle{a\to 0}$''}} \Biggl[
   \int_0^a d(\Delta t) \> F_d(\Delta t)
   +
   \int_a^\infty d(\Delta t) \> {\cal D}_2(\Delta t)
   \Biggr]
\\
   +
   \int_0^\infty d(\Delta t) \>
      \bigl[ F_2(\Delta t) - {\cal D}_2(\Delta t) \bigr] 
   + O(\eps) ,
\label {eq:Fsplit}
\end {multline}
where $F_d(\Delta t)$ is the integrand in dimensional regularization
for $d{=}2{-}\eps$ transverse spatial dimensions.  Above,
${\cal D}_2(\Delta t)$ is any convenient function
that
\begin {itemize}
\item
  matches the divergence of $F_2(\Delta t)$ as
  $\Delta t \to 0$;
\item
  falls off fast enough as $\Delta t \to \infty$ so that
  $\int_a^\infty d(\Delta t) \> {\cal D}_2(\Delta t)$ will
  converge for non-zero $a$;
\item
  is simple enough that $\int_a^\infty d(\Delta t) \> {\cal D}_2(\Delta t)$
  can be performed analytically.
\end {itemize}
The last integral in (\ref{eq:Fsplit}) is convergent and can be performed
numerically.  The first term can be found analytically by simplifying
the otherwise complicated integrand $F_d(\Delta t)$ by expanding
it in small $\Delta t$.
The scare quotes around the limit ``$a{\rightarrow}0$'' in
(\ref{eq:Fsplit}) mean that $a{\to}0$ should be taken
after the $\eps{\to}0$ limit.
The exact choice of ${\cal D}_2$ does not matter: the total
(\ref{eq:Fsplit}) will be the same.

Let's focus on the $x y \bar y \bar x$ diagram in fig.\ \ref{fig:crossed}.
The $d{=}2$ integrand for $2\Re(x y \bar y \bar x)$, corresponding
to $F_2(\Delta t)$ above, can be taken from ref.\ \cite{2brem} and
corresponds to a piece of our eqs.\ (\ref{eq:summaryA}--\ref{eq:summaryB}):
\begin {equation}
  F_2(\Delta t) = 2\Re C({-}1,y,z,x,\alpha,\beta,\gamma,\Delta t) ,
\label{eq:F2xyYbXb}
\end {equation}
with $C$ given by (\ref{eq:summaryC}--\ref{eq:summaryD}).
The small $\Delta t$ behavior of this result is given by
eq.\ (5.46) of ref.\ \cite{2brem} as
\begin {multline}
   F_2 (\Delta t) \simeq
   2\Re
   \Biggr\{
     \frac{i \CA^2 \alphas^2}{16\pi^2 \, \Delta t} \, 
       \bigl( \Omega_\ix \sgn M_\ix + \Omega_\fx \sgn M_\fx \bigr)
\\ \times
       \hat x_1^2 \hat x_2 \hat x_3^2 \hat x_4
       (\hat x_1+\hat x_4)^2(\hat x_3+\hat x_4)^2
       \left[
         (\alpha + \beta)
         - \frac{(\alpha + \gamma) \hat x_2 \hat x_4}
                {(\hat x_1 {+} \hat x_4)(\hat x_3 {+} \hat x_4)}
       \right]
  \Biggl\} ,
\label{eq:xyYbXbSmallDt}
\end {multline}
where
\begin {equation}
  (\hat x_1,\hat x_2,\hat x_3,\hat x_4)=(-1,y,z,x) .
\end {equation}
Following similar choices made in ref.\ \cite{QEDnf},
we could now take ${\cal D}_2(\Delta t)$
to be, for example,
the right-hand side of (\ref{eq:xyYbXbSmallDt}) with
the replacements%
\footnote{
  This is what we do to obtain the diagram-by-diagram numerical results
  that were used to determine the non-boldface entries of table
  \ref{tab:limits}.
}
\begin {equation}
   \frac{\Omega}{\Delta t}
   \longrightarrow
   \Omega^3\,\Delta t\,\csc^2(\Omega\,\Delta t),
\label {eq:F2toD2}
\end {equation}
which has the same small-$\Delta t$ behavior but falls off faster as
$\Delta t \to \infty$.
However, for the presentation in this paper, it will be less cumbersome
to just wait until we have assembled all the other elements for the
sum of crossed virtual diagrams and then choose a single overall
${\cal D}_2$ appropriate to that sum.

The information about $2\Re(x y \bar y \bar x)$ we will keep track
of for now is (i) the $d{=}2$ formula (\ref{eq:F2xyYbXb}) for its 
integrand and (ii) the first integral in (\ref{eq:Fsplit}), which
integrates over small times.  The latter, dimensionally-regulated integral
is given by ref.\ \cite{dimreg}:%
\footnote{
  Specifically, see eqs.\ (4.36) and (4.37) of ref.\ \cite{dimreg},
  except that the latter must be multiplied by $(\mu/E)^{2\eps}$,
  as discussed in appendix F.3 of ref.\ \cite{QEDnf}.
}
\begin {equation}
   2\Re\left[ \frac{d\Gamma}{dx\,dy} \right]_{xy\bar y\bar x}^{(\Delta t < a)}
   =
   2\Re\left[ \frac{d\Gamma}{dx\,dy} \right]_{xy\bar y\bar x}^{(\Delta t < a)[1]}
   +
   2\Re\left[ \frac{d\Gamma}{dx\,dy} \right]_{xy\bar y\bar x}^{(\Delta t < a)\rm[2]}
\label {eq:xyyxPole}
\end {equation}
with
\begin {align}
   \left[ \frac{d\Gamma}{dx\,dy} \right]_{xy\bar y\bar x}^{(\Delta t<a)[1]}
   &=
   \frac{\CA^2 \alphas^2}{16\pi^2}
   \left[ \frac{2}{\eps} + \ln\bigl(\frac{\mu^4 a}{E^2}\bigr) + c_1 \right]
     \bigl[ (i\Omega_\ix \sgn M_\ix)^{d/2} + (i\Omega_\fx \sgn M_\fx)^{d/2} \bigr]
\nonumber\\ &\qquad\times
     \hat x_1^2 \hat x_2 \hat x_3^2 \hat x_4
     (\hat x_1{+}\hat x_4)^2 (\hat x_3{+}\hat x_4)^2
       \left[
         (\alpha {+} \beta)
         - \frac{(\alpha {+} \gamma) \hat x_2 \hat x_4}
                {(\hat x_1{+}\hat x_4)(\hat x_3{+}\hat x_4)}
       \right] ,
\label {eq:IxyYbXb}
\\
   \left[ \frac{d\Gamma}{dx\,dy} \right]_{xy\bar y\bar x}^{(\Delta t<a)[2]}
   &=
   - \frac{i\CA^2 \alphas^2}{16\pi^2}
     \bigl[ \Omega_\ix \sgn M_\ix + \Omega_\fx \sgn M_\fx \bigr]
     \hat x_1^2 \hat x_2 \hat x_3^2 \hat x_4
     (\hat x_1+\hat x_4)^2 (\hat x_3+\hat x_4)^2
\nonumber\\ &\qquad\times
     \biggl\{
       \left[
         (\alpha + \beta)
         - \frac{(\alpha + \gamma) \hat x_2 \hat x_4}
                {(\hat x_1+\hat x_4)(\hat x_3+\hat x_4)}
       \right]
       \ln(\hat x_1 \hat x_2 \hat x_3 \hat x_4)
       - 2\gamma 
\nonumber\\ &\hspace{17em}
       - \frac{2 (\alpha+\gamma) \hat x_2 \hat x_4}
                {(\hat x_1+\hat x_4)(\hat x_3+\hat x_4)}
    \biggr\}
\label {eq:IIxyYbXb}
\end {align}
(up to terms that vanish as $\eps{\to}0$),
and
\begin {equation}
   c_1 \equiv 1 + \ln(2\pi^2) .
\end {equation}

As discussed in section 6 of ref.\ \cite{2brem}, the other
crossed $g{\to}ggg$ diagrams can be obtained by various
substitutions:
\begin {subequations}
\label {eq:CrossedTransforms}
\begin {align}
  x y \bar y \bar x \to x \bar y y \bar x
  &:~~
  (\hat x_1, \hat x_2, \hat x_3, \hat x_4) \to
      \bigl( -(1{-}y),{-}y,1{-}x,x \bigr)
    ~~\mbox{and}~~
    \alpha\leftrightarrow\beta ,
\\
  x y \bar y \bar x \to x \bar y \bar x y
  &:~~
  (\hat x_1, \hat x_2, \hat x_3, \hat x_4) \to
      \bigl( {-}y,-(1{-}y),x,1{-}x \bigr)
    ~~\mbox{and}~~
    (\alpha,\beta,\gamma) \leftrightarrow (\gamma,\alpha,\beta),
\label{eq:GETxyyx}
\end {align}
\end {subequations}
where the changes to $(\hat x_1,\hat x_2,\hat x_3,\hat x_4)$ are also
applied to our formulas (\ref{eq:OMif}) defining
$(\Omega,M)_\ix$ and $(\Omega,M)_\fx$.


\subsection {Crossed Virtual Diagrams}

We now obtain results for the crossed virtual diagrams from the
preceding expressions by using
the transformations of fig.\ \ref{fig:relateI}.


\subsubsection {$(\Delta t < a)[1]$ terms}

Let's first focus on the ``$(\Delta t < a)[1]$'' terms, which trace back
to (\ref{eq:IxyYbXb}), using (\ref{eq:CrossedTransforms}) when
relevant.  We find
\begin {align}
   2\Re&\left[ \frac{d\Gamma}{dx\,dy} \right]_{\virtIc}^{\Delta t{<}a[1]}
   =
\nonumber\\ &
   - \frac{\CA^2 \alphas^2}{8\pi^2}
     x y z (1{-}x)(1{-}y)
     \Re \biggl\{
       \left[ \frac{2}{\eps} + \ln\bigl(\frac{\mu^4 a}{E^2}\bigr) + c_1 \right]
         ( H_{yx\bar x y} +  H_{y\bar x xy} + H_{\bar x yxy} + H_{yxy\bar x} )
\nonumber\\ & \hspace{20em}
       - 4 H_{yxy\bar x} \ln(1{-}y)
     \biggr\}
     ,
\label {eq:poleX0}
\end {align}
where the contributions from individual diagrams (in some cases
complex conjugated) are
\begin {subequations}
\label {eq:H}
\begin {align}
  H_{yx\bar x y} &=
       \bigl[ (i\Omega_{-1,y,1-y})^{d/2} + (i\Omega_{-(1-x),y,z})^{d/2} \bigr]
       \bigl[
         (\alpha{+}\beta) z(1{-}x)(1{-}y)
         + (\alpha{+}\gamma) x y z
       \bigr] ,
\\
  H_{y\bar x xy} &=
       \bigl[ (i\Omega_{-1,y,1-y})^{d/2} + (i\Omega_{-(1-x),y,z})^{d/2} \bigr]
\nonumber\\ & \qquad\qquad \times
       \bigl[
         - (\alpha{+}\beta) z(1{-}x)(1{-}y)
         + (\beta{+}\gamma) x y (1{-}x)(1{-}y)
       \bigr] ,
\\
  H_{\bar x yxy} &=
     \bigl[ (i\Omega_{-1,x,1-x})^{d/2} + (i\Omega_{-(1-x),y,z})^{d/2} \bigr]
       \bigl[
         - (\alpha{+}\gamma) x y z
         - (\beta{+}\gamma) x y (1{-}x)(1{-}y)
       \bigr] ,
\\
  H_{yxy\bar x} &=
     \bigl[ (i\Omega_{-1,y,1-y})^{d/2} + (i\Omega_{-1,x,1-x})^{d/2} \bigr]
       \bigl[
         - (\alpha{+}\gamma) x y z
         - (\beta{+}\gamma) x y (1{-}x)(1{-}y)
       \bigr] .
\label {eq:Hyxyx}
\end {align}
\end {subequations}
The $1/\eps$ pieces of these formulas are the divergences
(\ref{eq:crossedXpieces}) presented in the main text.
The subscript ``$\virtIc$'' in (\ref{eq:poleX0}) stands for
``virtual crossed diagrams'' (which are all a type of Class I diagram),
as in (\ref{eq:virtIsummary}).

The $yxy\bar x$ is a little different from the other diagrams above
because it is the only one that involves a front-end transformation.
Fig.\ \ref{fig:relateI} shows that
$2\Re(yxy\bar x)$ is given by a front-end transformation of
$2\Re(x\bar y\bar x y)$ followed by $x\leftrightarrow y$.
The initial front-end transformation (\ref{eq:frontend})
takes
\begin {equation}
     (x,y,E) \longrightarrow
     \Bigl( \frac{{-}x}{1{-}x} \,,\, \frac{y}{1{-}x} \,,\, (1{-}x)E \Bigr) .
\label {eq:FExyE}
\end {equation}
One can check from the explicit formulas (\ref{eq:abc}) for
$(\alpha,\beta,\gamma)$ that this transformation maps
\begin {equation}
  (\alpha,\beta,\gamma) \longrightarrow (1{-}x)^{10} (\beta,\alpha,\gamma) .
\label {eq:FEabc}
\end {equation}
We have used this plus the fact that $(\alpha,\beta,\gamma)$ are symmetric
under $x\leftrightarrow y$ in deriving (\ref{eq:Hyxyx}).

The other special feature of the front-end transformation (\ref{eq:frontend})
is that it introduces an overall factor of $(1{-}x)^{-\eps}$.  To see what
happens to this, focus on the factor
$[2/\eps + \ln(\mu^4a/E^2) + c_1]$ in (\ref{eq:IxyYbXb}) for
$xy\bar y\bar x$.  By (\ref{eq:GETxyyx}), this factor is the same for
$x\bar y \bar x y$.  The front-end transformation (\ref{eq:FExyE})
of $E$ together with the overall front-end transformation factor
$(1{-}x)^{-\eps}$, followed by the switch of variables $x\leftrightarrow y$,
then takes
\begin {multline}
  \left[ \frac{2}{\eps} + \ln\bigl(\frac{\mu^4 a}{E^2}\bigr) + c_1 \right]
  \longrightarrow
  (1{-}y)^{-\eps} \left[ \frac{2}{\eps}
                   + \ln\bigl(\frac{\mu^4 a}{[(1{-}y)E]^2}\bigr) + c_1 \right]
\\
  =
  \left[ \frac{2}{\eps} + \ln\bigl(\frac{\mu^4 a}{E^2}\bigr) + c_1 \right]
  - 4\ln(1{-}y)
  + O(\eps) .
\label {eq:this}
\end {multline}
The extra $-4\ln(1{-}y)$ term above is responsible for the last term
in (\ref{eq:poleX0}), and we will see later that it conspires
in a natural way with similar logarithms in the ``$(\Delta t{<}a)[2]$''
piece of $2\Re(yxy\bar x)$ that we will derive from
(\ref{eq:IIxyYbXb}).

When the four terms (\ref{eq:H}) are added together in (\ref{eq:poleX0}),
all but the two $(i\Omega_{-1,x,1-x})^{d/2}$ terms cancel in pairs.
Expanding those in $\eps$, we find
\begin {align}
   2\Re&\left[ \frac{d\Gamma}{dx\,dy} \right]_{\virtIc}^{\Delta t{<}a[1]}
   =
\nonumber\\ &
   \frac{\CA^2 \alphas^2}{4\pi^2}
     x y z (1{-}x)(1{-}y)
\nonumber\\ & \qquad \times
     \Re\biggl\{
       \biggl(
         \left[ \frac{2}{\eps} + \ln\bigl(\frac{\mu^4 a}{i\Omega_0E^2}\bigr)
         + c_1 \right]
         i\Omega_0
       \biggr)
       \bigl[
         (\alpha{+}\gamma) x y z
         + (\beta{+}\gamma) x y (1{-}x)(1{-}y)
       \bigr]
\nonumber\\ & \hspace{20em}
       + 2 H_{yxy\bar x}^{(d=2)} \ln(1{-}y)
     \biggr\}
     ,
\label {eq:poleX1}
\end {align}
where $\Omega_0 \equiv \Omega_{-1,x,1-x}$.


\subsubsection {$(\Delta t < a)[2]$ terms}

Similarly combining (\ref{eq:IIxyYbXb}), (\ref{eq:CrossedTransforms})
and fig.\ \ref{fig:relateI}, we find the remaining contributions
from the dimensionally-regulated integration over $\Delta t < a$ are
\begin {align}
   2\Re\left[ \frac{d\Gamma}{dx\,dy} \right]_{\virtIc}^{\Delta t{<}a[2]}
   = {}&
   \frac{\CA^2 \alphas^2}{8\pi^2}
     x y z (1{-}x)(1{-}y)
     \Re \biggl\{
         ( H_{yx\bar x y}^{(d=2)} + H_{yxy\bar x}^{(d=2)} ) \ln(e^{-i\pi} x y z)
\nonumber\\ & \qquad
         + ( H_{y\bar x xy}^{(d=2)} + H_{\bar x yxy}^{(d=2)} )
              \ln\bigl(x y (1{-}x)(1{-}y) \bigr)
         - 4 H_{yxy\bar x}^{(d=2)} \ln(1{-}y)
\nonumber\\ & \qquad
         + h_{yx\bar x y} + h_{y\bar x xy} + h_{\bar x yxy} + h_{yxy\bar x}
     \biggr\}
     ,
\label {eq:poleX2}
\end {align}
with
\begin {subequations}
\label {eq:dGcrossedXII}
\begin {align}
   h_{yx\bar x y}
   &=
   (i\Omega_{-1,y,1-y} + i\Omega_{-(1-x),y,z})
     \bigl[
       - 2\gamma z (1{-}x)(1{-}y)
       + 2(\alpha+\gamma) x y z
     \bigr] ,
\\
   h_{y\bar x xy}
   &=
   (i\Omega_{-1,y,1-y} + i\Omega_{-(1-x),y,z})
     \bigl[
       2\gamma z (1{-}x)(1{-}y)
       + 2(\beta+\gamma) x y (1{-}x)(1{-}y)
     \bigr] ,
\\
   h_{\bar x yxy}
   &=
   (i\Omega_{-1,x,1-x} + i\Omega_{-(1-x),y,z})
     \bigl[
       2\beta x y z
       - 2(\beta+\gamma) x y (1{-}x)(1{-}y)
     \bigr] ,
\\
   h_{yxy\bar x}
   &=
   (i\Omega_{-1,y,1-y} + i\Omega_{-1,x,1-x})
     \bigl[
       2\alpha x y (1{-}x)(1{-}y)
       - 2 (\alpha+\gamma) x y z
     \bigr] .
\end {align}
\end {subequations}
The individual contributions from each diagram to (\ref{eq:poleX2})
can be identified by the subscripts.  The phase $e^{-i\pi}$ in a
logarithm should be interpreted as
\begin {equation}
   \ln(e^{-i\pi} xyz) = \ln(xyz) - i\pi ,
\end {equation}
and the selection of this branch cut is explained in
section 4.6 of ref.\ \cite{dimreg}.


\subsubsection {${\cal D}_2(\Delta t)$}

We are now in a position to choose ${\cal D}_2(\Delta t)$ of
(\ref{eq:Fsplit}) for the entire sum of crossed virtual
diagrams.  The $1/\eps$ divergence in (\ref{eq:poleX1}) represents
the dimensional regularization of a $\Delta t{\to}0$
divergent integral
$\int_0^a d(\Delta t)/(\Delta t)$.  We
may use this as a convenient short-cut to read off the
$\Delta t{\to}0$ behavior of the integrand from (\ref{eq:poleX1}),
using the observation of eq.\ (4.35) of ref \cite{dimreg} that the regulated
UV divergence is
\begin {equation}
   \int_0^a \frac{d(\Delta t)}{(\Delta t)^{d/2}}
   = \frac{2}{\eps} + \ln a + O(\eps) .
\end {equation}
From the $1/\eps$ terms of (\ref{eq:poleX1}), we then see that the
small $\Delta t$ behavior $F_2(\Delta t)$
of the $d{=}2$ integrand for the sum of
virtual crossed diagrams is
\begin {align}
   \frac{\CA^2 \alphas^2}{4\pi^2} &
     x y z (1{-}x)(1{-}y) \,
       \frac{\Re(i\Omega_0)}{\Delta t} \,
       \bigl[
         (\alpha{+}\gamma) x y z
         + (\beta{+}\gamma) x y (1{-}x)(1{-}y)
       \bigr]
     .
\label {eq:F2smalldt}
\end {align}
(Alternatively, one could explicitly extract the $\Delta t{\to}0$
behavior of each diagram and add them up to get the same answer.)
Applying the replacement (\ref{eq:F2toD2}) to (\ref{eq:F2smalldt})
yields our choice
of ${\cal D}_2$, given in (\ref{eq:D2virtIc}).

One of the terms we need in our split (\ref{eq:Fsplit}) of analytic
vs.\ numerical integration is an analytic integral of ${\cal D}_2(\Delta t)$.
Integrating (\ref{eq:D2virtIc}) using
\begin {equation}
   \int_a^\infty d(\Delta t) \> \Omega^3 \,\Delta t \csc^2(\Omega\,\Delta t)
   = -\Omega\bigl[ \ln(2i\Omega a) - 1 \bigr]
     + O(a)
\end {equation}
for small $a$, and combining with (\ref{eq:poleX1}) and (\ref{eq:poleX2}),
gives a result for what we call
\begin {equation}
   A^{\rm pole}_\virtIc(x,y) \equiv
   \lim_{\mbox{\small``$\scriptstyle{a\to 0}$''}} \Biggl[
   \int_0^a d(\Delta t) \> F_d(\Delta t)
   +
   \int_a^\infty d(\Delta t) \> {\cal D}_2(\Delta t)
   \Biggr]_\virtIc
\label {eq:ApoleIcdef}
\end {equation}
for this aspect of the sum of virtual crossed diagrams.
Our result for $A^{\rm pole}_\virtIc$ is given in (\ref{eq:ApoleIc}).


\subsubsection {$F_2(\Delta t)$}

From (\ref{eq:F2xyYbXb}) for $2\Re(xy\bar y\bar x)$, combined with
(\ref{eq:CrossedTransforms}) to get other crossed $g\to ggg$
diagrams, combined with fig.\ \ref{fig:relateI} to relate them to
crossed virtual diagrams, we have
\begin {align}
   \bigl[ F_2(\Delta t) \bigr]_\virtIc
   & =
\nonumber\\ 
     2 \Re\Bigl( &
       - C({-}1,x,z,y,\alpha,\beta,\gamma,\Delta t)
       - C\bigl({-}(1{-}x),{-}x,1{-}y,y,\beta,\alpha,\gamma,\Delta t\bigr)
\nonumber\\ &
       - C\bigl({-}y,{-}(1{-}y),x,1{-}x,\gamma,\alpha,\beta,\Delta t\bigr)
\nonumber\\ &
       - \Bigl\{\bigl[
           C\bigl({-}y,{-}(1{-}y),x,1{-}x,\gamma,\alpha,\beta,\Delta t\bigr)
         \bigr]_{(x,y,E)\to(\frac{-x}{1-x},\frac{y}{1-x},(1{-}x)E)}
         \Bigr\}_{x\leftrightarrow y}
     \Bigr) .
\label {eq:F2virtIc}
\end {align}
One can simplify the last $-C$ term.
Using (\ref{eq:FEabc}) and the fact that every term in the formulas
(\ref{eq:summaryC},\ref{eq:summaryD}) determining $C$ is proportional to one of
$(\alpha,\beta,\gamma)$, the
last term in (\ref{eq:F2virtIc}) is equivalent to
\begin {equation}
  - (1{-}y)^{10}
  \bigl[
     C\bigl(
       {-}\tfrac{x}{1-y},{-}\tfrac{z}{1-y},-\tfrac{y}{1-y},\tfrac{1}{1-y},
       \gamma,\beta,\alpha,\Delta t
      \bigr)
  \bigr]_{E \to (1-y)E} \,.
\label {eq:Cfuss1}
\end {equation}
From (\ref{eq:summaryC}) and (\ref{eq:summaryD}) [and keeping track of
all $E$'s hidden inside of definitions of $\Omega$'s and $M$'s and
thence inside $(X,Y,Z)$'s and $I$'s], one may verify the
scaling property that
\begin {equation}
  \bigl[
    C( \lambda x_1,\lambda x_2,\lambda x_3,\lambda x_4,
       \alpha,\beta,\gamma,\Delta t )
  \bigr]_{E\to E/\lambda}
  = \lambda^{10} C(x_1,x_2,x_3,x_4,\alpha,\beta,\gamma) .
\label {eq:Cscale}
\end {equation}
So (\ref{eq:Cfuss1}) may be rewritten as
\begin {equation}
  - C({-}x,{-}z,{-}y,1,\gamma,\beta,\alpha,\Delta t) .
\label {eq:Cfuss2}
\end {equation}
Since we take $2\Re( \cdots )$ in (\ref{eq:F2virtIc}), we may replace
the above by its complex conjugate.  In our formalism, conjugating
diagrams is equivalent to negating the values of all the $x_i$,
and so the conjugate of (\ref{eq:Cfuss2}) is
$-C(x,z,y,{-}1,\gamma,\beta,\alpha,\Delta t)$.
This is the version we have used for our final rewriting of (\ref{eq:F2virtIc}),
which is presented as $2\Re B_\virtIc$ in eqs.\ (\ref{eq:AvirtIc}) and
(\ref{eq:BvirtIc}).  Following (\ref{eq:Fsplit}), this
$(F_2)_\virtIc = 2\Re B_\virtIc$ is combined with
$({\cal D}_2)_\virtIc$ and $A_\virtIc^{\rm pole}$ (\ref{eq:ApoleIcdef})
to give our final total result (\ref{eq:AvirtIc})
for the crossed virtual diagrams.


\subsection {Sequential Virtual Diagrams}
\label {sec:DeriveVirtSeq}

The sum of Class I sequential virtual diagrams
($xy\bar xy$, $x\bar xyy$, and $x\bar x\bar y\bar y$ from
fig.\ \ref{fig:virtI} plus conjugates)
are, by fig.\ \ref{fig:relateI}, just the back-end transformation of
the sum of the three $g{\to}ggg$ sequential diagrams shown in the first
line of fig.\ \ref{fig:seq} plus conjugates.
The latter, computed previously \cite{seq}, is
${\cal A}_\seq(x,y) + {\cal A}_\seq(x,z)$ with
${\cal A}_\seq$ given by (\ref{eq:AseqBE}),
where the separate terms ${\cal A}_\seq(x,y)$ and
${\cal A}_\seq(x,z)$ correspond to two different large-$\Nc$ color
routings of the diagrams.%
\footnote{
  See section 2.2.1 of ref.\ \cite{seq}.
}
The back-end transformation
just introduces an overall minus sign, and we must include a loop symmetry
factor of $\tfrac12$ for the amplitude (blue) or conjugate amplitude (red)
loops in the resulting virtual
diagrams, giving
\begin {equation}
   - \tfrac12 \bigl[ {\cal A}_\seq(x,y) + {\cal A}_\seq(x,z) \bigr] .
\end {equation}
This result is summarized in (\ref{eq:AseqBE}).

Similarly, as depicted in fig.\ \ref{fig:relateII},
a front-end transformation of
${\cal A}_\seq(x,y) + {\cal A}_\seq(x,z)$ followed by $x{\leftrightarrow}y$
gives the sum
$2\Re[\bar y x \bar y \bar x + \bar y \bar y x \bar x + y y x \bar x]$
of three Class II sequential virtual diagrams:
\begin {equation}
  - \tfrac12
  \bigl[
    {\cal A}_\seq\bigl(\tfrac{-y}{1-y},\tfrac{x}{1-y}\bigr)
    +
    {\cal A}_\seq\bigl(\tfrac{-y}{1-y},\tfrac{1-x}{1-y}\bigr)
  \bigr]_{E \to (1{-}y)E} \,.
\label {eq:AseqFE0}
\end {equation}
Since rates for all of these processes (as well as for leading-order
$g\to gg$) ultimately depend on $\hat q$ and $E$ as
$\sqrt{\hat q/E}$, we can rewrite (\ref{eq:AseqFE0}) as%
\footnote{
   Note that, unlike (\ref{eq:this}),
   we do not need the overall $(1{-}y)^{-\eps}$
   factor arising from the front-end transformation.
   That's because ${\cal A}_\seq$
   is finite as $\eps{\to}0$.
}
\begin {equation}
  - \tfrac12 (1{-}y)^{-1/2}
  \bigl[
    {\cal A}_\seq\bigl(\tfrac{-y}{1-y},\tfrac{x}{1-y}\bigr)
    +
    {\cal A}_\seq\bigl(\tfrac{-y}{1-y},\tfrac{1-x}{1-y}\bigr)
  \bigr] .
\label {eq:AseqFE1}
\end {equation}
This is the result summarized in (\ref{eq:AseqFE}).

A small advantage of (\ref{eq:AseqFE1}) over (\ref{eq:AseqFE0})
for numerical work is that one may work throughout in units
where $\hat q{=}1$ and $E{=}1$ to get numerical results
for rates in units of $\sqrt{\hat q/E}$.  Alternatively,
one could implement the original (\ref{eq:AseqFE0}) by making
$E$ itself an additional
argument of all the functions in section \ref{app:seqsummary}.

For analytic work, there is a potential conceptual confusion
associated with (\ref{eq:AseqFE1}) concerning the meaning
of the integration variable $\Delta t$ in the definition
(\ref{eq:Aseq}) of ${\cal A}_\seq$.
In all the previous discussion in this paper, $\Delta t$
has represented the difference in time between the
middle two splitting vertices of interference diagrams
like figs.\ \ref{fig:crossed}, \ref{fig:seq}, \ref{fig:virtI},
and \ref{fig:virtII}.  However, if one steps through in mathematical detail
how the explicit formulas of section \ref{app:seqsummary}
produce equivalence of (\ref{eq:AseqFE0}) and (\ref{eq:AseqFE1}),
one finds that $\widetilde{\Delta t} = (1{-}y)^{-1/2} \Delta t$,
where $\widetilde{\Delta t}$ represents
the time integration variable associated here
with the formula for (\ref{eq:AseqFE1}).
In terms of our earlier scaling argument for (\ref{eq:AseqFE1})
from (\ref{eq:AseqFE0}), which bypassed looking at
details of the formulas for ${\cal A}_\seq$, this rescaling of
the meaning of the $\Delta t$ integration variable
reflects the fact that formation times scale
like $\sqrt{E/\hat q}$.

Finally, we mention that (\ref{eq:FEabc}) showed how the
combinations $(\alpha,\beta,\gamma)$ of helicity-dependent DGLAP
splitting functions mapped into each other under front-end
transformation, but there is no similar relation for
the combinations $(\bar\alpha,\bar\beta,\bar\gamma)$ that
appear in formulas like (\ref{eq:Bseq}) for
sequential diagrams.  But we have checked that
front-end transformation (\ref{eq:FExyE}) takes
\begin {equation}
  (\bar\alpha,\bar\beta,\bar\gamma) \longrightarrow
  (1{-}x)^{10} \, (\bar{\bar\alpha},\bar{\bar\beta},\bar{\bar\gamma}) ,
\label {eq:FEabcbar}
\end {equation}
where $(\bar{\bar\alpha},\bar{\bar\beta},\bar{\bar\gamma})$ are the
combinations of splitting functions that would be obtained by {\it directly}
evaluating front-end sequential virtual diagrams instead of
using our short-cut method of front-end transforming
previously known $g\to ggg$ sequential diagrams.
In detail, (\ref{eq:FEabcbar}) gives
$(\bar{\bar\alpha},\bar{\bar\beta},\bar{\bar\gamma})$
in terms of eqs.\ (\ref{eq:abcbar}) for $(\bar\alpha,\bar\beta,\bar\gamma)$
as
\begin {equation}
  \bar{\bar\alpha}(x,y)
  =
  (1{-}x)^{-10} \, \bar\alpha
  \bigl( \tfrac{-x}{1{-}x}, \tfrac{y}{1{-}x} \bigr)
  ,~~\mbox{etc.}
\label {eq:abcbarbar}
\end {equation}
Unlike $(\bar\alpha,\bar\beta,\bar\gamma)$, the
$(\bar{\bar\alpha},\bar{\bar\beta},\bar{\bar\gamma})$ turn out to
be symmetric in $x{\leftrightarrow}y$ and so are unaffected by
that step of the transformation of $g{\to}ggg$ diagrams into
Class II sequential virtual diagrams in fig.\ \ref{fig:relateII}.


\subsection {\boldmath$2\Re(xyy\bar x)$}
\label {app:methodFund}

As mentioned in the main text, we can obtain the result for the
$xyy\bar x$ diagram of fig.\ \ref{fig:virtI} by adapting the
results \cite{QEDnf} for the similar QED diagram of
fig.\ \ref{fig:QEDfund}.  To go from QED to QCD, we need
the following modifications.

\begin {itemize}
  \item
    To account for QCD group factors at the vertices, we need
    to replace
    $\Nf \alphaqed^2 \to
     \dA^{-1} \alphas^2 \tr(T^a_{\rm A} T^b_{\rm A} T^b_{\rm A} T^a_{\rm A})
     = \CA^2 \alphas^2$
    overall.
    However, there are two different large-$\Nc$ color routings
    of the QCD diagram, similar to the discussion of color routings
    of sequential diagrams in section 2.2.1 of ref.\ \cite{seq}.
    So the overall $\CA^2 \alphas^2$ corresponds to a factor of
    $\tfrac12 \CA^2 \alphas^2$ per large-$\Nc$ color routing.
  \item
    $P_{e\to e}$ and $P_{\gamma\to e}$ are both replaced by
    $P_{g\to gg}/\CA$, where the $\CA$ is taken out because each
    $\CA$ in a $P_{g\to gg}$ is already explicitly accounted for in the
    $\Nf \alphaqed^2 \to \CA^2 \alphas^2$ translation above.
    Similarly, one should use the (gluonic) QCD formulas of
    (\ref{eq:abcbar}) for the combinations
    $(\bar\alpha,\bar\beta,\bar\gamma)$ of helicity-dependent splitting
    functions that are needed for this diagram.  (See appendix
    \ref{app:abcdim} for an explanation of why $d{=}2{-}\eps$
    versions are not needed.)
  \item
    Unlike the electron self-energy loop in fig.\ \ref{fig:QEDfund},
    the corresponding gluon self-energy loop comes with a loop
    symmetry factor of $\frac12$.
  \item
    Use the (gluonic) QCD formulas of appendix
    \ref{app:gggSummary} for complex frequencies
    $\Omega$ and matrices $a$ of normal modes.
  \item
    Unlike QED, $\Omega_-$ is non-zero, and so, for example,
    prefactors such
    as $\Omega_+ \csc(\Omega_+ \Delta t) / \Delta t$ in the QED calculation
    will revert to the more general form
    $\Omega_+ \csc(\Omega_+\,\Delta t) \, \Omega_- \csc(\Omega_-\,\Delta t)$
    in the QCD calculation.
\end {itemize}  

Let $A_{\rm new}(x,y)$ represent a single color routing {\it not} including
the loop symmetry factor $\tfrac12$.  By the same arguments given in section
2.2.1 of ref.\ \cite{seq}, the two color routings are related by
$y\leftrightarrow z$, and so
\begin {equation}
  2\Re \left[ \frac{d\Gamma}{dx\, dy} \right]_{x y y\bar x}
  =
  \tfrac12 \bigl[ A_\new(x,y) + A_\new(x,z) \bigr] ,
\end {equation}
which is (\ref{eq:dGfund}).
For $A_\new$, we can then copy various formulas from ref.\ \cite{QEDnf}
with  $\Nf\alphaqed \to \tfrac12 \CA^2 \alphas^2$ and other modifications
listed above.  The $[d\Gamma/d\xe\,d\ye]^{\rm(subtracted)}_{xyy\bar x}$
of eq.\ (A.43) of ref.\ \cite{QEDnf} corresponds to the
$\int_0^\infty d(\Delta t) \> [ F_2(\Delta t) - {\cal D}_2(\Delta t) ]$
of our (\ref{eq:Fsplit}) and translates to what we call
$\int_0^\infty d(\Delta t) \> 2\Re[B_{\rm new}(x,y,\Delta t)]$
in our (\ref{eq:Anew}--\ref{eq:D2Isummary}).

All that remains is the pole piece, which we will package a little
differently in this paper than in ref.\ \cite{QEDnf}.
Similar to our analysis of crossed virtual diagrams, the pole piece
corresponds to the
$\int_0^a d(\Delta t) \> F_d(\Delta t) + \int_a^\infty {\cal D}_2(\Delta t)$
term in (\ref{eq:Fsplit}).
The QCD formula we need can be obtained by starting from
eq.\ (F.42) of ref.\ \cite{QEDnf} for QED:%
\footnote{
  Here we have accounted for an overall sign error that appeared
  in the original published version of eq.\ (F.42) in ref.\ \cite{QEDnf}.
}
\begin {align}
  \lim_{\mbox{\small``$\scriptstyle{a\to 0}$''}} \Biggl\{
&
    \left[ \frac{d\Gamma}{d\xe} \right]_{xyy\bar x}^{(\Delta t < a)}
    +
    \left[ \frac{d\Gamma}{d\xe} \right]_{xyy\bar x}^{({\cal D}_2)}
  \Biggr\}
\nonumber\\ &
  =
  \left( \frac{\mu^2}{E} \right)^{\eps}
  \frac{d \Nf \alphaqed^2}{2^{d+2} \pi^{2d}} \,
  \Gamma^2( \tfrac12{+}\tfrac{d}{4} ) \,
  \frac{ P^{(d)}_{e\to e}(\xe) }
       { \xe^{\eps/2}(1-\xe)^\eps } 
  \int_0^1 d\yfrake
    \frac{ P^{(d)}_{\gamma\to e}(\yfrake) }
         { \bigl[ \yfrake(1-\yfrake) \bigr]^{\eps/2} }
\nonumber\\ &\qquad \times
  2\pi^2 i \barOmega_\ix^{d-1} \Bigl[
    - \Bigl( \frac{2}{\eps} - \gammaE + \ln(4\pi) \Bigr)
    + 4\ln2 + 3\ln\pi - 1
  \Bigr]
  + O(\eps) ,
\label{eq:QEDdivagogo}
\end{align}
where $\yfrake \equiv \ye/(1{-}\xe)$.  Here we also adopt the shorthand
notation of ref.\ \cite{QEDnf} that
\begin {equation}
  \barOmega \equiv \Omega \sgn M .
\end {equation}
In the QED case, it was possible to explicitly perform the
integral over $\yfrake$ above to get
the pole piece of $[d\Gamma/dx]_{x y y \bar x}$.%
\footnote{
  In the QED case, the pole piece of $[d\Gamma/dx]_{x y y \bar x}$
  corresponds to everything other than the subtracted term in
  eq. (A.41) of ref.\ \cite{QEDnf}.
}
For the QCD translation, the corresponding $y$ integral will
be IR divergent because, unlike $P_{\gamma\to e}(\yfrak)$,
gluon splitting $P_{g\to g}(\yfrak)$ diverges
in the soft limit $\yfrak{\to}0$.
We could do the integral explicitly using our IR regulator
$\delta$, but,
for our various discussions of cancellations of QCD power-law
IR divergences
in this paper, it has been very useful to work in terms
of $d\Gamma/dx\,dy$ for virtual diagrams instead of directly with
the IR-regulated $d\Gamma/dx$ for each diagram.
So we'll instead translate the unintegrated version of
(\ref{eq:QEDdivagogo}),
\begin {align}
  \lim_{\mbox{\small``$\scriptstyle{a\to 0}$''}} \Biggl\{
&
    \left[ \frac{d\Gamma}{d\xe\,d\ye} \right]_{xyy\bar x}^{(\Delta t < a)}
    +
    \left[ \frac{d\Gamma}{d\xe\,d\ye} \right]_{xyy\bar x}^{({\cal D}_2)}
  \Biggr\}
\nonumber\\ &
  =
  \left( \frac{\mu^2}{E} \right)^{\eps}
  \frac{d \Nf \alphaqed^2}{2^{d+2} \pi^{2d}} \,
  \Gamma^2( \tfrac12{+}\tfrac{d}{4} ) \,
  \frac{ P^{(d)}_{e\to e}(\xe) }
       { \xe^{\eps/2}(1-\xe)^\eps } \,
  \frac{1}{(1-\xe) \vphantom{\xe^{\eps/2}}} \,
    \frac{ P^{(d)}_{\gamma\to e}(\yfrake) }
         { \bigl[ \yfrake(1-\yfrake) \bigr]^{\eps/2} }
\nonumber\\ &\qquad \times
  2\pi^2 i \barOmega_\ix^{d-1} \Bigl[
    - \Bigl( \frac{2}{\eps} - \gammaE + \ln(4\pi) \Bigr)
    + 4\ln2 + 3\ln\pi - 1
  \Bigr]
  + O(\eps) ,
\end{align}
from QED to
\begin {align}
  A^{\rm pole}_{x y y \bar x} &=
  \left( \frac{\mu^2}{E} \right)^{\eps}
  \frac{d \alphas^2}{2^{d+3} \pi^{2d}} \,
  \Gamma^2( \tfrac12{+}\tfrac{d}{4} ) \,
  \frac{ P^{(d)}(x) }
       { x^{\eps/2}(1-x)^\eps } \, 
  \frac{1}{(1-x)} \,
    \frac{ P^{(d)}(\yfrak) }
         { [ \yfrak(1-\yfrak) ]^{\eps/2} }
\nonumber\\ &\qquad \times
  2\pi^2 i \barOmega_\ix^{d-1} \Bigl[
    - \Bigl( \frac{2}{\eps} - \gammaE + \ln(4\pi) \Bigr)
    + 4\ln2 + 3\ln\pi - 1
  \Bigr]
  + O(\eps)
\label{eq:divagogo}
\end{align}
for QCD.
$\Omega_\ix$ for the QCD diagram is equal to $\Omega_0 \equiv \Omega_{-1,x,1-x}$.
Now (i) use the fact that the $g{\to}gg$ splitting function $P(x)$ is
independent of dimension (see appendix \ref{app:abcdim}),
(ii) fully expand the above in $\eps$ and drop terms that
vanish as $\eps\to 0$, and (iii) use $\yfrak \equiv y/(1{-}x)$.
This gives
\begin {equation}
   A^{\rm pole}_{x y y \bar x}(x,y) =
    \frac{\alphas^2}{4\pi^2} \,
    \frac{P(x)\,P(\frac{y}{1-x})}{1-x} \,
       i \barOmega_0
       \Bigl[
         - \tfrac{1}{\eps}
         - \ln\bigl(
               \tfrac{\pi\mu^2}{\barOmega_0E\vphantom{\bar{\bar\Omega}}}
           \bigr)
         + \tfrac12 \ln(x y z)
       \Bigr]
\label {eq:ApoleFund} .
\end {equation}
Taking $2\Re(\cdots)$ to add in the conjugate diagram gives
what we call $A^{\rm pole}_{\rm new}$ in (\ref{eq:Apolenew}).


\subsection {\boldmath$2\Re(x\bar y\bar y\bar x)$}
\label{sec:DeriveSEII}

We obtain $2\Re(x\bar y\bar y\bar x)$ from $2\Re(xyy\bar x)$ by
combined front-end and back-end transformation as depicted in
fig.\ \ref{fig:relateFund}.
The non-pole (subtracted) piece
$\int_0^\infty d(\Delta t) \> 2\Re B_{\rm new}(x,y,\Delta t)$ of
$2\Re(xyy\bar x)$ transforms to
\begin {equation}
  \int_0^\infty d(\Delta t) \>
  2\Re
  \bigl[
    B_{\rm new}\bigl(\tfrac{-x}{1-x},\tfrac{y}{1-x},\Delta t\bigr)
  \bigr]_{E \to (1{-}x)E} \,.
\label {eq:BnewWithEsub}
\end {equation}

For the same reasons as described for ${\cal A}_\seq$ in
(\ref{eq:AseqFE1}), the above can be rewritten as
\begin {equation}
  \int_0^\infty \frac{d(\widetilde{\Delta t})}{(1{-}x)^{1/2}} \>
  2\Re
  \bigl[
    B_{\rm new}\bigl(\tfrac{-x}{1-x},\tfrac{y}{1-x},\widetilde{\Delta t}\bigr)
  \bigr] .
\end {equation}
This result appears as the integral in (\ref{eq:Abarnew}).
Here we have called the integration variable $\widetilde{\Delta t}$
instead of $\Delta t$ for reasons similar to those described in
section \ref{sec:DeriveVirtSeq}, but here the relation is
\begin {equation}
   \widetilde{\Delta t} = (1{-}x)^{-1/2} \Delta t .
\label {eq:dttilde}
\end {equation}

For the pole piece, we could do the same thing, but we prefer to do
the transformation by hand in order to be careful about
issues concerning branch cuts.  Using fig.\ \ref{fig:relateFund},
the pole piece
(\ref{eq:ApoleFund}) for $xyy\bar x$ transforms to
\begin {align}
   A^{\rm pole}_{x \bar y \bar y \bar x}(x,y) ={} &
    (1{-}x)^{-\eps}
    \frac{\alphas^2}{4\pi^2} \,
    P(x)\,P(y) \,
    \biggl\{
       -i \Omega_0^*
       \Bigl[
         - \tfrac{1}{\eps}
         - \ln\bigl(\tfrac{\pi\mu^2}{-\Omega_0^*(1{-}x)E} \bigr)
         + \tfrac12 \ln\bigl( \tfrac{-xy(1{-}y)}{(1{-}x)^3} \bigr)
       \Bigr]
    \biggr\}^*
\nonumber\\
   ={} &
    \frac{\alphas^2}{4\pi^2} \,
    P(x)\,P(y) \,
       i \Omega_0
       \Bigl[
         - \tfrac{1}{\eps} - \ln\bigl(-\tfrac{\pi\mu^2}{\Omega_0 E} \bigr)
         + \tfrac12 \ln\bigl( -x y (1{-}x) (1{-}y) \bigr)
       \Bigr]
    .
\label {eq:ApoleFund2}
\end {align}

The arguments of the above logarithms have minus signs, and we need
to decide which branch of the logarithms they land us on.
The QED discussion given in appendix H.1 of ref.\ \cite{QEDnf}
applies equally well here.  That discussion tracks the origin
of the complex phases in direct calculations of what we
would call here $x \bar y \bar y \bar x$
relative to $x y y \bar x$.
The result is that
the $x \bar y \bar y \bar x$ diagram should have a phase of
$i^d$ relative to the $x y y \bar x$ diagram, which means
$\pm i^{-\eps}$ since the discussion did not keep track of overall signs.
Since
\begin {equation}
   i^{-\eps} \times \tfrac{1}{\eps}
   = \tfrac{1}{\eps} - \tfrac{i\pi}{2} ,
\end {equation}
this means that the branch-cut ambiguity in (\ref{eq:ApoleFund2})
resolves as
\begin {equation}
   A^{\rm pole}_{x \bar y \bar y \bar x}(x,y) =
    \frac{\alphas^2}{4\pi^2} \,
    P(x)\,P(y) \,
       i \Omega_0
       \Bigl[
         - \tfrac{1}{\eps} - \ln\bigl(\tfrac{\pi\mu^2}{\Omega_0 E} \bigr)
         + \tfrac12 \ln\bigl( x y (1{-}x) (1{-}y) \bigr)
         + \tfrac{i\pi}{2}
       \Bigr]
    .
\label {eq:ApoleFund2b}
\end {equation}
Finally, taking $2\Re(\cdots)$  of (\ref{eq:ApoleFund2b}) gives
what we call $\Abar^{\rm pole}_{\rm new}$ in (\ref{eq:Abarpolenew}).

There are alternative ways one could write our result for
$2\Re(x\bar y\bar y\bar x)$ that may be useful for some purposes.
If one wants a formula in terms of the actual duration $\Delta t$
of the self-energy bubble, one can make the change of variables
(\ref{eq:dttilde}) in (\ref{eq:Abarnew}) to write
\begin {equation}
  \Abar_\new(x,y) =
  \Abar_\new^{\rm pole}(x,y)
  + \int_0^\infty \frac{d(\Delta t)}{(1{-}x)} \>
   2\Re\bigl[
   B_\new\bigl(\tfrac{-x}{1-x},\tfrac{y}{1-x},(1{-}x)^{-1/2} \Delta t\bigr) \bigr]
  .
\end {equation}

Alternatively, going back to (\ref{eq:BnewWithEsub}), one could
use (\ref{eq:FEabcbar}) and scaling arguments similar to
(\ref{eq:F2virtIc}--\ref{eq:Cscale}) to write (\ref{eq:Abarnew})
as
\begin {equation}
  \Abar_\new(x,y) =
  \Abar_\new^{\rm pole}(x,y)
  + \int_0^\infty d(\Delta t) \>
      2\Re\bigl[
         D_\new(1{-}x,{-}y,-(1{-}y),x,
            \bar{\bar\alpha},\bar{\bar\beta},\bar{\bar\gamma},\Delta t)
      \bigr] .
\end {equation}
We have checked that this is the form one would get by directly
evaluating the $x\bar y\bar y\bar x$ diagram using our methods
\cite{2brem,seq,dimreg,QEDnf} instead of taking
our shortcut of front- and back-end transforming the $x y y \bar x$
diagram.


\section {Power-law IR divergences diagram by diagram}
\label {app:IRcancel}

In the main text, we merely demonstrated numerically that power-law
IR divergences cancel to leave a double log divergence.  In this
appendix, we give detailed diagram-by-diagram information on
the size of power-law divergences and give (with caveats)
analytic formulas.


\subsection{Individual Results}

The diagram-by-diagram power-law divergences are given in
table \ref{tab:limits}, which requires a bit of explanation
about why only a subset of diagrams are included, the notation
used in the table, and how to combine the various cases shown
to see the cancellation of power-law divergences.

\def\sep{;}
\def\pre[#1]{{}^{\rm(#1)}_{\vphantom{()}} ~}
\def\ppre[#1][#2]{{}^{\rm(#1)}_{\rm(#2)} ~}
\def\halfpre[#1]{{}^{\frac12\rm(#1)}_{\vphantom{()}} ~}
\def\halfppre[#1][#2]{{}^{\frac12\rm(#1)}_{\frac12\rm(#2)} ~}

\def\mythickrule{\cmidrule[1.2pt]{1-4}}
\def\mythickruleB{\cmidrule[1.2pt]{2-4}}
\def\mythinrule{\cmidrule[0.4pt]{1-4}}
\def\mydblrule{\cmidrule[1pt]{1-4}\\[-16pt]\cmidrule[1pt]{1-4}}

\begin {table}[tp]

\setlength{\tabcolsep}{7pt}
\begin{tabular}{@{}r@{}rrr@{}}
\mydblrule
  & \multicolumn{1}{c}{ $\mathbf{y{\to}0}$ }
  & \multicolumn{1}{c}{ $\mathbf{z{\to}0}$ }
  & \multicolumn{1}{c}{ $\mathbf{x{\to}0}$ }
  \\ \mythickruleB
  \multicolumn{4}{l}{$\mathbf{g\to ggg}$ \bfseries Real Diagrams: }
\\ \mythickrule
  \multicolumn{4}{l}{crossed (a minimal subset): }
\\\mythinrule\\[-1.3em]
$2\Re(y x \bar x \bar y)$
  & $ {+}2 \sep 0
           \sep {+}( \bm{G}{\bm{-}}\tfrac{\bm{\pi}}{\bm{2}} ) $
  & $ {+}2 \sep 0
           \sep {+}( \bm{G}{\bm{-}}\tfrac{\bm{\pi}}{\bm{2}} ) $
  & $\ppre[A3][B3] 0 \sep {+}1
           \sep {+}\bm{D} $
   \\ 
$2\Re(y \bar x x \bar y)$
  & $ {+}2 \sep 0
           \sep {+}( G{+}\tfrac{\pi}{2} ) $
  & $ \sim z^{-5/2} $
  & $\ppre[C1][C2] 0 \sep {+}1
           \sep {+}( \bm{D}{\bm{+}}\tfrac{\bm{\pi}}{\bm{2}} ) $
\\ 
$2\Re(x \bar y \bar x y)$
  & $ {-}2 \sep 0
           \sep {-}( G{+}\tfrac{\pi}{2} ) $
  & $ \sim z^{-5/2} $
  & $ {-}2 \sep 0
       \sep {-}( G{+}\tfrac{\pi}{2} ) $
\\[0.5em]
~~sum $2\Re(y \bar x x \bar y + x \bar y \bar x y)$
  &
  & $ {-}2 \sep {+}1 \sep {-}X $
  &
\\[0.5em]
\multicolumn{4}{l}{
  sequential: 
  one color routing of
    $2\Re(x y \bar x \bar y + x \bar x y \bar y + x \bar x \bar y y)$
}
\\ \mythinrule
${\cal A}_{\rm seq}(x,y)$
  & $ 0 \sep {-}1
        \sep {-}( \bm{D}{\bm{-}}\tfrac{\bm{\pi}}{\bm{4}} ) $
  & $ 0 \sep {-}1
        \sep {-}( \bm{D}{\bm{-}}\tfrac{\bm{\pi}}{\bm{4}} ) $
  & $ 0 \sep {-}1
        \sep {-}( D{-}\tfrac{\pi}{4} ) $
\\[0.5em] 
\multicolumn{4}{l}{ \bfseries Virtual Diagrams (Class I): }
\\ \mythickrule
\multicolumn{4}{l}{ crossed: }
\\ \mythinrule
$2\Re(y x \bar x y)$
  & $ {-}2 \sep 0
           \sep {-}( \bm{G}{\bm{-}}\tfrac{\bm{\pi}}{\bm{2}} ) $
  & $ {-}2 \sep 0
           \sep {-}( \bm{G}{\bm{-}}\tfrac{\bm{\pi}}{\bm{2}} ) $
  & $ 0 \sep {-}1 \sep {-}\bm{D} $
\\
$2\Re(y \bar x x y)$
  & $ {-}2 \sep 0
       \sep {-}( G{+}\tfrac{\pi}{2} ) $
       & $ \sim z^{-5/2} $
  & $ 0 \sep {-}1
       \sep {-}( \bm{D}{\bm{+}}\tfrac{\bm{\pi}}{\bm{2}} ) $
\\
$2\Re(\bar x y x y)$
  & $ {+}2 \sep 0
           \sep {+}( G{+}\tfrac{\pi}{2} ) $
           & $ \sim z^{-5/2}$
  & $ {+}2 \sep 0
           \sep {+}( G{+}\tfrac{\pi}{2} ) $
\\ 
$2\Re(y x y \bar x)$
  & $ {+}2 \sep 0
           \sep {+}( G{-}\tfrac{\pi}{2} ) $
  & $\ppre[A2][B1] 0 \sep {+}1 \sep {+}D$
  & $ {+}2 \sep 0
           \sep {+}( G{-}\tfrac{\pi}{2} ) $
\\[0.5em]
~~sum $2\Re(y \bar x x y + \bar x y x y)$
  &
  & $ {+}2 \sep {-}1 \sep {+}X $
  &
  \\[0.5em] 
\multicolumn{4}{l}{
  back-end sequential:
}
\\ \mythinrule
$- \tfrac12 [ {\cal A}_\seq(x,y) + {\cal A}_\seq(x,z) ]$
  & $0 \sep {+}1
       \sep {+}( \bm{D}{\bm{-}}\tfrac{\bm{\pi}}{\bm{4}} ) $
  & $0 \sep {+}1
       \sep {+}( \bm{D}{\bm{-}}\tfrac{\bm{\pi}}{\bm{4}} ) $
  & $0 \sep {+}1
       \sep {+}( D{-}\tfrac{\pi}{4} ) $
\\[0.5em] 
\multicolumn{4}{l}{ virtual $x y y \bar x$: }
\\ \mythinrule \\[-1.3em]
$2\Re(x y y \bar x)$
  & $\halfppre[A1][B2] 0 \sep {-}1 \sep {-}D $
  & $\halfppre[A1][B2] 0 \sep {-}1 \sep {-}D $
  & $ {-}2 \sep 0
           \sep {-}( G{-}\tfrac{\pi}{2} ) $
\\[0.7em] 
\multicolumn{4}{l}{
  \bfseries Virtual Diagrams (Class II):
}
\\ \mythickrule
\multicolumn{4}{l}{
  front-end sequential:
}
\\ \mythinrule
  see eq.\ (\ref{eq:AseqFE})
  & $0 \sep {+}1
       \sep {+}( {\bm D}{\bm +}\tfrac{\bm{3\pi}}{\bm 4} ) $
  & $\sim z^0$
  & $0 \sep {+}1
       \sep {+}( \bm{D}{\bm{-}}\tfrac{\bm{\pi}}{\bm{4}} ) $
\\[1em] 
\multicolumn{4}{l}{ virtual $x \bar y \bar y \bar x$: }
  \\ \mythinrule \\[-1.3em]
$2\Re(x \bar y \bar y \bar x)$
  & $\quad\halfpre[C3] 0 \sep {-}1
           \sep {-}( D{+}\tfrac{\pi}{2} ) $
  & $ \sim z^0$
  & $ {-}2 \sep 0
           \sep {-}( G{+}\tfrac{\pi}{2} ) $
   \\[0.5em]
\mydblrule
\end{tabular}
\caption{
   \label{tab:limits}
   The limiting behaviors of different diagrams.
   Format explained in the text.
}
\end{table}

An entry of the form ``$T\sep U\sep V$'' means that the limiting behavior
of the unrenormalized result is
\begin {equation}
   \frac{ \CA \alphas^2 P(\xi) }
        { 4\pi^2 (\texttt{small})^{3/2}  }
   \sqrt{\frac{\qhatA}{E}} \,
   \left[
     T \left(
       \frac{1}{\eps} + \ln\Bigl(\frac{\mu^2}{(\qhatA E)^{1/2}}\Bigr)
     \right)
     +
     \tfrac12 \, U \, \ln(\texttt{small})
     +
     V(\xi)
   \right]
\label {eq:form}
\end {equation}
where ``\texttt{small}'' is the variable that is going to zero (e.g.\ $y$),
and $\xi$ is either of the two other variables, which are being held
fixed (e.g.\ $\xi=x$ fixed as $y\to 0$, causing $z \to 1{-}x$).
All of the limits shown in the table turn out to be symmetric under
$\xi \to 1{-}\xi$.%
\footnote{
   This is a special feature of the power-law IR divergences.
   There is no similar diagram-by-diagram $\xi \to 1{-}\xi$
   symmetry for the IR log divergences, as demonstrated by
   the lack of such symmetry for the circles in
   fig.\ \ref{fig:SingleLogs}.
}
The function designations $V$ in the table are \textbf{bolded} if
we've worked them out analytically and not just numerically.
Those that are not bolded indicate
cases where we have not taken the time to derive an analytic result
but have instead extracted the function $V$ with numerics to
roughly 5 digit precision for the specific case $\xi=0.3$ and
noticed that $V$ is numerically the same as a bolded case.
The functions $V$ listed in the table include
\begin {align}
  D(\xi) &=
  \tfrac12 \ln\bigl( \tfrac{1}{\xi} {+} \tfrac{1}{1-\xi} {-} 1 \bigr)
  + \ln(2\pi)
  - \gammaE ,
\label {eq:Dxi}
\\
  G(\xi) &=
  -\ln\bigl( \xi(1{-}\xi) \bigr)
  + 2\ln2
  + 3\ln\pi
  - \gammaE .
\label {eq:Gxi}
\end {align}
For $X(\xi)$, we have not yet derived an analytic formula.
At the moment, we only know that in leading-log approximation
for $\texttt{small} \ll \xi \ll 1$ (or, symmetrically,
for $\texttt{small}\!{}\ll 1{-}\xi \ll 1$), it is
\begin {align}
  X(\xi) &\approx -\tfrac12 \ln\bigl[\xi(1-\xi)\bigr] .
\label {eq:X}
\end {align}

Some individual entries are {\it more} divergent than the
$(\texttt{small})^{-3/2}$ of (\ref{eq:form}), but these more severe
divergences cancel between pairs of diagrams, leaving behind
a net $(\texttt{small})^{-3/2}$ divergence.
For example, the $z{\to}0$ limit of
$2\Re(y \bar x x \bar y)$ and $2\Re(x \bar y \bar x y)$ are marked
in the table as each diverging like $(\texttt{small})^{-5/2}$,
but we give a separate line in the table showing the net
divergence of their sum.

The table explicitly shows as ``$\pm\tfrac{m\pi}{n}$''
all contributions that arise from logs of complex phases, which
are commented on in section \ref{sec:error}.

The annotations (A1), (B3), etc.\ on some entries are just
comments to connect to the soft limits of those diagrams
considered in previous leading-log analyses of overlap
effects.  See section \ref{sec:ABC123} below for an explanation.


\subsection {Assembling \boldmath$y{\to}0$ limit of
             \boldmath$v(x,y) + \tfrac12 r(x,y)$}

In the table, we have entries for only three of the crossed
$g{\to}ggg$ diagrams (plus their conjugates).
The full set of crossed $g{\to}ggg$ diagrams (fig.\ \ref{fig:crossed})
consists of these three entries plus all possible permutations of the three
daughters $(x,y,z)$.
But those other cases can be read off from permutations of
what is included in the table.
For instance, the $y{\to}0$ limit of $2\Re(xy\bar y\bar x)$,
which is not listed in the table, corresponds by permutation symmetry
to the $x{\to}0$ limit of $2\Re(yx\bar x\bar y)$, which is listed.
We have chosen 
$y x \bar x\bar y$, $y\bar x x\bar y$, and $x \bar y\bar x y$
to be our three representative entries in the table
in order to highlight their direct back-end relation to the
virtual-diagram table entries for $y x\bar x y$, $y \bar x x y$,
and $\bar x y x y$: the corresponding rows of the table are
just the negative of each other.

The single table entry for $g{\to}ggg$ sequential diagrams shows
the ${\cal A}_{\rm seq}$ of (\ref{eq:Aseq}).
As discussed in ref.\ \cite{seq}, this corresponds to {\it one} of two
large-$\Nc$ color routings for the sum of the three diagrams shown
explicitly in the top line of fig.\ \ref{fig:seq} (plus their
conjugates).  The complete set of sequential diagrams and
color routings corresponds \cite{seq} to summing ${\cal A}_{\rm seq}$ over
all possible permutations of $(x,y,z)$, as made explicit in
(\ref{eq:dGammaseq}).

The total differential rate $\R(x,y)$ for $g \to ggg$ (\ref{eq:Rdef})
corresponds to
the sum over all six permutations of the table entries discussed above.
Because of the relationship between limits of those permutations,
the $y{\to}0$ limit of $\R(x,y)$ is then twice 
the sum of the results listed in all three columns
$y{\to}0$, $z{\to}0$, and $x{\to}0$ of the subset of $g{\to}ggg$
results given in the table.
Adding the $g{\to}ggg$ table entries together then gives
\begin {equation}
  \tfrac12 \R(x,y) \underset{y{\to}0}{\simeq} {}
  \bigl[ 0 \sep 0 \sep G{-}D{-}X{-}\tfrac{\pi}{4} \bigr]
  =
  \frac{ \CA \alphas^2 P(x) }
        { 4\pi^2 y^{3/2}  }
  \sqrt{\frac{\qhatA}{E}} \,
  \left[
    G(x) - D(x) - X(x) - \tfrac{\pi}{4}
  \right] .
\label {eq:Rlim}
\end {equation}

Now turn to the virtual diagrams listed in the table.
The Class I virtual crossed diagrams in the table correspond to
{\it all} of the virtual crossed diagrams (top line of
fig.\ \ref{fig:virtI} plus conjugates) --- there are no
permutations to add.
The Class I and Class II virtual sequential diagrams
are related by back-end and front-end
transformation to the $g{\to}ggg$ sequential diagrams discussed
above.  See section \ref{sec:DeriveVirtSeq} for a discussion.
Again there are no permutations to add, and the same is true
for the remaining virtual diagram entries $2\Re(xyy\bar x)$ and
$2\Re(x\bar y\bar y\bar x)$.

Because of the addition of ``$(y\leftrightarrow z)$'' in the
definition (\ref{eq:Vdef}) of $\V(x,y)$, the $y{\to}0$ limit
of $\V(x,y)$ will sum both the $y{\to}0$ and $z{\to}0$ (but not
$x{\to}0$) columns of the virtual diagram entries of the table,
with result%
\footnote{
  The contributions of just Class I diagrams or just Class II
  diagrams to (\ref{eq:Vlim})
  are $[0;0; {-}G{+}D{+}X]$ and
  $[0;0;{+}\tfrac{\pi}{4}]$ respectively.
}
\begin {equation}
  \V(x,y) \underset{y{\to}0}{\simeq} {}
  \bigl[ 0 \sep 0 \sep {-}G{+}D{+}X{+}\tfrac{\pi}{4} \bigr] ,
\label {eq:Vlim}
\end {equation}
which is the negative of (\ref{eq:Rlim}).
This is in detail how power-law IR divergences cancel in the combination
$\V(x,y) + \tfrac12 \R(x,y)$ presented in (\ref{eq:VRlimit}).

Note that we never made use of the $x{\to}0$ column for the virtual
diagrams.  Those entries do not add to zero.
These divergences (and the related $1{-}x{\to}0$ divergences for
class II diagrams) correspond to the blue lines in fig.\ \ref{fig:Gxylims}.
They do not cause divergences in the applications we have
discussed for the reasons described in section \ref{sec:xNLOdivs}.


\subsection {The diagrams responsible for double logs in earlier papers}
\label {sec:ABC123}

\begin {figure}[t]
\begin {center}
  \includegraphics[scale=0.6]{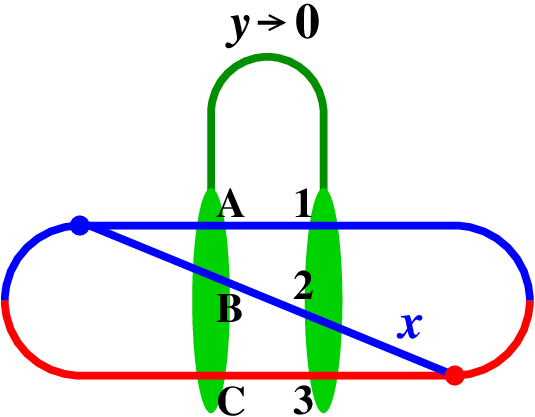}
  \caption{
     \label{fig:dbllog}
     Double-log diagrams.  The labels A, B, C, 1, 2, and 3 for where
     the soft $y$ gluon might connect the three harder gluons are
     provided for the sake of later naming individual diagrams.
     The green color of the soft gluon line is used to indicate that the
     line could be either blue or red depending on how it is
     connected. (See figs.\ \ref{fig:dbllogA} and \ref{fig:dbllog1},
     for example).
  }
\end {center}
\end {figure}

The diagrams that were analyzed in earlier papers \cite{Blaizot,Iancu,Wu}
that found the IR double logarithm correspond to the subset of 9 diagrams
(A1, A2, ..., C3)%
\footnote{
  This naming convention for these diagrams can be made to agree
  with that used by
  ref.\ \cite{Blaizot} if our names $xE$ and $yE$ for gluon
  energies are translated to their $zE$ and $\omega'$.
  In their notation, ref.\ \cite{Blaizot} works mostly in the limit
  $\omega' \ll \omega \equiv (1{-}z)E \ll E$.
}
depicted by fig.\ \ref{fig:dbllog}, where $y$ represents the softest
gluon in the process.
Here we comment on why our IR power-law divergences were absent
in their analysis.

The $y{\to}0$ limit of each of these diagrams
corresponds to the entries of table \ref{tab:limits} correspondingly
marked (A1), (A2), etc.
Some entries in the table correspond to more than one of these diagrams:
for example, the $x{\to}0$ limit of $2\Re(yx\bar x\bar y)$ is listed as
both (A3) and (B3).  That's because permutation symmetries relate this
to the $y{\to}0$ limit of both ${\rm A3} = xy\bar y\bar x$ and
${\rm B3} = zy\bar y\bar z$.  In other places, an entry may be listed as
giving only half of the corresponding contribution.  For example,
the table entries for both the $y{\to}0$ and $z{\to}0$ limits of
$2\Re(xyy\bar x)$ are listed as half of the $y{\to}0$ limit of
(A1).  That's just a combinatoric issue arising from our
labeling the two internal lines of the gluon
self-energy loop in the $xyy\bar x$ diagram in fig.\ \ref{fig:virtI}
as $y$ and $z=1{-}x{-}y$,
and in our table there are divergences associated with either becoming
soft.  In fig.\ \ref{fig:dbllog}, however, $y$ is by definition whichever
one of the two is softest.

The resulting $y{\to}0$ divergences for the diagrams
of fig.\ \ref{fig:dbllog} are collected in table \ref{tab:ABC123}.

\begin{table}[t]

\setlength{\tabcolsep}{7pt}
\begin{tabular}{lccc}
\hline\hline
  & 1 & 2 & 3
  \\ \cline{2-4}
A & $ 0 \sep {-}2 \sep {-}2D $
  & $ 0 \sep {+}1 \sep {+}D $
  & $ 0 \sep {+}1 \sep {+}D $
\\
B & $ 0 \sep {+}1 \sep {+}D $
  & $ 0 \sep {-}2 \sep {-}2D $
  & $ 0 \sep {+}1 \sep {+}D $
\\
C & $ 0 \sep {+}1 \sep {+}( D{+}\tfrac{\pi}{2} ) $
  & $ 0 \sep {+}1 \sep {+}( D{+}\tfrac{\pi}{2} ) $
  & $ 0 \sep {-}2 \sep {-}2( D{+}\tfrac{\pi}{2} ) $
  \\[0.5em]
\hline\hline
\end{tabular}
\caption{
   \label {tab:ABC123}
   The power-law IR
   divergences of diagrams A1, A2, $\cdots$, C3 extracted from table
   \ref{tab:limits}.
}
\end{table}

\begin {figure}[t]
\begin {center}
  \includegraphics[scale=0.33]{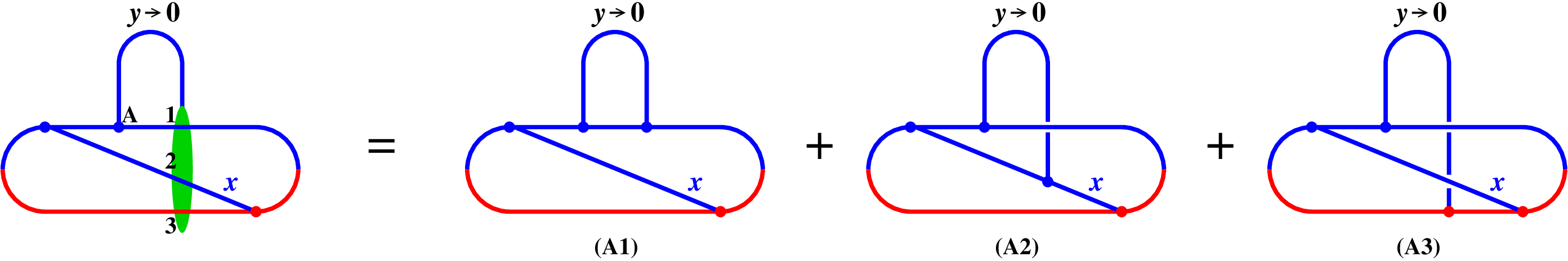}
  \caption{
     \label{fig:dbllogA}
     An example ${\rm A1}+{\rm A2}+{\rm A3}$
     of three diagrams whose power-law small-$y$
     behaviors cancel.
  }
\end {center}
\end {figure}

\begin {figure}[t]
\begin {center}
  \includegraphics[scale=0.4]{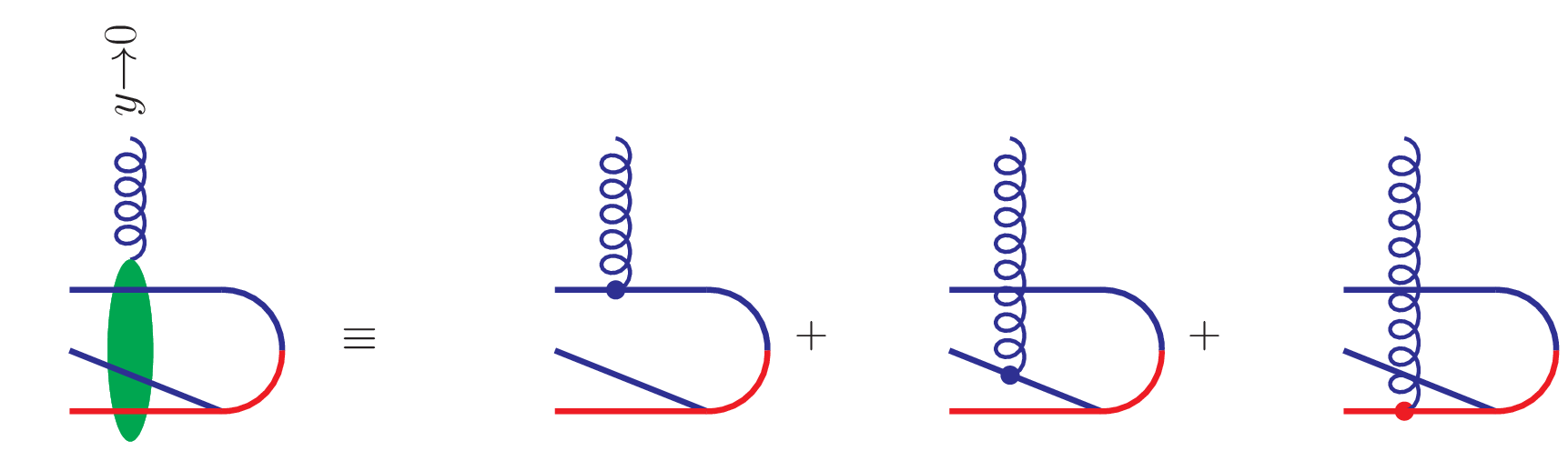}
  \caption{
     \label{fig:smally}
     Diagram elements whose leading $y{\to}0$ behaviors cancel in
     the small-$y$ limit.
  }
\end {center}
\end {figure}

Each row of table \ref{tab:ABC123} sums to zero.  Consider, for example,
the sum ${\rm A1}+{\rm A2}+{\rm A3}$ shown in fig.\ \ref{fig:dbllogA}.
The reason
for this cancellation is that the diagrams are identical except for
which line the blue $y{\to}0$ gluon couples to on the right-hand side,
and so the sum is proportional to the sum of those couplings, shown
in fig.\ \ref{fig:smally}.  Because the three hard particles form a
color singlet on the right-hand side of this diagram, the coupling
of the small-$y$ gluon to the collection of all three will be
suppressed compared to its coupling to any individual particle, which is
why the leading IR behavior (the power-law divergences) cancel
among these diagrams.

\begin {figure}[t]
\begin {center}
  \includegraphics[scale=0.33]{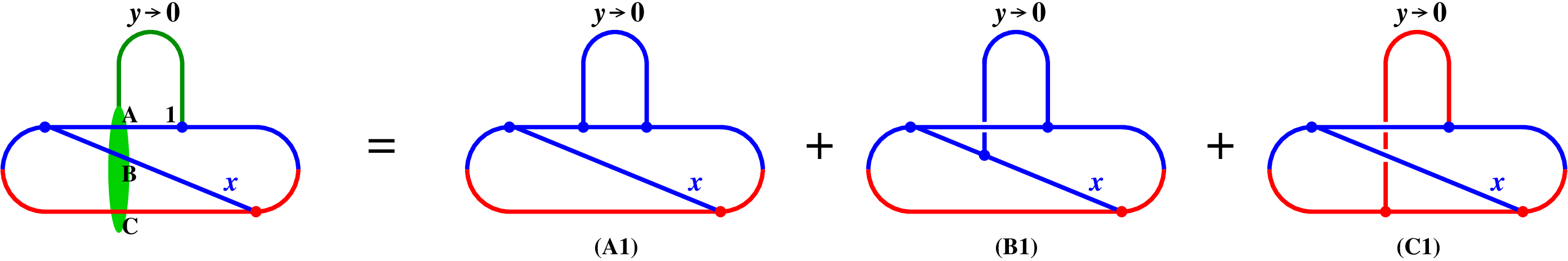}
  \caption{
     \label{fig:dbllog1}
     An example ${\rm A1}+{\rm B1}+{\rm C1}$
     of three diagrams whose leading small-$y$ behaviors
     do not cancel.
  }
\end {center}
\end {figure}

In contrast, it's interesting to note that
the columns of table \ref{tab:ABC123} do not sum individually to zero.
Consider, for example, the sum ${\rm A1}+{\rm B1}+{\rm C1}$
shown in fig.\ \ref{fig:dbllog1}.
They differ not only by which line the $y{\to}0$ gluon couples to
on the left-hand side of each diagram
but {\it also} by whether the $y{\to}0$ gluon
corresponds to a particle propagating in the amplitude (blue line) or
conjugate amplitude (red line), which changes the overall time evolution of
the diagram.
For this reason, one cannot simply factorize out the sum over vertex
couplings as we did for ${\rm A1}+{\rm A2}+{\rm A3}$,
and so there is no reason for
this particular sum of diagrams to be suppressed.

Regardless, the cancellation of each row of table \ref{tab:ABC123}
is sufficient to guarantee that there will be no power-law IR divergences
in the sum of all nine diagrams of fig.\ \ref{fig:dbllog}, which is
why earlier leading-log analyses did not need to address such
divergences.


\subsection {Derivation of \boldmath$D(\xi)$}
\label {app:Dxi}

Here we will give an example of the derivation of one of the
boldfaced $\bm D$'s in table \ref{tab:limits}.  We will focus on
the entry for the $x{\to}0$ limit of $2\Re(y x\bar x\bar y)$.  This is
the same, by permutation, as the $y{\to}0$ limit of $2\Re(x y\bar y\bar x)$,
to which we now turn since $x y\bar y\bar x$ is the canonical crossed
diagram presented in earlier work \cite{2brem,dimreg}.


\subsubsection {Spurious $y^{-2}$ divergence of $2\Re(xy\bar y\bar x)$}

Let's look first at the $\Delta t$ integral associated with the
$x y\bar y\bar x$ diagram, which is the term
\begin {equation}
  \int_0^\infty d(\Delta t) \>
  2\Re \bigl[ C({-}1,y,z,x,\alpha,\beta,\gamma,\Delta t) \bigr]
\label {eq:Dxiint}
\end {equation}
of (\ref{eq:summaryA}) and (\ref{eq:summaryB}), where $C$ is given
by (\ref{eq:summaryC}) in terms of the $D$ of (\ref{eq:summaryD}).
One finds that the integral is dominated by $\Delta t \sim y$ for
small $y$.%
\footnote{
\label {foot:loglinear}%
  A quick, initial way to figure out the scaling of the dominant contribution
  is to make a numerical log-linear plot of $\Delta t$ times the
  integrand vs.\ $\Delta t$
  for two extremely small values of $y$ and see how the most
  prominent feature of the plot scales with $y$.
  Because of large round-off error associated with delicate
  subtractive cancellations in our formulas for small $\Delta t$,
  we found this method requires using much higher precision numerics
  than standard machine precision in order to get good results for
  the integrand at extremely small value of $y$ and $\Delta t$.
}
An analytic analysis of the integrand for $\Delta t \sim y \to 0$
yields%
\footnote{
  \label {foot:Dxi}%
  In particular, $D$ is dominated for $\Delta t \sim y$
  by the $2\gamma Z_{\yx\ybx} I_1$ and
  $\gamma \bar Y_{\yx\ybx} Y_{\yx\ybx} I_2$
  terms of (\ref{eq:summaryD});
  these $(X,Y,Z)$ are individually given by
  the $1/\Delta t$ terms shown
  in eq.\ (D.2) of ref.\ \cite{2brem}, but the combination
  $
   X_\yx X_\ybx - X_{\yx\ybx}^2 \simeq
   -\frac{x^2 y M_0 E}{(\Delta t)^2}
    \bigl[ 1 + \frac{2i\Omega_0(1{-}x)\,\Delta t}{xy} \bigr]
  $;
  $I_2 \simeq I_0$; and
  $\gamma \simeq 2 P(x) / x^2(1{-}x)^3y^3\CA$.
}
\begin {equation}
  D({-}1,y,z,x,\alpha,\beta,\gamma)
  \simeq
  D_{\rm approx}
\end {equation}
with
\begin {equation}
  D_{\rm approx} =
  - \frac{\CA\alphas^2\,P(x)}{4\pi^2 y(\Delta t)^2}
  \left[
    \ln\left( \frac{xy}{1{-}x} + 2i\Omega_0\,\Delta t \right)
    + \left( 1 + \frac{2i\Omega_0(1{-}x)\,\Delta t}{xy} \right)^{-1}
  \right] ,
\label {eq:Dapprox}
\end {equation}
where $\Omega_0 = \Omega_{-1,x,1-x}$ as in (\ref{eq:Om0def}).
Subtracting the vacuum ($\hat q \to 0$ and so $\Omega_0 \to 0$)
gives
$C({-}1,y,z,x,,\alpha,\beta,\gamma)
 \simeq C_{\rm approx}$
with
\begin {equation}
  C_{\rm approx} =
  - \frac{\CA\alphas^2\,P(x)}{4\pi^2 y(\Delta t)^2}
  \left[
    \ln\left( 1 + \frac{2i\Omega_0(1{-}x)\,\Delta t}{xy} \right)
    + \left( 1 + \frac{2i\Omega_0(1{-}x)\,\Delta t}{xy} \right)^{-1}
    - 1
  \right] .
\label {eq:Capprox}
\end {equation}
One can rewrite the above
as a total divergence,
\begin {equation}
  C_{\rm approx} =
  \frac{\CA\alphas^2\,P(x)}{4\pi^2 y} \,
  \frac{d}{d(\Delta t)}
  \left[
    \frac{
       \ln\left( 1 + \frac{2i\Omega_0(1{-}x)\,\Delta t}{xy} \right)
    }{
       \Delta t
    }
  \right] ,
\end {equation}
and so do the integral and then take $2\Re(\cdots)$ to find
the leading $y{\to}0$ behavior
\begin {equation}
  \int_0^\infty d(\Delta t) \> 2\Re C_{\rm approx}
  = 
  - \frac{\CA\alphas^2(1{-}x)\,P(x)}{\pi^2 x y^2} \,
  \Re(i \Omega_0)
\label{eq:Ddiv1}
\end {equation}
of the $\Delta t$ integral for $2\Re(xy\bar y\bar x)$.
This is a $y^{-2}$ divergence, which would dominate over the $y^{-3/2}$
divergences of table \ref{tab:limits} except that (\ref{eq:Ddiv1})
exactly cancels the $y{\to}0$ limit of the pole term for
$2\Re(xy\bar y\bar x)$.  This pole term \cite{dimreg} represents the portion
of $A^{\rm pole}$ (\ref{eq:Apole}) attributable to that diagram.
The piece of the pole term responsible for the $y^{-2}$ divergence
is $2\Re(\cdots)$ of the $-2\gamma$ term in (\ref{eq:IIxyYbXb}).

So we need not
worry about the canceling $y^{-2}$ divergences {\it except}\/
that (\ref{eq:Capprox})
hides a sub-leading $y^{-3/2}$ divergence of the integral.
(Such cancellations make us wonder whether there is some
more elegant analysis of diagrams that would give simpler
formulas that more directly reveal the physics of the
$y{\to}0$ limit.)


\subsubsection {The surviving $y^{-3/2}$ divergence}

The contributions $2\Re(C-C_{\rm approx})$
to the $\Delta t$ integrand that are not accounted for by
$2\Re C_{\rm approx}$ above are dominated%
\footnote{
  One may use the same method as footnote \ref{foot:loglinear}.
}
by $\Delta t \sim y^{1/2}$.
Physically, this
corresponds
to $\Delta t \sim t_{\rm form}(y)$, where $t_{\rm form}(y)$ is
the formation time associated with bremsstrahlung of a soft $y$
gluon.

Repeating the analysis of the small-$y$ expansion of $D$
but now for $\Delta t \sim y^{1/2}$
instead of $\Delta t \sim y$, we find%
\footnote{
  Not much changes from the previous derivation for
  $\Delta t \sim y$ except that (i) some of the terms that were
  important for $\Delta t \sim y$ can be ignored for
  $\Delta t \sim y^{1/2}$, and (ii) it is no longer possible to
  take the small-$\Delta t$ approximation
  to $\Omega_+ \csc(\Omega_+\,\Delta t)$ when calculating $Z_{\yx\ybx}$.
  In particular, we find that $\Omega_+$ is of order the inverse
  $y$-formation time for small $y$, so that $\Omega_+\, \Delta t \ll 1$
  for the previous
  case $\Delta t \sim y$ but $\Omega_+\,\Delta t \sim 1$
  for the $\Delta t \sim y^{1/2}$ case
  here.  This point only matters for $Z_{\yx\ybx}$ since we find that the
  small-$y$ limits of the relevant $(X,Y)$'s
  are not sensitive to $\Omega_+ \csc(\Omega_+ \Delta t)$.
}
\begin {equation}
  D \simeq
  - \frac{\CA\alphas^2\,P(x)}{4\pi^2 y} \,
  [\Omega_y \csc(\Omega_y\,\Delta t)]^2
  \ln(2 i \Omega_0\,\Delta t) ,
\end {equation}
where
\begin {equation}
  \Omega_y \equiv \sqrt{ \frac{-i \qhatA}{2 y E} } .
\label {eq:Omy}
\end {equation}
Comparing to the already-accounted-for $D_{\rm approx}$ of (\ref{eq:Dapprox}),
and remembering that now $\Delta t \sim y^{1/2}$,
\begin {equation}
  D \simeq
  D_{\rm approx} + \delta C
\end {equation}
with
\begin {equation}
  \delta C \equiv
  - \frac{\CA\alphas^2\,P(x)}{4\pi^2 y} \,
  \left(
    \bigl[\Omega_y \csc(\Omega_y\,\Delta t)\bigr]^2 - \frac{1}{(\Delta t)^2}
  \right)
  \ln(2 i \Omega_0\,\Delta t) .
\label {eq:deltaC}
\end {equation}
We've called it $\delta C$ instead of $\delta D$ because it already vanishes
in the vacuum limit $\hat q {\to} 0$, which takes both
$\Omega_y$ and $\Omega_0$ above to zero.  So the vacuum subtraction
has no effect on this contribution to $D$.

The $y^{-3/2}$ divergence of $2\Re(x y \bar y \bar x)$ will now come from
taking the integral over $\Delta t$ of $2\Re\delta C$.
By changing integration variable to $\tau \equiv i \Omega_y \, \Delta t$,
which runs from $0$ to $e^{i\pi/4} \infty$, and then arguing that one
can safely add a contour at infinity to deform the integral to be
from $0$ to $+\infty$, one gets
\begin {equation}
  2\Re\left[ \frac{d\Gamma}{dx\,dy} \right]_{x y \bar y\bar x}
  \simeq
  \frac{\CA\alphas^2\,P(x)}{2\pi^2 y} \,
  \Re\left[
    i\Omega_y \int_0^\infty d\tau \>
    \Bigl( \frac{1}{\tau^2} - \frac{1}{\sh^2\tau} \Bigr)
    \ln\Bigl( \frac{2\Omega_0}{\Omega_y} \, \tau \Bigr)
  \right] .
\end {equation}
The integral formula%
\footnote{
  See appendix \ref{app:details} for (\ref{eq:DxiIntegral}).
}
\begin {equation}
  \int_0^\infty d\tau \>
    \Bigl( \frac{1}{\tau^2} - \frac{1}{\sh^2\tau} \Bigr)
    \ln(a \tau)
  =
  \ln(\pi a) - \gammaE
\label {eq:DxiIntegral}
\end {equation}
then gives
\begin {equation}
  2\Re\left[ \frac{d\Gamma}{dx\,dy} \right]_{x y \bar y\bar x}
  \simeq
  \frac{\CA\alphas^2\,P(x)}{2\pi^2 y} \,
  \Re\left(
    i\Omega_y
    \left[
       \ln\Bigl( \frac{2\pi\Omega_0}{\Omega_y} \Bigr)
       - \gammaE
    \right]
  \right) .
\label {eq:A3integrate0}
\end {equation}
In the style of (\ref{eq:form}), this is
\begin {equation}
   \frac{ \CA \alphas^2 P(x) }
        { 4\pi^2 y^{3/2}  }
   \sqrt{\frac{\qhatA}{E}} \,
   \left[
     \tfrac12 \ln y
     +
     D(x)
   \right]
\label {eq:Dxiderive}
\end {equation}
with $D(\xi)$ determined in this derivation to be (\ref{eq:Dxi}).
Permuting $x\leftrightarrow y$ in (\ref{eq:Dxiderive}) gives the
entry in table \ref{tab:limits} for $2\Re(y x\bar x\bar y)$ as
$x\to 0$.


\subsection {Derivation of \boldmath$G(\xi)$}
\label {app:Gxi}

Now we give an example of the derivation of one of the
boldfaced $\bm G$'s in table \ref{tab:limits}.  We focus on
the entry for the $y{\to}0$ limit of $2\Re(y x\bar x\bar y)$, which by
permutation is the $x{\to}0$ limit of the same canonical crossed
diagram $2\Re(x y\bar y\bar x)$ analyzed in the previous subsection.


\subsubsection {Spurious $x^{-5/2}$ divergence of $2\Re(xy\bar y\bar x)$}

Similar to the $y{\to}0$ limit of $2\Re(xy\bar y\bar x)$ studied in
section \ref{app:Dxi}, the $\Delta t$ integral (\ref{eq:Dxiint}) also
generates a spurious dominant divergence in the $x{\to}0$ limit.
In this case, the integral is dominated by $\Delta t \sim x^{3/2}$,
for which%
\footnote{
  \label {foot:Gxi}%
  The situation is similar to footnote \ref{foot:Dxi} except that here
  $
   X_\yx X_\ybx - X_{\yx\ybx}^2 \simeq
   -\frac{x^3 y(1{-}y) E^2}{(\Delta t)^2}
    \bigl[ 1 + \frac{2i\Omega_x(1{-}y)\,\Delta t}{xy} \bigr]
  $
  and
  $\gamma \simeq 2 P(y) / y^2(1{-}y)^3x^3\CA$.
}
\begin {equation}
  D_{\rm approx} =
  - \frac{\CA\alphas^2\,P(y)}{4\pi^2 x(\Delta t)^2}
  \left[
    \ln\left( \frac{xy}{1{-}y} + 2i\Omega_x\,\Delta t \right)
    + \left( 1 + \frac{2i\Omega_x(1{-}y)\,\Delta t}{xy} \right)^{-1}
  \right] ,
\label {eq:DapproxG}
\end {equation}
where
\begin {equation}
  \Omega_x \equiv \sqrt{ \frac{-i \qhatA}{2 x E} }
\end {equation}
is the small-$x$ limit of $\Omega_0$.
Correspondingly,
\begin {align}
  C_{\rm approx} &=
  - \frac{\CA\alphas^2\,P(y)}{4\pi^2 x(\Delta t)^2}
  \left[
    \ln\left( 1 + \frac{2i\Omega_x(1{-}y)\,\Delta t}{xy} \right)
    + \left( 1 + \frac{2i\Omega_x(1{-}y)\,\Delta t}{xy} \right)^{-1}
    - 1
  \right]
\nonumber\\
  &=
  \frac{\CA\alphas^2\,P(y)}{4\pi^2 x} \,
  \frac{d}{d(\Delta t)}
  \left[
    \frac{
       \ln\left( 1 + \frac{2i\Omega_x(1{-}y)\,\Delta t}{xy} \right)
    }{
       \Delta t
    }
  \right]
\label {eq:CapproxGLO}
\end {align}
When integrated, this generates an $x^{-5/2}$ contribution to the
$\Delta t$ integral, which is canceled by a similar contribution
from the pole term.
The relevant piece of the pole term
again comes from the $-2\gamma$ term in (\ref{eq:IIxyYbXb}).


\subsubsection {The surviving $x^{-3/2}$ divergence}

In this case, the dominant contribution to $2\Re[C-C_{\rm approx}]$
comes from two places.  One is $\Delta t \sim x^{1/2}$, which physically
corresponds to $\Delta t \sim t_{\rm form}(x)$.  The other is
sub-leading corrections to the $\Delta t \sim x^{3/2}$ region we just
analyzed above.

Let's start with $\Delta t \sim x^{1/2}$.
In this region, we find
\begin {equation}
  D_{(\Delta t \sim x^{1/2})} \simeq
  - \frac{\CA\alphas^2\,P(y)}{4\pi^2 x (\Delta t)^2} \,
  \ln( 1-e^{-2i\Omega_x\,\Delta t} ) .
\end {equation}
The difference of this with the already-accounted-for
$D_{\rm approx}$ of (\ref{eq:DapproxG}) is
\begin {equation}
  \delta C_{(\Delta t \sim x^{1/2})} \simeq
  - \frac{\CA\alphas^2\,P(y)}{4\pi^2 x (\Delta t)^2} \,
  \ln \Bigl(
     \frac{1-e^{-2i\Omega_x\,\Delta t}}{2i\Omega_x\,\Delta t}
  \Bigr) .
\end {equation}
Taking $\Delta t \to 0$ above, $2\Re(\delta C)$ diverges as
\begin {equation}
  2\Re \delta C_{(\Delta t \sim x^{1/2})} \approx
  \frac{\CA\alphas^2\,P(y)}{2\pi^2 x \, \Delta t}
  \Re(i\Omega_x) ,
\label {eq:deltaCG2}
\end {equation}
and so we cannot simply integrate $2\Re \delta C$ to find the
result we are interested in.
In general, the $1/\Delta t$ divergence of
individual diagrams is what created
the need for analyzing what we call pole terms
of diagrams.  In the current case, this divergence shows up
at an order in $x$ that makes it relevant to the integral of
$2\Re \delta C$.  We will need to subtract out the $1/\Delta t$ divergence
to get a convergent integral and then add the subtraction back in as part of
the pole term, as in (\ref{eq:Fsplit}).
Following (\ref{eq:F2toD2}), at this order in $x$
we will choose%
\begin {equation}
  {\cal D}_2 \simeq
  \frac{\CA\alphas^2\,P(y)}{2\pi^2 x}
  \Re\bigl[ i\Omega_x^3 \Delta t\,\csc^2(\Omega_x\,\Delta t) \bigr],
\label {eq:calD2G}
\end {equation}
whose $\Delta t\to 0$ behavior matches (\ref{eq:deltaCG2}).
Eq.\ (\ref{eq:calD2G}) is the same as taking the small-$x$ limit of
applying (\ref{eq:F2toD2}) to the more general small-$\Delta t$
result (\ref{eq:xyYbXbSmallDt}) for $2\Re(xy\bar y\bar x)$.
Defining $\tau \equiv i\Omega_x\,\Delta t$, the integral we want
is
\begin {align}
  2\Re
  \left[ \frac{d\Gamma}{dx\,dy} \right]_{x y \bar y\bar x}^{(\Delta t\sim x^{1/2})}
  &\simeq \int_0^\infty dt \>
     \bigl[ 2\Re\bigl( \delta C_{(\Delta t\sim x^{1/2})} \bigr) - {\cal D}_2 \bigr]
\nonumber\\
  &\simeq
  - \frac{\CA\alphas^2\,P(y)}{2\pi^2 x} \Re(i\Omega_x)
  \int_0^\infty d\tau \>
    \left[
      \frac{1}{\tau^2}
      \ln \Bigl(
        \frac{1-e^{-2\tau}}{2\tau}
      \Bigr)
      +
      \frac{\tau}{\sinh^2\tau}
    \right]
\nonumber\\
  &\simeq
  \frac{\CA\alphas^2\,P(y)}{2\pi^2 x} \Re(i\Omega_x)
  \bigl[ \ln(2\pi)-\gammaE \bigr] .
\label {eq:GxiIntegral1}
\end {align}
(See appendix \ref{app:details} for the last integral.)

Now turn back to $\Delta t \sim x^{3/2}$.  Carrying out the expansion of
$D$ to next order in $x$
(including the size of $\Delta t$ in the counting of order),
we find%
\footnote{
  We will not list intermediate steps here except to mention, as a checkpoint,
  that
  \[
    X_\yx X_\ybx - X_{\yx\ybx}^2 =
    -\frac{x^3 y z E^3}{(\Delta t)^2}
    \left[
      1 + (2+\altxi^{-1}) i \Omega_x \Delta t - \altxi^{-1} (\Omega_x \Delta t)^2
    \right]
    \bigl[ 1 + O(x^2) \bigr]
    ,
  \]
  which at leading order in $x$ matches the simpler formula of
  footnote \ref{foot:Gxi}.
} 
\begin {multline}
  D_{(\Delta x \sim x^{3/2})} =
  -\frac{\CA^2\alphas^2}{8\pi^2} \,
  (x y z)^2 (1{-}x)(1{-}y) \, \frac{\gamma}{(\Delta t)^2}
\\ \times
  \left[
    (1{+}\altxi)
    \ln\left( \frac{xy R}{(1{-}x)(1{-}y)} \right)
    + \frac{ (1 - 2i\Omega_x\,\Delta t) }{R}
  \right]
  \bigl[ 1 + O(x^2) \bigr]
 ,
\label {eq:DapproxGNLO}
\end {multline}
where
\begin {equation}
  R \equiv R_0 + \delta R ,
  \qquad
  R_0 \equiv 1 + \frac{i \Omega_x \Delta t}{\altxi} \,,
  \qquad
  \delta R \equiv \frac{(\Omega_x \Delta t)^2}{\altxi} \,,
  \qquad
  \altxi \equiv \frac{xy}{2z} \,.
\end {equation}
Note that $R_0$ is $O(1)$, but $\delta R$ and $\altxi$ are $O(x)$ and so small.
We could have more thoroughly written out the $x$ expansion of what
is shown explicitly in (\ref{eq:DapproxGNLO}), but keeping it in its
current form will be convenient.  For example, not explicitly expanding
$\gamma$ (\ref{eq:abc}) will make it simpler to see what parts of
this calculation eventually cancel with the pole terms at this
order in $x$.
Subtracting the vacuum limit from (\ref{eq:DapproxGNLO}) gives
\begin {multline}
  C_{(\Delta x \sim x^{3/2})} =
  -\frac{\CA^2\alphas^2}{8\pi^2} \,
  (x y z)^2 (1{-}x)(1{-}y) \, \frac{\gamma}{(\Delta t)^2}
\\ \times
  \left[
    (1{+}\altxi) \ln R
    + \frac{ (1 - 2i\Omega_x\,\Delta t) }{R}
    - 1
  \right]
  \bigl[ 1 + O(x^2) \bigr]
 .
\label {eq:CapproxGNLO}
\end {multline}
At leading order in $x$, this reproduces (\ref{eq:CapproxGLO}),
but (\ref{eq:CapproxGNLO}) correctly accounts for the next order
in $x$ as well.
At that order, ${\cal D}_2$ (\ref{eq:calD2G}) is relevant, and its subtraction
must be included as well, in order for the $\Delta t{\to}0$ integration to
converge.
It's convenient to use the leading-order conversion%
\footnote{
  We've used
  the leading-order relation
  $\gamma \simeq 2 P(y) / y^2(1{-}y)^3x^3\CA$
  of footnote \ref{foot:Gxi}.
}
\begin {equation}
   \frac{\CA^2\alphas^2}{4\pi^2}
   (x y z)^2(1{-}x)(1{-}y)\gamma
   =
   \frac{\CA\alphas^2\,P(y)}{2\pi^2 x}
   \bigl[ 1 + O(x) \bigr]
\end {equation}
to rewrite (\ref{eq:calD2G}) as
\begin {equation}
    {\cal D}_2 \simeq
    \frac{\CA^2\alphas^2}{4\pi^2}
    (x y z)^2(1{-}x)(1{-}y)\gamma
    \Re\bigl[ i\Omega_x^3 \Delta t\,\csc^2(\Omega_x\,\Delta t) \bigr] .
\label {eq:calD2G2}
\end {equation}
[The leading-order conversion is adequate because ${\cal D}_2$ is
already a sub-leading effect to our calculation of
$\int d(\Delta t) \> 2\Re(C-{\cal D}_2)$.]
For $\Delta t \sim x^{3/2}$, the argument of the $\csc$ is small,
so we may approximate
\begin {equation}
    {\cal D}_2 \simeq
    \frac{\CA^2\alphas^2}{4\pi^2 \, \Delta t}
    (x y z)^2(1{-}x)(1{-}y)\gamma
    \Re(i\Omega_x)
\label {eq:calD2G3}
\end {equation}
This matches the $1/\Delta t$
divergent behavior of $2\Re(C_{(\Delta x \sim x^{3/2})})$, as ${\cal D}_2$ should.
It is also convenient to switch from the $\Delta t$ variable, which is
$O(x^{3/2})$ in (\ref{eq:CapproxGNLO}), to the $O(1)$ variable
\begin {equation}
   \altT \equiv \frac{i\Omega_x\,\Delta t}{\altxi} \,,
\end {equation}
in terms of which
\begin {align}
  d(\Delta t) \>
  \bigl[ 2&\Re\bigl( C_{(\Delta x \sim x^{3/2})} \bigr) - {\cal D}_2 \bigr] 
  \simeq
\nonumber\\ &
  -\frac{\CA^2\alphas^2}{4\pi^2} \,
  (x y z)^2 (1{-}x)(1{-}y) \gamma
\nonumber\\ & \qquad \times
  \Re \left\{
    \frac{i\Omega_x}{\altxi} \, \frac{d\altT}{\altT^2}
    \left[
      (1{+}\altxi) \ln(1+\altT-\altxi \altT^2)
      + \frac{ (1 - 2\altxi \altT) }{1+\altT-\altxi \altT^2}
      - 1
      + \altxi \altT
    \right]
  \right\}
\nonumber\\ & \qquad \times
  \bigl[ 1 + O(\altxi^2) \bigr]
  .
\end {align}
Expansion in $x$ is now equivalent to expansion in $\altxi$.
Expanding explicitly to NLO in $\altxi$, we find
that we can rewrite the
argument of $\Re$ above as
\begin {equation}
  \frac{i\Omega_x}{\altxi} \, d\altT \times 
  \frac{d}{d\altT}
  \left[
    - \frac{(1+\altxi)}{\altT} \ln(1+\altT)
    + \frac{\altxi \altT}{(1+\altT)}
    \right] .
\end {equation}
Integration then gives
\begin {multline}
  2\Re
  \left[ \frac{d\Gamma}{dx\,dy} \right]_{x y \bar y\bar x}^{(\Delta t\sim x^{3/2})}
  \simeq
  \int_0^\infty d(\Delta t) \>
  \bigl[ 2\Re\bigl( C_{(\Delta x \sim x^{3/2})} \bigr) - {\cal D}_2 \bigr] 
\\
  =
  -\frac{\CA^2\alphas^2}{4\pi^2} \,
  (x y z)^2 (1{-}x)(1{-}y) \gamma
  \Re(i\Omega_x)
  \left( \frac{1}{\altxi} + 2 \right)
\label {eq:GxiIntegral2}
\end {multline}
through $O(x^{-3/2})$.

The last element we need is to extend analysis of the $O(x^{-5/2})$
pole terms to $O(x^{-3/2})$.  Since we have had to make the ${\cal D}_2$
subtraction above, we also need to add the ${\cal D}_2$ term back
to the pole terms as in (\ref{eq:Fsplit}).
Using (\ref{eq:calD2G2}) and $2\Re(\cdots)$ of
(\ref{eq:xyyxPole}--\ref{eq:IIxyYbXb}),
and expanding in $x$, we find
\begin {align}
  \lim_{\mbox{\small``$\scriptstyle{a\to 0}$''}} \Biggl\{ &
    2\Re \left[ \frac{d\Gamma}{dx\,dy} \right]_{xyy\bar x}^{(\Delta t < a)}
    +
    \int_a^\infty d(\Delta t) \> {\cal D}_2(\Delta t)
  \Biggr\}
  =
\nonumber\\ &
  \frac{\CA^2\alphas^2}{4\pi^2} \,
  (x y z)^2 (1{-}x)(1{-}y) \gamma
\nonumber\\ & \quad \times
  \Re\left\{
     \left[
       \frac{1}{\altxi}
       + \frac{2}{\eps}
       + 2\ln \Bigl( \frac{\mu^2}{i\Omega_x E} \Bigr)
       - \ln(e^{-i\pi} xyz)
       + 2 + 2 \ln\pi
     \right]
     i\Omega_x
   \right\}.
\end {align}
Adding this to the two contributions (\ref{eq:GxiIntegral1}) and
(\ref{eq:GxiIntegral2}) from $\int d(\Delta t) \> 2\Re[C-{\cal D}_2]$,
we see once again that the leading-order contributions
(represented here by the $1/\altxi$ terms) cancel, now leaving
the $O(x^{-3/2})$ result
\begin {multline}
  2\Re \left[ \frac{d\Gamma}{dx\,dy} \right]_{xyy\bar x}
  \simeq
  \frac{\CA^2\alphas^2}{4\pi^2} \,
  (x y z)^2 (1{-}x)(1{-}y) \gamma
  \Re\biggl\{
     \biggl[
       \frac{2}{\eps}
       + 2\ln \Bigl( \frac{\mu^2}{i\Omega_x E} \Bigr)
       - \ln(e^{-i\pi} xyz)
\\
       + \ln2 + 3 \ln\pi - \gammaE
     \biggr]
     i\Omega_x
   \biggr\}.
\end {multline}
Since the $O(x^{-5/2})$ pieces have canceled, we may now use
leading-order expressions for $z$ and $\gamma$ to get
\begin {multline}
  2\Re \left[ \frac{d\Gamma}{dx\,dy} \right]_{xyy\bar x}
  \simeq
  \frac{\CA\alphas^2\,P(y)}{2\pi^2 x} \,
  \Re\biggl\{
     \biggl[
       \frac{2}{\eps}
       + 2\ln \Bigl( \frac{\mu^2}{i\Omega_x E} \Bigr)
       - \ln\bigl(e^{-i\pi} xy(1{-}y)\bigr)
\\
       + \ln 2 + 3 \ln\pi - \gammaE
     \biggr]
     i\Omega_x
   \biggr\}.
\end {multline}
In the style of (\ref{eq:form}), this is
\begin {equation}
   \frac{ \CA \alphas^2 P(y) }
        { 4\pi^2 x^{3/2}  }
   \sqrt{\frac{\qhatA}{E}} \,
   \left[
     2 \left(
       \frac{1}{\eps} + \ln\Bigl(\frac{\mu^2}{(\qhatA E)^{1/2}}\Bigr)
     \right)
     +
     G(y)
     - \tfrac{\pi}{2}
   \right]
\label {eq:Gxiderive}
\end {equation}
with $G(\xi)$ determined in this derivation to be (\ref{eq:Gxi}).
Permuting $x\leftrightarrow y$ in (\ref{eq:Gxiderive}) gives the
entry in table \ref{tab:limits} for $2\Re(y x\bar x\bar y)$ as
$y\to 0$.

This has been a complicated derivation of $G(\xi)$.  Reassuringly, one
can confirm the final answer numerically by comparing to the
soft limit of our full numerical results for the diagram.


\section{Power-law IR cancellations in stopping distance formulas}
\label {app:lstop}

Consider moments $\langle \ell^n \rangle$ of the energy-weighted
distribution in distance $\ell$ of where energy is deposited by
a shower that stops in the medium.
Imagine also that splitting rates $d\Gamma$ scale with energy $E$
as some power $E^{-\nu}$, even though that is not precisely true
for NLO rates because of the double-log dependence in QCD.
Applied to our case of purely gluonic showers,
eqs.\ (A10) and (A12-A14) of ref.\ \cite{qedNfstop} give the
recursion relation
\begin {equation}
   \langle \tilde \ell^n \rangle =
   \frac{n \langle \tilde \ell^{n-1} \rangle}{M_{(n)}}
   \,,
\label {eq:recursionM}
\end {equation}
where
\begin {multline}
   M_{(n)} =
  \frac12 \int_0^1 dx\> \frac{d\tilde\Gamma_{g\to gg}}{dx} \,
    [ 1 - x^{1+n\nu} - (1{-}x)^{1+n\nu} ]
\\
  +
  \frac1{3!} \int_{y<1-x} dx\> dy\> \frac{d\tilde\Gamma_{g\to ggg}}{dx\,dy} \,
    [ 1 - x^{1+n\nu} - y^{1+n\mu} - z^{1+n\nu} ]
\label {eq:Mn}
\end {multline}
and $z \equiv 1{-}x{-}y$ in this presentation.
Above,
\begin {equation}
   \tilde\ell \equiv E^{-\nu} \ell,
   \qquad
   d\tilde\Gamma \equiv E^\nu \, d\Gamma ,
\label {eq:scale}
\end {equation}
and we do not notationally distinguish
$d\Gamma$ vs.\ $\Delta\,d\Gamma$.
We now show that
(\ref{eq:Mn}) can be written in terms of the $\R(x,y)$ and
$\V(x,y)$ defined in the main text and has the same
organization for the cancellation of power-law IR divergences.

Using the final-state permutation symmetries of $(x,1{-}x)$ for
$g \to gg$ and of $(x,y,z)$ for $g\to ggg$, (\ref{eq:Mn}) can
be rewritten as
\begin {equation}
   M_{(n)} =
  \frac12 \int_0^1 dx\> \frac{d\tilde\Gamma_{g\to gg}}{dx} \,
    [ 1 - 2 x^{1+n\nu} ]
  +
  \frac1{3!} \int_{y<1-x} dx\> dy\> \frac{d\tilde\Gamma_{g\to ggg}}{dx\,dy} \,
    [ 1 - 3 x^{1+n\nu} ]
\end {equation}
and thence
\begin {equation}
  M_{(n)} =
  \tilde\Gamma
  - \int_0^1 dx \> \left[ \frac{d\tilde\Gamma}{dx} \right]_{\rm net} x^{1+n\nu} ,
\label {eq:Mn2}
\end {equation}
where $\Gamma$ and $[d\Gamma/dx]_{\rm net}$ are given by
(\ref{eq:Gtot}) and (\ref{eq:dGnet}), here scaled by (\ref{eq:scale}).
The form (\ref{eq:Mn2}) is somewhat similar to the right-hand side
of (\ref{eq:Nevolve0}).  One may then mirror the steps from
(\ref{eq:Nevolve0}) to (\ref{eq:NevolveRV}) to obtain
\begin {subequations}
\label {eq:MnFinal}
\begin {equation}
  M_{(n)}
  =
  \hat{\cal S}^{\kern1pt\LObar} + \hat{\cal S}^{\kern1pt\NLObar}
\end {equation}
where
\begin {align}
  \hat{\cal S}^{\kern1pt\LObar}
  &=
  \frac12 \int_0^1 dx
    \left[ \frac{d\tilde\Gamma}{dx} \right]^{\LObar}
    \!\! [ 1 - x^{1+n\nu} - (1{-}x)^{1+n\nu} ]
\nonumber\\
  &=
  \int_0^1 dx
  \biggl\{
    \left[ \frac{d\tilde\Gamma}{dx} \right]^{\LObar}
    \!\! \theta(x > \tfrac12 )
    -
    \left[ \frac{d\tilde\Gamma}{dx} \right]^{\LObar}
    \!\! x^{1+n\nu}
  \biggr\}
\end {align}
and
\begin {multline}
  \hat{\cal S}^{\kern1pt\NLObar}
  =
  \int_0^1 dx \int_0^{1/2} dy
  \biggl\{
     \Bigl[
         \V(1{-}x,y) \, \theta(y<\tfrac{x}{2})
         + \tfrac12 \R(x,y) \, \theta(y<x) \, \theta(y<\tfrac{1-x}{2})
       \Bigr]
\\
     - \Bigl[
         \V(x,y) \, \theta(y<\tfrac{1-x}{2})
         + \V(1{-}x,y) \, \theta(y<\tfrac{x}{2})
         + \R(x,y) \, \theta(y<\tfrac{1-x}{2})
       \Bigr]
       x^{1+n\nu}
  \biggr\} .
\label {eq:SbarNLO}
\end {multline}
\end {subequations}
Similar to (\ref{eq:SevolveNLO}), the integrand in (\ref{eq:SbarNLO})
has no support for
$y{\to}1$ (with fixed $x$), and power-law divergences cancel as
$y{\to}0$.
Unlike (\ref{eq:SevolveNLO}), the integrand
in (\ref{eq:SbarNLO}) has support as $x{\to}0$.  However,
the terms that have such support are suppressed by the
$x^{1+n\nu}$ factor and so do not generate a divergent
$x$ integration.


\end {document}